\documentclass[floatfix,aps,prd,preprintnumbers,footinbib,bibnotes,superscriptaddress,letterpaper,twocolumn]{revtex4-1}
\usepackage{amsmath,amssymb,amsbsy,booktabs,array}
\usepackage{bm,comment}
\usepackage[pdftex]{graphicx,color}
\usepackage[colorlinks=true, linkcolor=blue, filecolor=blue, urlcolor=blue, citecolor=blue, pdftex=true, plainpages=false]{hyperref}
\usepackage{rotating}
\usepackage{amsmath,amssymb,braket,slashed,mathrsfs}
\usepackage[utf8]{inputenc}

\usepackage{array}
\newcolumntype{C}{>{$}c<{$}}
\newcolumntype{L}{>{$}l<{$}}

\makeatletter
\renewcommand{\p@subsection}{}
\renewcommand{\p@subsubsection}{}
\makeatother

\def\nn{\nonumber}
\def\be{\begin{equation}}
\def\ee{\end{equation}}
\def\bea{\begin{eqnarray}}
\def\eea{\end{eqnarray}}

\def\amu{a_\mu}
\def\amulohvp{a_\mu^\text{LO-HVP}}
\def\aelllohvp{a_\ell^\text{LO-HVP}}

\def\aelllohvp#1{a_{#1}^{\text{LO-HVP}}}
\def\aelllohvpf#1#2{a_{#1,#2}^{\text{LO-HVP}}}

\def\aelllohvpflat#1#2{a_{#1,#2,\text{lat}}^{\text{LO-HVP}}}
\def\dpert{\Delta^\text{pert}}
\def\dfv{\Delta^\text{FV}}

\def\Qmax{Q_\text{max}}

\def\reff#1{\ref{#1}}

\def\eq#1{Eq.~(\reff{#1})}
\def\eqs#1#2{Eqs.~(\reff{#1}) and (\reff{#2})}
\def\fig#1{Fig.~\reff{#1}}

\def\sec#1{Sec.~\reff{#1}}
\def\tab#1{Table~\reff{#1}}

\def\nn{\nonumber}

\def\mev{\mathrm{Me\kern-0.1em V}}
\def\gev{\mathrm{Ge\kern-0.1em V}}
\def\tev{\mathrm{Te\kern-0.1em V}}
\def\fm{\mathrm{fm}}
\def\re{\mathrm{Re}}
\def\im{\mathrm{Im}}

\def\msbar{{\overline{\mathrm{MS}}}}

\newcommand{\lsim}{ {\
\lower-1.2pt\vbox{\hbox{\rlap{$<$}\lower5pt\vbox{\hbox{$\sim$}}}}\ } }
\newcommand{\gsim}{ {\
\lower-1.2pt\vbox{\hbox{\rlap{$>$}\lower5pt\vbox{\hbox{$\sim$}}}}\ } }

\makeatletter
\DeclareRobustCommand{\text}{%
  \ifmmode\expandafter\text@\else\expandafter\mbox\fi}
\let\nfss@text\text
\def\text@#1{{\mathchoice
  {\textdef@\displaystyle\f@size{#1}}%
  {\textdef@\textstyle\f@size{#1}}%
  {\textdef@\textstyle\sf@size{#1}}%
  {\textdef@\textstyle \ssf@size{#1}}%
  \check@mathfonts
  }%
}
\def\textdef@#1#2#3{\hbox{{%
                    \everymath{#1}%
                    \let\f@size#2\selectfont
                    #3}}}
\makeatother

\newcommand{\marseille}{\affiliation{CNRS, Aix-Marseille Univ, Université de Toulon, CPT, UMR 7332, F-13288 Marseille, France}}
\newcommand{\wuppertal}{\affiliation{Department of Physics, Bergische Universit\"at Wuppertal, Gaussstrasse 20, D-42119 Wuppertal, Germany}}
\newcommand{\juelich}{\affiliation{J\"ulich Supercomputing Centre, Forschungszentrum J\"ulich, D-52425 J\"ulich, Germany}}
\newcommand{\budapest}{\affiliation{Institute for Theoretical Physics, E\"otv\"os University, P\'azm\'any P\'eter s\'et\'any 1/A, H-1117 Budapest, Hungary}}
\newcommand{\mdes}{\affiliation{CNRS, CEA, Maison de la Simulation, USR 3441, F-91191 Gif-sur-Yvette Cedex, France}}
\newcommand{\KMI}{\affiliation{Kobayashi-Maskawa Institute for the Origin of Particles and the Universe, Nagoya University, Nagoya 464-8602, Japan}}

\begin{document}
\bibliographystyle{apsrev}

\title{Hadronic vacuum polarization contribution to the anomalous magnetic moments of leptons from first
  principles}
\author{Sz.~Borsanyi}
\wuppertal
\author{Z.~Fodor}
\wuppertal
\budapest
\juelich
\author{C.~Hoelbling}
\wuppertal
\author{T.~Kawanai}
\juelich
\author{S.~Krieg}
\wuppertal
\juelich
\author{L.~Lellouch}
\marseille
\author{R.~Malak}
\marseille
\mdes
\author{K.~Miura}
\marseille
\KMI
\author{K.K.~Szabo}
\wuppertal
\juelich
\author{C.~Torrero}
\marseille
\author{B.C.~Toth}
\wuppertal
\collaboration{Budapest-Marseille-Wuppertal collaboration}\noaffiliation

\begin{abstract}

  We compute the leading, strong-interaction contribution to the
  anomalous magnetic moment of the electron, muon and tau using
  lattice quantum chromodynamics (QCD) simulations. Calculations
  include the effects of $u$, $d$, $s$ and $c$ quarks and are
  performed directly at the physical values of the quark masses and in
  volumes of linear extent larger than $6\,\fm$.  All connected and
  disconnected Wick contractions are calculated. Continuum limits are
  carried out using six lattice spacings. We obtain
  $\aelllohvp{e}=189.3(2.6)(5.6)\times 10^{-14}$,
  $\aelllohvp{\mu}=711.1(7.5)(17.4)\times 10^{-10}$ and
  $\aelllohvp{\tau}=341.0(0.8)(3.2)\times 10^{-8}$, where the first
  error is statistical and the second is systematic.

\end{abstract}

\maketitle


{\em Introduction.--} Ever since the discovery of the electron's spin
\cite{Gerlach:1922ur,Goudschmidt:1926ea}, the magnetic moments of
leptons have accompanied the development of quantum mechanics and
quantum field theory. This is particularly true of the small,
``anomalous,'' quantum corrections to these moments, $a_{\ell}$, where
$\ell$ denotes either the electron ($e$), the muon ($\mu$) or the tau
($\tau$) (see e.g. \cite{Jegerlehner:2009ry} for an
introduction). Today, $a_e$ is one of the most precisely measured
\cite{Hanneke:2008tm} and computed \cite{Aoyama:2012wk,Aoyama:2014sxa}
quantities in nature, with a total uncertainty below 1 ppb. Theory and
experiment agree and the measurement can be used to make the 
most precise determination of the fine-structure constant $\alpha$
\cite{Aoyama:2014sxa}.

In the case of the muon, the precision of the measurement
\cite{Bennett:2006fi} and of the standard model (SM) prediction
(e.g.\ \cite{Davier:2017zfy}) are roughly matched at around 0.5
ppm. However, theory and experiment disagree by more than
3 standard deviations. This is particularly enticing, because it could
be a sign of new, fundamental physics. The anomalous magnetic
moment of the muon is generically much more sensitive to new, massive
degrees of freedom than that of the electron. This is because, in many
extensions of the SM, the contributions of new particles are
proportional to the lepton mass squared, which is roughly $4\times 10^4$ times
larger for the muon. Moreover, a new experiment is beginning to take data at
Fermilab \cite{Holzbauer:2016cnd}, with the goal of reducing errors by
a factor of four, and another one is planned at J-PARC, with similar
objectives \cite{Otani:2015jra}.

The same argument should make $a_\tau$ even more interesting for new
physics searches: the $\tau$ mass is close to 17 times that of the
muon. However, its very short lifetime, of order $10^{-13}s$, has
meant that no direct measurement of $a_\tau$ has yet been made, though
a concrete proposal for doing so \cite{Fael:2013ij} is being
implemented \cite{Oberhof:2015hea}.

Theoretically, the leading source of uncertainty in the SM prediction
of $a_\mu$ is the leading order (LO) hadronic vacuum polarization
(HVP) contribution, $\amulohvp$, which is responsible for over 79\% of
the total error \cite{Davier:2017zfy}. This contribution also
dominates the uncertainty in $a_\tau$ \cite{Eidelman:2007sb} and for
$a_e$, has a total size of around 6.5 times the experimental error
\cite{Jegerlehner:2015stw}. Today this contribution is determined most
precisely using dispersion relations and the cross section of $e^+e^-$
to hadrons and/or the rate of hadronic $\tau$ decays
\cite{Eidelman:1995ny,Jegerlehner:2009ry,Davier:2010nc,Hagiwara:2011af,
  Miller:2012opa,Jegerlehner:2017lbd,Davier:2017zfy,Keshavarzi:2018mgv}. However,
since the pioneering work of \cite{Blum:2002ii}, lattice QCD
calculations of $\amulohvp$
\cite{Aubin:2006xv,Feng:2011zk,DellaMorte:2011aa,Burger:2013jya,Blum:2013qu,
  Gregory:2013taa,Malak:2015sla,Chakraborty:2014mwa,Blum:2015you,Chakraborty:2015ugp,
  Bali:2015msa,Blum:2016xpd,Chakraborty:2016mwy,Borsanyi:2016lpl,
  DellaMorte:2017dyu,Giusti:2017jof,Boyle:2017gzv,Chakraborty:2017tqp}
have made significant progress and provide a completely independent
cross-check that will become competitive in the coming years.

Here we present lattice QCD calculations of the LO-HVP contribution to
the anomalous magnetic moments of all three leptons. The calculations
include all contributions from $u$, $d$, $s$ and $c$ quarks, directly
at the physical values of their masses, in their quark-connected and
quark-disconnected configurations. Contributions from third generation
quarks can easily be estimated and are found to be much smaller than
our statistical errors, even for the $\tau$ that is most sensitive to them
(see e.g. \cite{Colquhoun:2014ica} for a calculation of the $b$
contribution to $\amulohvp$). Some previous lattice calculations of
$\aelllohvp{\ell}$, $\ell{=}e,\mu$
\cite{Burger:2013jya,Chakraborty:2015ugp}, and of
$\aelllohvp{\ell}$, $\ell{=}e,\mu,\tau$ \cite{Burger:2015oya}, included
all of these contributions, but involved difficult extrapolations to
the physical value of the average $u$-$d$ quark mass and only
estimated the disconnected parts. In the present Letter, we work
directly at the physical point and compute disconnected contributions
directly. Moreover, we implement a description of the lattice results
\cite{Bernecker:2011gh,Spraggs:2016jcx} that solves the small
virtuality issue \cite{Aubin:2012me} with finite-volume (FV) artifacts
that are exponentially suppressed in lattice size.

A unified treatment of the HVP contribution to the three lepton
anomalous magnetic moments provides important cross-checks that
validate the methods used. As the typical virtualities probed by these
moments are around $m_\ell^2/4$, the vast difference in the mass of
the leptons means that a large range of relevant scales are
checked. In particular, agreement between our results and
phenomenology in the electron case validates our understanding of
small virtualities, and of larger virtualities in the $\tau$ case. In
addition, the inclusion of all flavors up to the charm allows a
controlled matching onto perturbation theory. Thus, all energy scales
from zero to infinity are controlled in our calculation.

{\em Methodology.--} We consider the zero three-momentum, two-point function of
the quark electromagnetic current in Euclidean time $t$:
\be
\label{eq:Cmunu}
C_{\mu\nu}(t) = \frac1{e^2}\int d^3x\;\langle j_\mu(x) j_\nu(0)\rangle
\ ,\ee
with $e$ the positron charge, $x{=}(t,\vec{x})$ and
$j_\mu/e{=}\frac{2}{3} \bar{u}\gamma_\mu u - \frac{1}{3}
\bar{d}\gamma_\mu d - \frac{1}{3} \bar{s}\gamma_\mu s + \frac{2}{3}
\bar{c} \gamma_\mu c$. We work in the isospin limit,
$m_u{=}m_d$. Because $C_{\mu\nu}$'s flavor components are calculated
separately and have different statistical and systematic
uncertainties, it is useful to treat them separately. Physically, an
isospin separation is useful. Thus,
\bea
C_{\mu\nu}(t) &=& C_{\mu\nu}^{ud}(t) + C_{\mu\nu}^s(t) +  C_{\mu\nu}^c(t) +  C_{\mu\nu}^\text{disc}(t)\nn\\
& = &  C_{\mu\nu}^{I{=}1}(t) + C_{\mu\nu}^{I{=}0}(t)
\label{eq:cmunudef}
\ ,\eea
where in the top equality the first three terms correspond to the
quark-connected contractions of the light ($u$ and $d$ combined),
strange and charm quarks, and the fourth to the quark-disconnected
contractions of all four flavors. In the second
equality, the separation is made between isospin $I{=}1$ and $I{=}0$
contributions, given by $C_{\mu\nu}^{I{=}1}=\frac{9}{10}C_{\mu\nu}^{ud}$ and
$C_{\mu\nu}^{I{=}0}=\frac1{10}C_{\mu\nu}^{ud} +  C_{\mu\nu}^s + 
C_{\mu\nu}^c + C_{\mu\nu}^\text{disc}$.

It is
straightforward to obtain the corresponding LO-HVP contributions to the anomalous
magnetic moment of lepton $\ell$ from these 
correlation functions \cite{Lautrup:1971jf,deRafael:1993za,Blum:2002ii}:
\be
\label{eq:aell_int}
\aelllohvpf{\ell}{f} = \left(\frac{\alpha}{\pi}\right)^2\int_0^\infty 
\frac{dQ^2}{m_\ell^2}\;\omega\left(\frac{Q^2}{m_\ell^2}\right)\hat\Pi^f(Q^2)
\ ,\ee
with $\omega(r)=\pi^2\left[r+2-\sqrt{r(r+4)}\right]^2/\sqrt{r(r+4)}$, $\alpha=e^2/(4\pi)$
and where the scalar polarization function renormalized in the Thomson
limit is given by (see also \cite{Bernecker:2011gh})
\bea
\label{eq:pihat}
\hat\Pi^f(Q^2)&\equiv&\Pi^f(Q^2)-\Pi^f(0)\\
&=& \frac13\sum_{i=1}^3\int_0^\infty dt\;\left[t^2-\frac4{Q^2}\sin^2\left(\frac{Qt}2\right)\right]\re\,C_{ii}^f(t)\nn
\ .\eea
In \eqs{eq:aell_int}{eq:pihat}, the superscript $f$ can stand for
$ud,s,c,\text{disc},I{=}1,I{=}0$ and $\text{\textvisiblespace}$ where
the ``\textvisiblespace'' indicates that this equation also applies to
the full LO-HVP contribution. \eq{eq:pihat} implicitly includes the
subtraction of the polarization tensor $\Pi_{\mu\nu}(Q{=}0)$, which
was shown in \cite{Malak:2015sla} to be critical for reducing FV
effects and, through the factor $t^2$, the subtraction of the
polarization scalar $\Pi^f(0)$.

On a $T\times L^3$ lattice with spacing $a$, the integral over $t$ in
\eq{eq:pihat} is replaced by a sum, in increments of $a$, that runs up
to $T/2$, once the correlator $C_{ii}^f(t)$ has been averaged with
$C_{ii}^f(T-t)$. Moreover, the integral over $Q$ in \eq{eq:aell_int}
should, in principle, be replaced by a sum from $0$ to $\pi/a$ in
steps of $2\pi/T$. Here we keep the integral, but cut it off at a
value $Q{=}\Qmax$, chosen much smaller than $\pi/a$, so as to keep
discretization errors under control, but above which perturbation
theory can be applied. Then we decompose the anomalous magnetic
moments of the leptons into three terms:
\bea
\aelllohvpf{\ell}{f} &=& \aelllohvpf{\ell}{f}(Q{\le}\Qmax)
+ \gamma_\ell(\Qmax)\;\hat\Pi^f(\Qmax^2)\nn\\
\label{eq:aellsep}
&&+\dpert\aelllohvpf{\ell}{f}(Q{>}\Qmax)
\ ,\eea
where the low momentum contribution,
$\aelllohvpf{\ell}{f}(Q{\le}\Qmax)$, is obtained from the lattice as
described above, and where the last term is the high-momentum,
contribution renormalized at $\Qmax$ and computed in perturbation
theory \cite{SMPRL17}. The second term in \eq{eq:aellsep} is required
to shift the renormalization point from $\Qmax$ to $Q{=}0$. It is
obtained with lattice results for $C_{ii}^f(t)$, through \eq{eq:pihat}
with $Q{=}\Qmax$. $\gamma_\ell(\Qmax)$ is a known kinematical factor
\cite{SMPRL17}. In obtaining \eq{eq:aellsep}, it is assumed that
$\dpert\aelllohvpf{\ell}{f}(Q{>}\Qmax)$ is equal to the value that it
would have nonperturbatively. We check this by studying the dependence
of our results on the choice of $\Qmax$.

The replacement of the FV sum over $Q$ by the corresponding integral
is our choice of interpolation for the HVP function $\hat\Pi(Q^2)$. It
constitutes an alternative to e.g.\ the Padé approximation proposed in
\cite{Aubin:2012me}, the Marichev interpolation advocated in
\cite{deRafael:2017gay} or the finite-energy sum rule approach of
\cite{Dominguez:2017omw}. The integrand in \eq{eq:aell_int} has no
singularities in the region of integration. Therefore, an application
of Poisson's summation theorem guarantees that the corrections
entailed in replacing the sum by an integral over $Q$ are
exponentially small in $T$. These corrections will be accounted for in
our estimate of FV uncertainties.

{\em Lattice details.--} We employ a tree-level improved Symanzik
gauge action \cite{Luscher:1984xn} and a fermion action for
$N_f{=}2+1+1$ flavors of stout-smeared \cite{Morningstar:2003gk},
rooted, staggered quarks.  We have generated 15 ensembles at six
values of the bare coupling, $\beta$, corresponding to lattice
spacings ranging from $0.064$ to $0.134\,\fm$.  The average up and
down quark mass and the strange quark mass are tuned to around the
physical mass point defined using the Goldstone pion and kaon
masses. The charm quark mass is fixed in units of the strange mass to
$m_c/m_s{=}11.85$ \cite{Davies:2009ih}. The spatial dimensions of our
lattices are in the range $6.1$-$6.6\,\fm$ and the temporal ones in
the interval $8.6$-$11.3\,\fm$. The lattice spacing is fixed with
the pion leptonic decay constant, $f_\pi$. At each value of $\beta$,
between 450 and 3500 configurations, separated by 10 unit length
rational hybrid Monte Carlo (RHMC)~\cite{Clark:2006fx} trajectories,
are used. Details are given in \cite{SMPRL17} and more information
about the simulations can be found in \cite{Bellwied:2015lba}.

For the electromagnetic current correlator, we use the conserved
lattice current at the source and sink so that no renormalization is
necessary. We calculate the connected contributions to the correlators
using point sources.  We use the all-mode-averaging (AMA) technique of
\cite{Blum:2012uh} and 768 random source positions on each
configuration for the light quarks, 64 sources for the strange and 4
for the charm.  To compute the quark-disconnected contributions, we
apply AMA again, and exploit the approximate SU(3) flavor symmetry on
around 6000 stochastic sources
\cite{Francis:2014hoa,Blum:2015you}. These are random, four-volume
sources with which we compute the zero-momentum, time propagators,
correcting for bias. For the disconnected contribution of the charm we
apply a hopping parameter expansion.

{\em Analysis.--} Even with our high statistics, the signal
deteriorates quickly with increasing distance in our light and
disconnected correlators. Thus, in implementing
\eqs{eq:aell_int}{eq:pihat}, we introduce cuts, $t_c$, in time beyond
which we replace the correlator by the average of an upper and a lower
bound \cite{Borsanyi:2016lpl,SMPRL17}. $t_c$ is chosen such that the upper and
lower bounds agree well within statistical errors and where these
errors are not too large. The upshot is that our result for the light
contribution to $\aelllohvp{\ell}$ is obtained by summing the
integrand in \eq{eq:pihat} with $C_{ii}^{ud}(t)$ given by our lattice
data up to $t_c$ and performing the rest of the sum from $t{>}t_c$ to
$T/2$ with $C_{ii}^{ud}(t)$ replaced by the bound average. The results
of this procedure for $t_c$ in the range of $(3.000\pm 0.134)\,\fm$
are averaged to account for possible statistical fluctuations in the
correlator at a given $t_c$. The disconnected contribution to
$\aelllohvp{\ell}$ is obtained in an identical fashion, but with $t_c$
in the range $(2.600\pm 0.134)\,\fm$.

We limit the integral over $Q$ in \eq{eq:aell_int} to $\Qmax$, and use
perturbation theory to obtain the complement. We consider
$\Qmax^2{=}1,2,\cdots, 5\,\gev^2$. In what follows, quantities with
the subscript ``lat'' correspond to lattice results obtained in a
given simulation. Their dependence on lattice spacing and quark masses
will be left implicit. To extrapolate our results
$\aelllohvpflat{\ell}{f}(Q{\le} \Qmax)$ to the continuum limit and to
interpolate them to the physical mass point, we fit them to a function
which depends on the Goldstone pion and kaon masses squared, on the
$\eta_c$ mass and on the lattice spacing squared~\cite{SMPRL17}.
Since the simulations are performed close to the physical mass point,
a constant or linear dependence in the mass parameters is always
sufficient. Moreover, for all flavor contributions, good fit qualities
can be achieved with a linear $a^2$ dependence for all three leptons
and all values of $\Qmax$ considered here. Because taste violations
play an important role in the continuum extrapolation of
$\aelllohvpf{\ell}{ud}$, we have also tried correcting for these
effects using one-loop staggered chiral perturbation theory before
performing a continuum extrapolation \cite{Chakraborty:2016mwy}. While
the continuum extrapolation is significantly milder, the
continuum limit results obtained are consistent with the ones presented
here. Our continuum extrapolations are discussed in detail and
examples are shown in \cite{SMPRL17}. Here we emphasize that with
simulations at six lattice spacings down to $0.064\,\fm$, we have full
control over the continuum extrapolations.

This analysis yields the continuum extrapolated flavor quantities,
$\aelllohvpf{\ell}{f}(Q{\le} \Qmax)$, for the five values of $\Qmax$
considered. For each value of $\Qmax$, we sum the appropriate flavor
quantities, to get the corresponding $I{=}1,I{=}0$ and total, low-$Q$
contributions to the lepton anomalous magnetic moments. The results
for these and the individual flavor contributions are given in
\cite{SMPRL17}, with statistical and systematic errors obtained as
described below. To these contributions we add the corresponding
complements given in \eq{eq:aellsep}. These complements require the
computation of $\hat\Pi^f(\Qmax^2)$. This is done using \eq{eq:pihat}
and requires a continuum limit and physical mass point interpolation
very similar to that performed for $\aelllohvpflat{\ell}{f}(Q{\le}
\Qmax)$ \cite{SMPRL17}.

The perturbative contributions,
$\dpert\aelllohvpf{\ell}{f}(Q{>}\Qmax)$, are computed from results for
$\Pi^f(Q^2)$, with terms up to $O(\alpha_s^4)$, obtained using the
code \texttt{rhad} \cite{Harlander:2002ur}, as explained in
\cite{SMPRL17}. These corrections are below our statistical errors for
the $e$ and $\mu$, which have very little sensitivity to large $Q$,
but are significant for the $\tau$. In \cite{SMPRL17} we study the
$\Qmax$ dependence of our results for $\aelllohvpf{\ell}{f}$. The fact
that they are independent of $\Qmax\ge \sqrt2\,\gev$ within errors, in
particular for $\ell{=}\tau$, indicates that our continuum-limit,
lattice results are consistent with five-loop perturbation theory for
momenta, $Q$, above that value.

\begin{table*}
    \centering
    \begin{ruledtabular}
	\begin{tabular}{LCCC}
	    & \ell=e\; \mbox{(units of $10^{-14}$)} & \ell=\mu\; \mbox{(units of $10^{-10}$)} & \ell=\tau\; \mbox{(units of $10^{-8}$)}\\
	    \hline\\[-0.3cm]
            \aelllohvpf{\ell}{I{=}1} & 156.9(2.4)(2.1)(0.0)(0.0)(1.2)(4.6)        & 582.9(6.7)(7.2)(0.1)(0.0)(4.5)(13.5)    & 253.2(0.7)(1.4)(0.0)(0.1)(1.2)(1.8) \\
            \aelllohvpf{\ell}{I{=}0} & 30.7(1.2)(1.0)(0.1)(0.0)(0.2)              & 120.5(3.4)(3.5)(0.2)(0.0)(1.0)          & 84.4(0.4)(0.7)(0.0)(1.1)(3.4) \\
	    \aelllohvp{\ell}         & 189.3(2.6)(2.3)(0.1)(0.0)(1.5)(4.6)(1.6) & 711.1(7.5)(8.0)(0.2)(0.0)(5.5)(13.5)(5.1) & 341.0(0.8)(1.6)(0.0)(1.1)(1.5)(1.8)(1.1) \\
	\end{tabular}
    \end{ruledtabular}
    \caption
    { 
      \label{tab:final} LO-HVP contribution to
      the anomalous magnetic moments of the $e$, $\mu$ and $\tau$
      leptons.  The first two lines give our results for the
      $I{=}1$ and $I{=}0$ contributions.  The $I{=}1$ results include
      the FV corrections, which are negligible in the $I{=}0$ case.
      The last line displays our results for the total LO-HVP
      contribution. In addition to the terms included in the $I{=}1$
      and $I{=}0$ components, this total also accounts for QED and
      $m_d{\ne} m_u$ corrections.
      The first error on all results is statistical,
      the second is associated with the continuum extrapolation, 
      the third with our bounding procedure,
      the fourth with the matching to perturbation theory, the fifth with the 
      the lattice spacing uncertainty and,
      where applicable, the sixth with the FV correction and the seventh
      with the IB correction.  }
\end{table*}

{\em Systematic errors and results.--} The procedure described above
yields $\aelllohvpf{\ell}{f}$ for all $f$ and for all three
leptons. In our physical fits, the errors associated with the small
interpolations in mass are negligible. Those associated with the
continuum extrapolations are not. To estimate them, we impose four
cuts on the lattice spacing: no cut, and $a{\le} 0.118$, $0.111$,
$0.095\,\fm$. This number is reduced to three in the disconnected case
for which we have no results at $a=0.064\,\fm$. The systematic error
associated with the matching to perturbation theory is determined from
the results with $\Qmax^2{=}2,\cdots,5\,\gev^2$, covering the range
safely accessible to our lattice calculations and to perturbation
theory. The one associated with the time cut is determined by
considering $t_c$ ranges shifted by $-0.134\,\fm$ compared to those
given above. Over the total range of $t_c$ considered, the two-pion
bounds change by a factor close to 3. The final central value is
the unweighted average of all results. Each systematic error component
is chosen to cover all of the central values resulting from the
variation, over the ranges described above, of the variable associated
with that component. Furthermore, we add a $0.8\%$ systematic error to
our results for $\aelllohvpf{\ell}{f}(Q{\le}\Qmax)$ due to the
uncertainty in our determination of the lattice spacing
\cite{SMPRL17}. The statistical error is the jackknife error of the
central value over jackknife samples with bins of length 10
configurations. The results for the individual flavor contributions to
the magnetic moments of all three leptons are given in \cite{SMPRL17}.

In the absence of a systematic study with simulations in a variety of
volumes, only model estimates of FV effects can be made. As argued in
\cite{Aubin:2015rzx,Francis:2013qna}, for large volumes those effects
will be governed by pion contributions that can be computed in chiral
perturbation theory ($\chi$PT) \cite{Aubin:2015rzx}. Since the $I{=}0$
channel is dominated by three-pion exchange, the FV effects are
expected to be smaller than those of the $I{=}1$ contribution, which
are already small. Thus we consider only the latter. Our computation
of these effects is summarized in \cite{SMPRL17} and the appropriate
corrections are added to our $I{=}1$ and total results. They are
$4.6(4.6)\times 10^{-14}$ for the $e$ and $13.5(13.5)\times 10^{-10}$
for the $\mu$ with negligible $\Qmax$ dependence in the range of
interest. For the $\tau$ they range from $9.4(9.4)\times 10^{-9}$ to
$1.6(1.6)\times 10^{-8}$ for $\Qmax=1\div\sqrt5\,\gev$. We associate
with these corrections a 100\% uncertainty included in our error
budget.

Compared to phenomenological determinations of $\amulohvp$
\cite{Jegerlehner:2017lbd,Davier:2017zfy,Keshavarzi:2018mgv},
our $m_d{=}m_u$ calculation without QED is missing isospin-breaking (IB)
effects. These are detailed in \cite{SMPRL17}. Here we note that the
corrections to be added are $(1.7\pm 1.6)\times 10^{-14}$
for the $e$, $(7.8\pm 5.1)\times 10^{-10}$ for the $\mu$ and
$(3.4\pm 1.1)\times 10^{-8}$ for the $\tau$.

We quote our final results for $\aelllohvp{\ell}$ for all three
leptons in \tab{tab:final}. Combining all errors in quadrature, we
obtain $\aelllohvp{e}$ with an uncertainty of $3.3\%$,
$\aelllohvp{\mu}$ of $2.7\%$ and $\aelllohvp{\tau}$ of $1.0\%$. Not
surprisingly, the relative error increases with the sensitivity of the
anomalous moment to long-distance physics.

{\em Discussion.--} It is interesting to compare these results with
those in the literature. There are only two lattice QCD determinations
of the LO-HVP contribution to the muon anomalous moment which include
the contributions of quarks up to the charm
\cite{Burger:2013jya,Chakraborty:2016mwy}.  Compared to those, our
calculation is the only one in which the continuum extrapolation is
performed directly at the physical mass point and which includes a
reliable determination of the quark-disconnected contribution. There
exist also many precise phenomenological determinations of
$\amulohvp$, as discussed in the Introduction. Here we consider three
recent ones
\cite{Jegerlehner:2017lbd,Davier:2017zfy,Keshavarzi:2018mgv}.

\begin{figure}
    \centering
    \includegraphics[width=0.9\columnwidth]{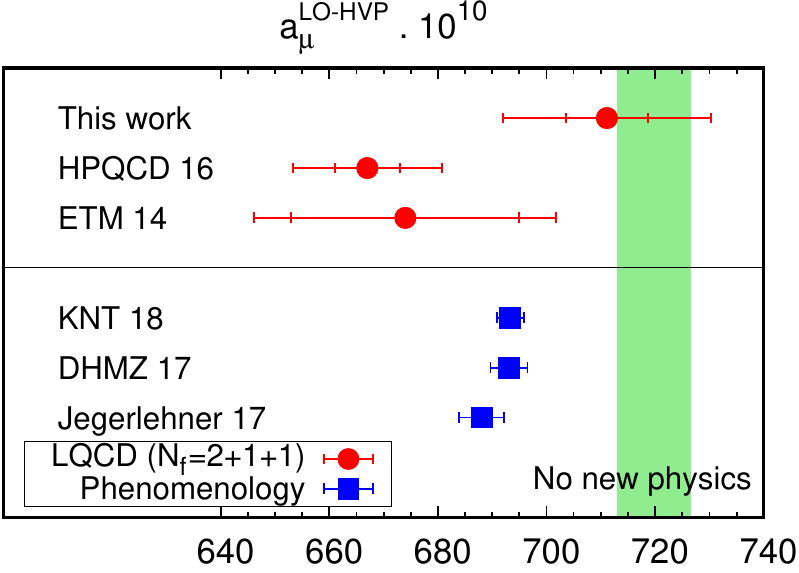}
    \caption
        {\label{fig:amu_compare} Comparison of
          our result for $\amulohvp$ with the only other two
          $N_f= 2+1+1$ lattice QCD calculations
          \cite{Burger:2013jya,Chakraborty:2016mwy} and with recent
          ones obtained from phenomenology
          \cite{Jegerlehner:2017lbd,Davier:2017zfy,Keshavarzi:2018mgv}. For
          the lattice results, the first error is statistical and the
          second is the total error, including systematics. The shaded
          region is the value that $\aelllohvp{\mu}$ would have to
          have to explain the experimental measurement of $\amu$,
          assuming no new physics.}
\end{figure}
            
We plot all of these results in \fig{fig:amu_compare}, together with
ours. Also shown on this plot is the value that $\aelllohvp{\mu}$
would have to have to explain the experimental measurement of $\amu$
\cite{Bennett:2006fi}, assuming that all other SM contributions are
unchanged, i.e. assuming no new physics (NP). Using the SM
contributions summarized in \cite{Davier:2017zfy}, we find
$\aelllohvp{\mu,\text{noNP}}=(720.0\pm 6.8) \times 10^{-10}$. The
errors on the lattice results, which are in the range of $2.0\%$ to
$4.1\%$ are substantially larger than those of the phenomenological
approach. Our result for $\aelllohvp{\mu}$ is larger than those of the
other lattice calculations and in slight tension with the one
from HPQCD \cite{Chakraborty:2016mwy} which is $1.9\,\sigma$ away. A
more detailed flavor-by-flavor comparison is given in
\cite{SMPRL17}. However, our result is consistent with those from
phenomenology within about 1 standard deviation, as well as with
$\aelllohvp{\mu,\text{noNP}}$. Thus, one will have to wait for the
next generation of lattice QCD calculations to confirm or
infirm the larger than $3\,\sigma$ deviation between the measurement
of $\amu$ and the prediction of the SM based on phenomenology.

Regarding $\aelllohvp{e}$, there are two other lattice calculations
\cite{Burger:2015oya,Chakraborty:2016mwy} and only one concerning
$\aelllohvp{\tau}$ \cite{Burger:2015oya}. The results in
\cite{Burger:2015oya} are $\aelllohvp{e}=1.782(64)(86)\times 10^{-12}$
and $\aelllohvp{\tau}=3.41(8)(6)\times 10^{-6}$ and in
\cite{Chakraborty:2016mwy}, $\aelllohvp{e}=1.779(39)\times
10^{-12}$. From the point of view of phenomenology, a dispersive
analysis similar to the one implemented for the muon gives
$\aelllohvp{e}=1.846(12)\times 10^{-12}$ \cite{Jegerlehner:2015stw}
and $\aelllohvp{\tau}=3.38(4)\times 10^{-6}$
\cite{Eidelman:2007sb}. Comparing these results to ours in
\tab{tab:final}, we find the following. The result of
\cite{Chakraborty:2016mwy} for $\aelllohvp{e}$ displays a tension with
ours which is slightly smaller than the one for $\aelllohvp{\mu}$. On
the other hand, our results are fully compatible with the
phenomenological ones, indicating that we control the physics of the
HVP over full range of $Q^2$. In addition, our result for
$\aelllohvp{e}$ has an error which is about half that of the lattice
result of \cite{Burger:2015oya} and for $\aelllohvp{\tau}$, it is
approximately 3 times more precise. In fact, our result for the
latter is more precise than the phenomenological one.

{\em Acknowledgments.--} We thank S.~D\"urr for contributions in the
early stages of this project and Z.~Zhang for noticing a typo in our
final result for $\aelllohvp{e}$. In addition, L.~L. thanks M.~Benayoun,
C.~Davies, F.~Jegerlehner, M.~Knecht, C.~Lehner, E.~de Rafael and
R.~Van de Water for informative discussions. Computations were
performed on JUQUEEN and JUROPA at Forschungszentrum J\"ulich, on
Turing at the Institute for Development and Resources in Intensive
Scientific Computing (IDRIS) in Orsay, on SuperMUC at Leibniz
Supercomputing Centre in M\"unchen, on Hermit at the High Performance
Computing Center in Stuttgart.  This project was supported, in part,
by the OCEVU Laboratoire d’excellence (ANR-11-LABX-0060) and the
A*MIDEX Project (ANR-11-IDEX-0001-02), which are funded by the
``Investissements d’avenir'' French government program and managed by
the ``Agence nationale de la recherche'' (ANR), by the DFG Grant
SFB/TR55, by the Gauss Centre for Supercomputing e.V and by the
GENCI-IDRIS supercomputing Grant No. 52275. R.~M. was supported in part
by a joint Ph.D.\ fellowship from the Centre national de la recherche
scientifique (CNRS) and the Commissariat à l’énergie atomique et aux
énergies alternatives (CEA).

{\em Note added.--} After submission of this Letter for publication,
  a preprint reporting on an $N_f=2+1$ lattice QCD calculation
  of $\aelllohvp{\mu}$ appeared \cite{Blum:2018mom}. That calculation includes
  a lattice computation of many isospin breaking effects. Its result
  for $\aelllohvp{\mu}$ is in excellent agreement with ours.


  
\newpage

$ $

\newpage


\section*{Supplemental Material}

\setcounter{figure}{0}
\setcounter{table}{0}
\setcounter{equation}{0}
\renewcommand\thefigure{S\arabic{figure}}
\renewcommand\thetable{S\arabic{table}}
\renewcommand{\theequation}{S\arabic{equation}}

\subsection{Lattice parameters}

The work presented here is based on 15, $N_f=2+1+1$ simulations
performed around the physical mass point and at six values of the
bare coupling, on lattices of spatial extents larger than
$6\,\fm$. The parameters of these simulations are summarized in
\tab{tab:conf}, for each of the six bare coupling values. Because we
are using staggered fermions, our results suffer from taste
artefacts. The corresponding splittings in the pion spectrum are also
given in \tab{tab:conf}.

\begin{figure}[t]
    \includegraphics[width=0.9\columnwidth]{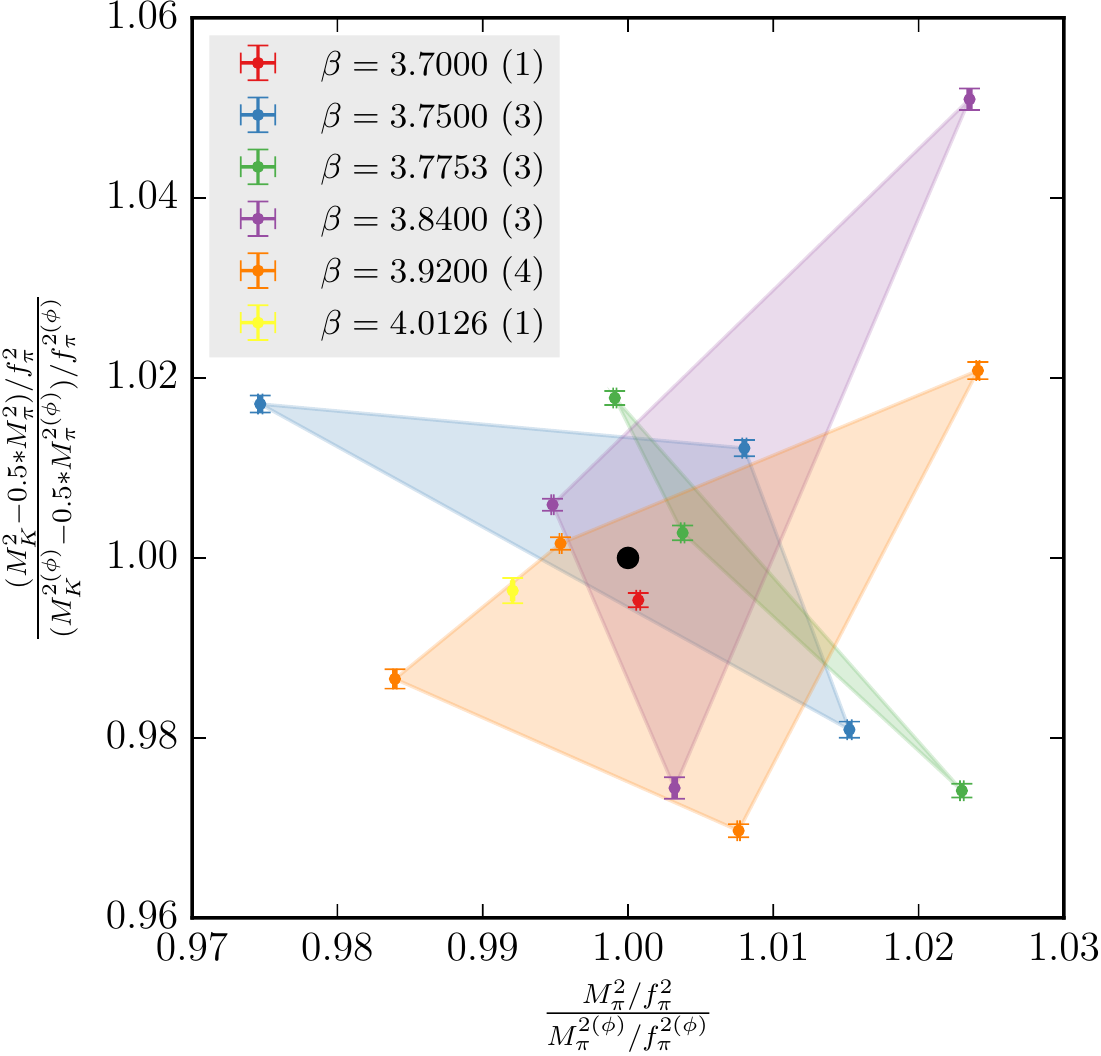}
    \caption
    {\label{fig:landscape}
      Positions of our 15 simulations in the
      $m_{ud}$-$m_s$ plane, as represented by $(M_\pi/f_\pi)^2$ and
      $(M_K^2-M_\pi^2/2)/f_\pi^2$, respectively, normalized by their
      physical values.  Here $M_\pi$ and $M_K$ are the masses of the
      Goldstone taste partners. At a given value of $\beta$, the
      polygons delimited by the points
      $(M_\pi^2,M_K^2-M_\pi^2/2)/f_\pi^2$ indicate whether these points
      bracket the physical point. The numbers in parentheses, next to
      the values of $\beta$, indicate the number of independent
      simulations at that $\beta$.
    }
\end{figure}

All of our simulations are performed around the physical mass point in
the isospin limit, with $m_d=m_u$ and $\alpha=0$. In \fig{fig:landscape}
we show the positions of our 15 simulations in the $m_{ud}$-$m_s$
plane, as represented by $(M_\pi/f_\pi)^2$ and
$(M_K^2-M_\pi^2/2)/f_\pi^2$, respectively. Here $M_\pi$ and $M_K$ are
the masses of the Goldstone taste partners. For four of our six
lattice spacings, we have three or more different values of $m_{ud}$
and $m_s$. As the figure shows, in most cases these simulations
bracket the physical point and the values of $(M_\pi/f_\pi)^2$ and
$(M_K^2-M_\pi^2/2)/f_\pi^2$ are never more than 3, respectively 6,
percent away from their physical values.

To obtain a precise signal for the connected-light and
quark-disconnected contributions to the current-current correlator at
the large distances required for the computation of the light lepton
magnetic moments, a significant statistics is needed. Thus, we have
generated a large number of configurations, separated by 10 unit
length RHMC trajectories, and have performed the computations using a
large number of sources on each configuration. The number of sources
required for the strange contribution is significantly smaller,
because of the better quality of its signal. All of these numbers are
also specified in \tab{tab:conf}. Moreover, as mentioned in the main
text, the precision of the connected-charm contribution is so high,
that we have chosen to evaluate it on a reduced set of configurations,
to bring its statistical error more in line with those of the other
contributions and to allow for fits with good fit qualities. We have
found that 40 configurations, maximally separated in molecular
dynamics time, and 4 sources are sufficient.

\begin{table*}[t]
    \begin{ruledtabular}
	\begin{tabular}{cccccc}
	    $\beta$ & $a$ $[\fm]$ & $(T{\times}L/a^2)$ & $a^2\Delta_\mathrm{taste}$ & \#conf-$ud$/\#conf-$s$/\#conf-$c$/\#conf-disc & \#src-$ud$/\#src-$s$/\#src-$c$/\#src-disc\\
	    \hline
	    3.7000 & 0.134 & $ 64\times48$ & 0.01809(4) & 1000/1000/40/1000    &  768/64/4/9000\\
	    3.7500 & 0.118 & $ 96\times56$ & 0.00992(3) & 1500/1500/40/1500    & 768/64/4/6000\\
	    3.7753 & 0.111 & $ 84\times56$ & 0.007378(16)$^*$ & 1500/1500/40/1500    & 768/64/4/6144\\
	    3.8400 & 0.095 & $ 96\times64$ & 0.00337(2) & 2500/2500/40/1500    & 768/64/4/3600\\
	    3.9200 & 0.078 & $128\times80$ & 0.001090(14) & 3500/3500/40/1000  & 768/64/4/6144\\
	    4.0126 & 0.064 & $144\times96$ & 0.000327(15)&  450/450/40/-   &  768/64/4/- \\
	\end{tabular}
        \flushleft{\footnotesize $^*$ Obtained by interpolation.}
    \caption{\label{tab:conf} Parameters of the 15 simulations
      performed. For a given bare coupling, $\beta$, the simulations
      differ only by a slightly distinct choice of quark masses around
      their physical value. Here we list the $\beta$, the lattice
      spacings, $a$, the sizes $T\times L$, the taste splittings,
      $a^2\Delta_\mathrm{taste}=(aM_{i5})^2-(aM_\pi)^2$ (with $M_{5i}$
      the mass of the $\xi_{5i}$ taste partner and $M_\pi$, the mass
      of the Golstone), the number of configurations used and the
      number of sources per configuration used for each flavor
      contribution. Note that for the $ud$ and $s$ contributions, all
      15 simulations are used while for the charmed one, 2
      simulations for $\beta=3.9200$ are omitted and, in addition, the one
      at $\beta=4.0126$ for the quark-disconnected one.}
    \end{ruledtabular}
\end{table*}

\subsection{Time cuts and bounds for the light and disconnected contributions}
\label{sec:tcut}

\begin{figure}
    \includegraphics[width=0.9\columnwidth]{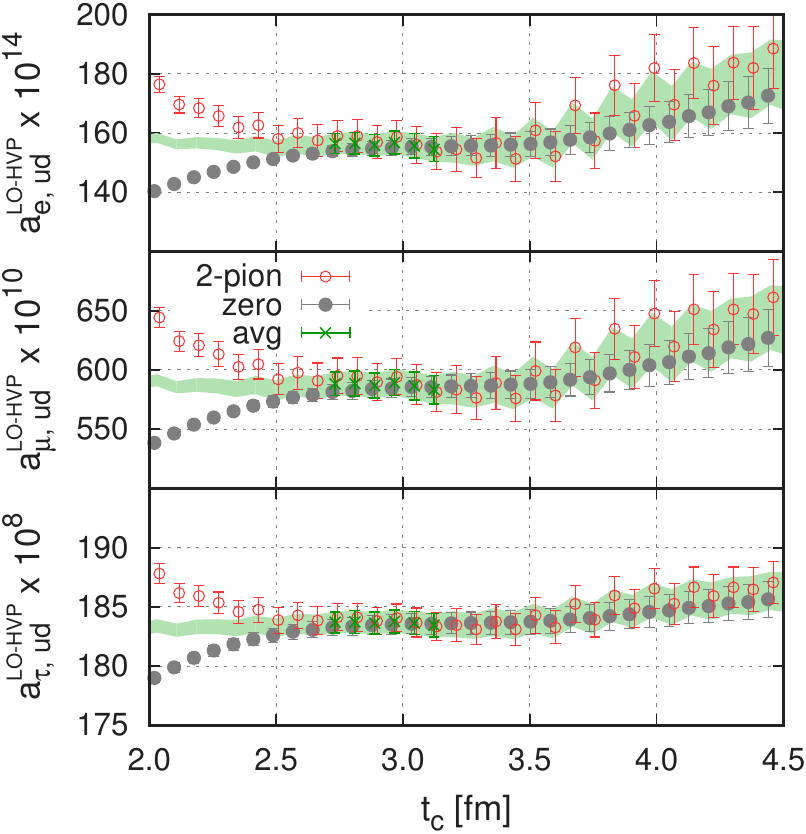}
    \caption
    {
      \label{fig:aell_ud_bounds}
      Upper/lower bounds on $\aelllohvpf{\ell}{ud}(Q{\le} 2\,\gev)$ as a
      function of the time cut, $t_c$. From top to bottom, the plots
      correspond to $\ell=e,\mu,\tau$. In calculating
      $\aelllohvpf{\ell}{ud}(Q{\le} 2\,\gev)$ from
      \eqs{eq:aell_int}{eq:pihat} of the main text, the correlator $C_{ii}^{ud}(t)$,
      $i=1,2,3$, is used up to and including $t=t_c$ in the sum over
      $t$ in \eq{eq:pihat}. For $t{>}t_c$, $C_{ii}^{ud}(t)$ is replaced
      by the upper and lower bounds of \eq{eq:bound1}. The red open
      circles correspond to the upper bound, obtained by assuming that the
      correlator falls off exponentially with the lowest two-pion
      energy, for $t{>}t_c$. The grey closed circles are the lower bound obtained
      by setting $C_{ii}^{ud}(t)$ to zero for $t{>}t_c$. Points are slightly shifted around $t_c$ so that
      they do not overlap. The green band represents the average of
      the two bounds, with errors. The green crosses indicate the
      values of $t_c$ and the corresponding bound-averages that we
      consider for obtaining $\aelllohvpf{\ell}{ud}$. All results are for an ensemble at
      $\beta=3.9200$.  }
\end{figure}

In the case of the light and disconnected correlators,
$C_{ii}^{ud}(t)$ and $C_{ii}^{\text{disc}}(t)$, $i=1,2,3$, the signal
deteriorates quickly with increasing time. Thus, we introduce a cut
$t_c$ in time, beyond which we replace the correlator by an upper and
a lower bound~\cite{Borsanyi:2016lpl}. In the continuum limit,
$C_{ii}^{ud}(t)$ is a sum of falling exponentials with positive
coefficients. Therefore, it is bounded below by zero. Moreover,
because it is proportional to the isospin triplet correlator, whose
lowest-energy contribution comes from a two-pion state, it is bounded
above by that contribution. Therefore, in the continuum limit and the
limit of infinite statistics, the correlator satisfies
\be
    \label{eq:bound1}
    0 \leq C_{ii}^{ud}(t) \leq C_{ii}^{ud}(t_c)\frac{\varphi(t)}{\varphi(t_c)}\ ,
\ee
where $\varphi(t)= \cosh\left[E_{2\pi}(T/2-t)\right]$ and $E_{2\pi}$
is the energy of two pions, each with the smallest nonvanishing
lattice momentum, for which we use $2\pi/L$. \eq{eq:bound1} holds up
to higher-order, wrap-around contributions which are suppressed
exponentially in nonvanishing multiples of $TE_{2\pi}$. In
\fig{fig:aell_ud_bounds}, for an ensemble at $\beta=3.9200$ and for
all three leptons, we plot the resulting upper and lower bounds on
$\aelllohvpf{\ell}{ud}(Q{\le} 2\,\gev)$ as a function of $t_c$. These
are obtained by inserting the bounds of \eq{eq:bound1} into
\eqs{eq:aell_int}{eq:pihat} from the main text, for $t{>}t_c$. For
$t{\le} t_c$, it is the correlator $C_{ii}^{ud}(t)$ obtained directly
in the simulation which is used. As the figure shows, the upper and
lower bounds typically agree for $t_c\gtrsim 3\,\fm$, for all three
leptons. In our analyses, we consider the ranges $t_c=(3.000\pm
0.134)\,\fm$ and $t_c=(2.866\pm 0.134)\,\fm$ for the light connected,
timelike correlators. For each $t_c$ in these intervals, we replace
$\aelllohvpf{\ell}{ud}(Q{\le} 2\,\gev)$, obtained directly from the
lattice correlation function by the mean of its upper and lower
bounds, and then average this mean over each $t_c$ interval. The two
intervals are used to estimate the systematic uncertainty associated
with our procedure, as explained in the main text.

The reason for averaging over $t_c$ is to dampen the effect of
possible statistical fluctuations in the value of the correlator from
from one $t_c$ to the next. This is particularly useful with staggered fermions,
for which even and odd times are mostly uncorrelated. Thus, for each
lattice spacing, we round the number of time slices in these $t_c$
intervals to the nearest even integer. For $\beta\ge 3.9200$, this
leads to four time slices in each $t_c$ interval and to a shift of two
time slices between intervals. For $\beta<3.9200$, these figures are
two and one. While this average is not strictly necessary, because the
bound means between two neighboring $t_c$'s in the relevant interval
agree statistically, it does reduce overall the fluctuations in
central values, leading to better $\chi^2/\text{dof}$, e.g.  in the
continuum extrapolations.

\begin{figure}
    \includegraphics[width=0.9\columnwidth]{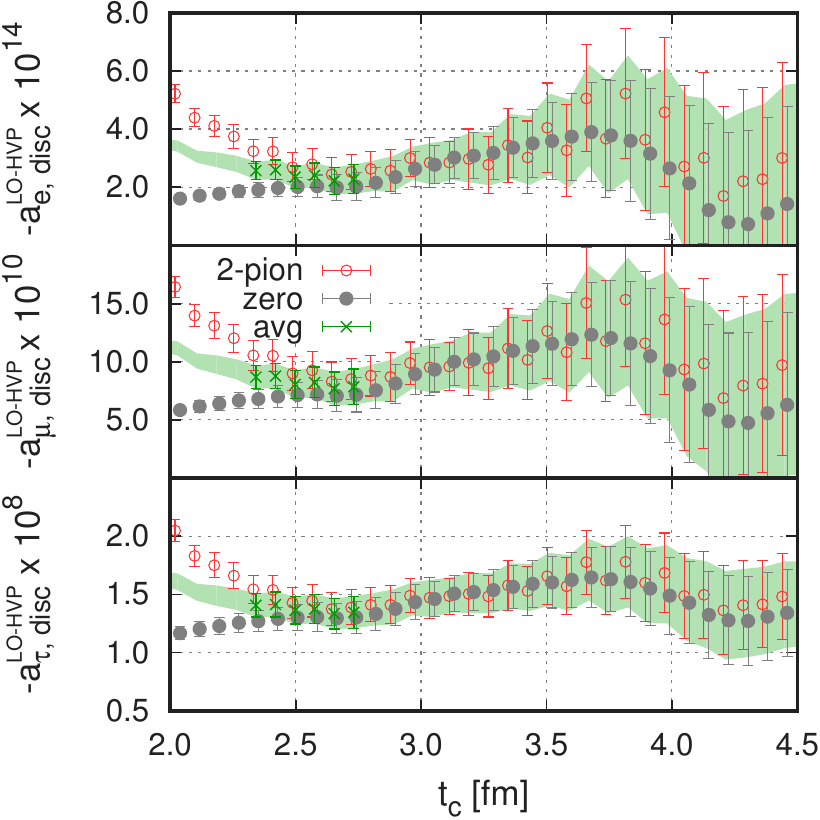}
    \caption
    {
      \label{fig:aell_disc_bounds}
      Same a \fig{fig:aell_ud_bounds}, but for $\aelllohvpf{\ell}{\text{disc}}(Q{\le} 2\,\gev)$ vs $t_c$.  }
\end{figure}

The disconnected contribution alone can be constrained for large
enough time separations, where the isospin singlet channel, dominated
by three-pion states, can be neglected compared to the triplet one,
dominated by two-pion states. Here we have, for large enough $t_c$,
\be
    \label{eq:bound2}
    0 \leq -C_{ii}^\text{disc}(t) \leq
    \frac1{10}C_{ii}^{ud}(t_c)\frac{\varphi(t)}{\varphi(t_c)}+C_{ii}^{s}(t)+C_{ii}^{c}(t)
    \ ,\ee
up to higher-order, wrap-around contributions.  At large $t$, the
connected strange and charm contributions in \eq{eq:bound2} are
exponentially suppressed, and their presence does not make a
difference when determining $t_c$, so we neglect them.  As done in
\fig{fig:aell_ud_bounds} for $\aelllohvpf{\ell}{ud}(Q{\le} 2\,\gev)$, in
\fig{fig:aell_disc_bounds} we plot the resulting upper and lower
bounds on $\aelllohvpf{\ell}{\text{disc}}(Q{\le} 2\,\gev)$ as a function
of $t_c$. Here, the $t_c$ time ranges are taken to be $t_c=(2.600\pm
0.134)\,\fm$ and $t_c=(2.466\pm 0.134)\,\fm$, as suggested by the
region of the merging of the bounds in the figure.

Pion-pion interactions change the smallest two-pion
momentum from $2\pi/L$ in that channel.  Using the model of
\cite{Luscher:1991cf} and neglecting four-pion contributions, we
determine the change in the two-pion energy to be around 2\%. We checked
that such a reduction of the momentum changes the result on
$\aelllohvpf{\ell}{ud}$ and $\aelllohvpf{\ell}{\text{disc}}$ by a small fraction of the statistical error.

\subsection{Physical point and lattice spacing uncertainty}
\label{sec:phypt}

We define the physical mass point by using the isospin corrected pion
and kaon masses, $\bar M_\pi= 134.8(3)\,\mev$ and $\bar
M_K=494.2(3)\,\mev$, from \cite{Aoki:2016frl}, as well as the
electromagnetically corrected $\eta_c$ mass,
$M_{\eta_c}=2.9863(27)\,\gev$ of \cite{Chakraborty:2014aca}. To
convert the lattice results into physical units, we use the pion decay
constant obtained from pion leptonic decays, which is free of
electromagnetic corrections and, to very good accuracy, equals the
decay constant in the $m_d=m_u$ limit \cite{Gasser:2010wz}. This
yields a well defined physical point in the isospin limit. In
intermediate steps of the analysis, we use the Wilson-flow-based
\cite{Luscher:2010iy} $w_0$-scale \cite{Borsanyi:2012zs} as described
below.

As noted in \cite{Bell:1996md,Bernecker:2011gh}, and generalized here for individual flavor contributions,
\be
\label{eq:limmellaell}
\lim_{m_\ell\to 0}\frac{\aelllohvpf{\ell}{f}}{m_\ell^2}=\frac{4\pi^2}3\left(\frac{\alpha}{\pi}\right)^2\frac{d\Pi^f(Q^2)}{dQ^2}
\Big\vert_{Q^2=0} \ .
\ee
Thus, for very light leptons, the leading dependence of
$\aelllohvpf{\ell}{f}$ on lattice spacing is quadratic. In practice,
we find this to be almost exactly true for the electron. For the
$\mu$, this dependence ranges from quadratic to about 20\% weaker than
quadratic, depending on the flavor contribution. Only for the $\tau$
can it be significantly less for some contributions, as low as
$a^{0.8}$. Because it maximizes the error bar due to the lattice
spacing uncertainty, we will assume a quadratic dependence on $a$ for
all leptons and for all flavor contributions. This means that, for a
given simulation, the relative error on
$\aelllohvpflat{\ell}{f}(Q{\le}\Qmax)$, due to the scale uncertainty,
will be taken to be twice that on the lattice spacing. While this
error may be modified some by the combined mass interpolations and
continuum extrapolations, it represents the dominant effect on our
results for $\aelllohvpf{\ell}{f}(Q{\le}\Qmax)$.

This strong dependence on $a$ means that uncertainties on the lattice
spacing must be estimated with care. As noted above, we fix $a$ with
$f_\pi$, through $a/w_0$ and $w_0 f_\pi$. $a/w_0$ is obtained to
sub-permil precision for each of our simulations. On a subset of these
simulations, we determine $w_0 f_\pi$ in the continuum limit with a
total 0.4\% error. Using $f_\pi=130.50(1)(3)(13)\,\mev$ from the PDG
\cite{Rosner:2015wva,Patrignani:2016xqp}, this means that we have a
0.4\% error on our lattice spacings~\footnote{This analysis yields a
  value of $w_0$ which is compatible with the value used by HPQCD in
  \cite{Chakraborty:2016mwy}, $w_0=0.1715(9)\,\fm$
  \cite{Dowdall:2013rya}, within less than one combined standard
  deviation. Thus, the origin of the tension with HPQCD on
  $\aelllohvpf{\mu}{ud}$, discussed in Sec.~X of the SM, cannot
  attributed to a different value of $w_0$. In fact, this is confirmed
  by our excellent agreement with HPQCD on the connected strange and
  charm contributions to the HVP, which have the same sensitivity to
  lattice spacing as $\aelllohvpf{\mu}{ud}$. The reporting of a
  complete analysis of $w_0$ is left for a future publication.}. As
discussed above, this translates into a 0.8\% uncertainty on our
results for $\aelllohvpf{\ell}{f}(Q{\le}\Qmax)$, $\ell{=}e,\mu,\tau$,
$f{=}ud,s,c,\text{disc}$, which we carry over to our final results for
$\aelllohvpf{\ell}{f}$,
$f{=}(I{=}1),(I{=}0),\text{\textvisiblespace}$.  It is important to
note that this relative error on the individual flavor contributions
to $\aelllohvp{\ell}$ is 100\% correlated across these
contributions. Thus, it is the same for the individual contributions
as it is for their sum.

\subsection{Continuum limit and mass interpolation of $\aelllohvpflat{\ell}{f}(Q{\le}\Qmax)$}
\label{sec:aell_contlim}

\begin{figure*}[t]
  \centering
  \includegraphics[width=0.32\linewidth]{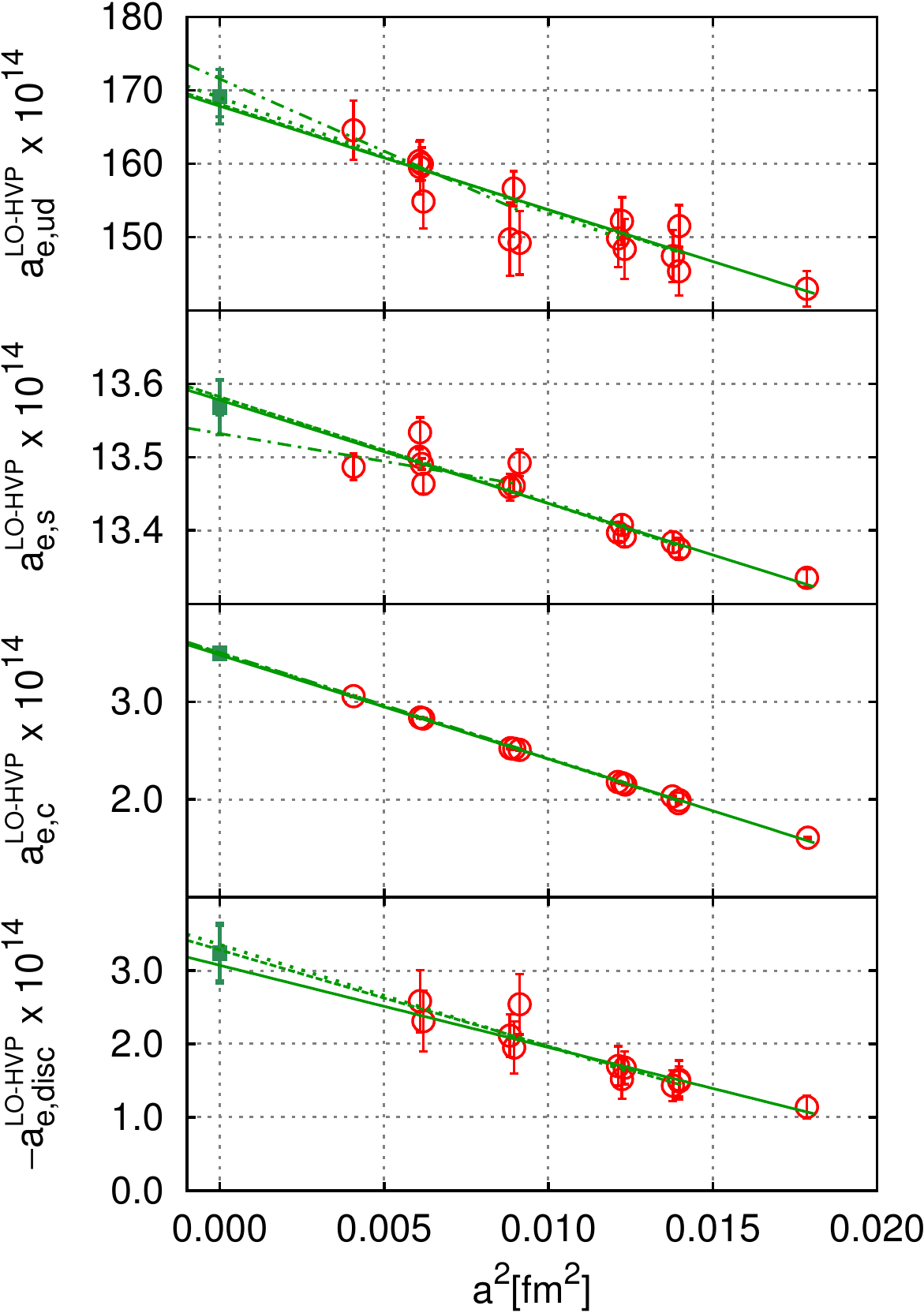}
\includegraphics[width=0.32\linewidth]{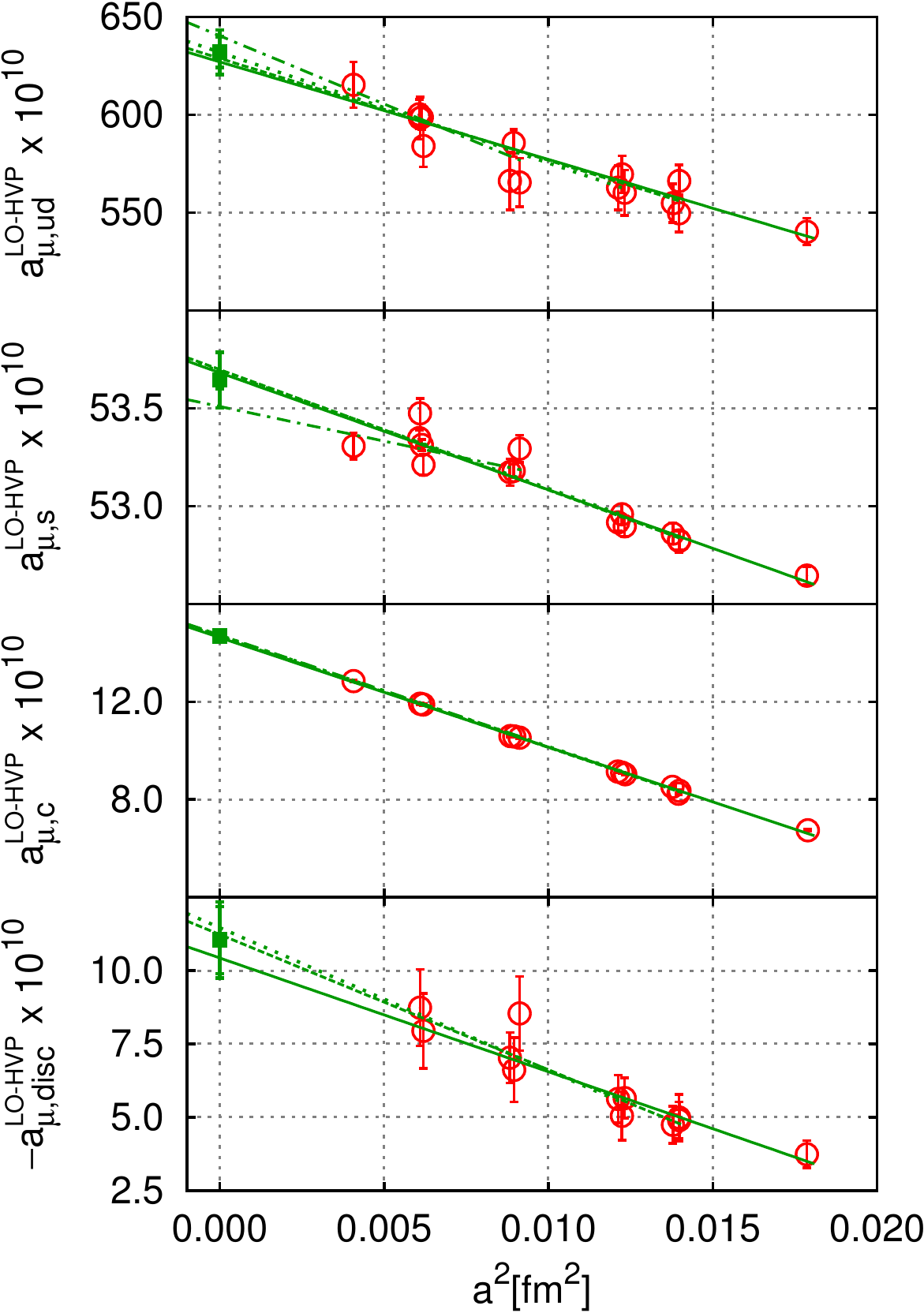}
  \includegraphics[width=0.32\linewidth]{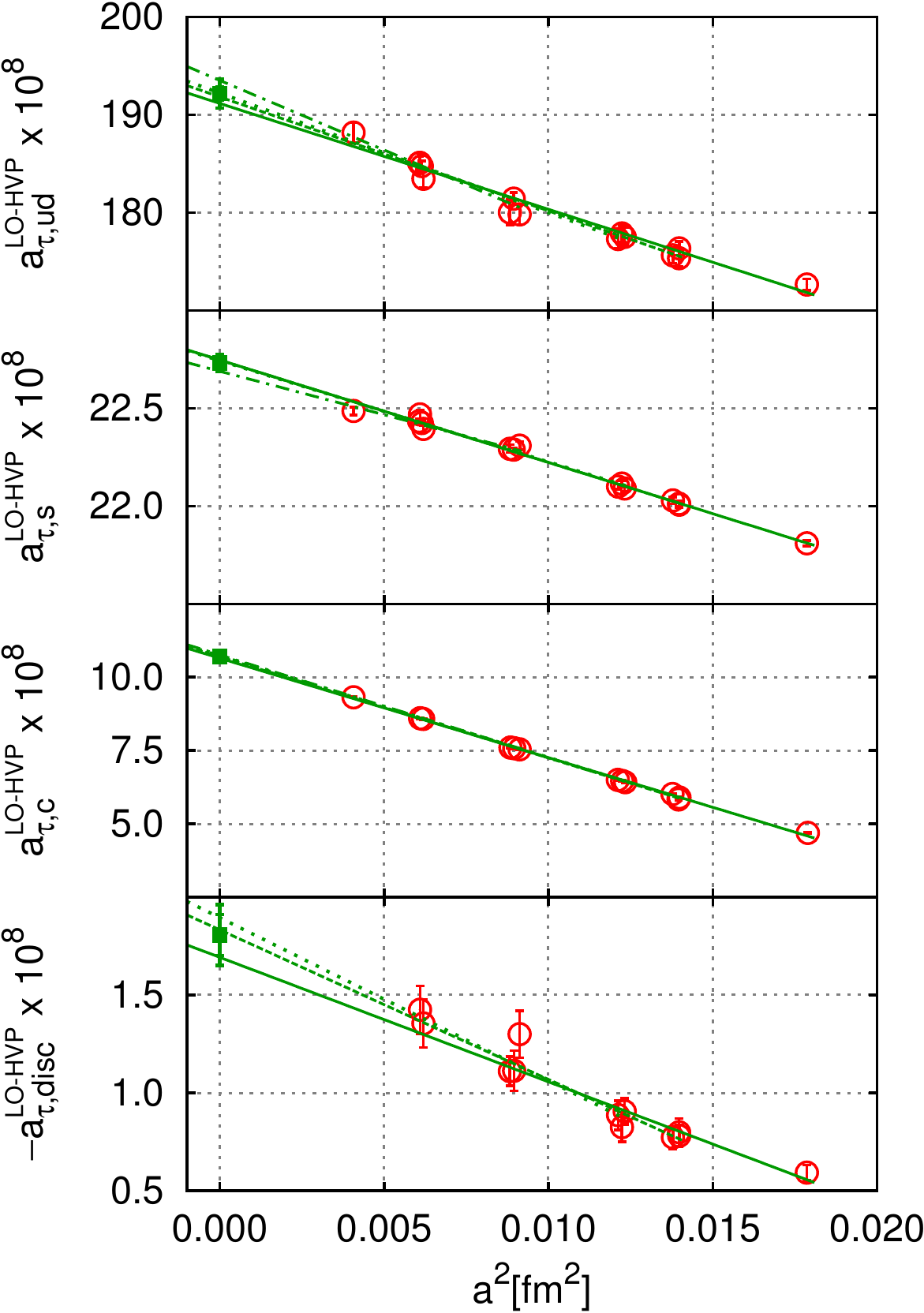}
    \caption
        {\label{fig:af_cont_extrap} Continuum extrapolation of the
          various flavor contributions to $\aelllohvp{\ell}(Q{\le}
          2\,\gev)$ obtained using $t_c=(3.000\pm 0.134)\,\fm$ for the $ud$
          contribution and $t_c=(2.600\pm 0.134)\,\fm$ for the
          disconnected one. From left to right, $\ell=e$, $\mu$,
          $\tau$. From top to bottom, the connected light, strange,
          charm, and disconnected contributions. The red open circles
          with errors are the results from our 15 simulations for the
          $ud$ and $s$, 13 for the charm and 12 for the disconnected
          contributions, with statistical uncertainties. These points
          have been interpolated to the physical mass point using the
          fits to all lattice spacings (solid lines). The different
          lines represent the fits obtained by imposing cuts in $a$
          (solid for no cut, dashed for $a\le0.118\,\fm$, dotted for
          $a\le0.111\,\fm$, dot-dashed for $a\le0.095\,\fm$). The fact
          that a few of the lines do not appear to fit the red points
          is due to the dependence on other lattice parameters in
          those fits, which is slightly different from the one
          corresponding to the solid line. The green squares are the
          continuum extrapolated results for the given $\Qmax$ and
          $t_c$, with statistical and continuum extrapolation errors
          only. }
\end{figure*}

\begin{table*}[t]
    \centering
    \begin{ruledtabular}
	\begin{tabular}{LCCC}
	    (f,\,\Qmax^2)& \ell=e\; \mbox{(units of $10^{-14}$)} & \ell=\mu\; \mbox{(units of $10^{-10}$)} & \ell=\tau\; \mbox{(units of $10^{-8}$)}\\
  	  \hline
          (ud,\,1\,\gev^2)                        & 169.2(2.6)(2.3)(0.0) & 627.1(7.5)(7.9)(0.1) & 89.1(0.4)(0.8)(0.0) \\
          (s,\,1\,\gev^2)                         & 13.6(0.0)(0.0)(-) & 53.0(0.0)(0.1)(-) & 9.2(0.0)(0.0)(-) \\
          (c,\,1\,\gev^2)                         & 3.5(0.0)(0.0)(-) & 14.4(0.0)(0.1)(-) & 3.2(0.0)(0.0)(-) \\
          (\text{disc},\,1\,\gev^2)               & -3.3(0.3)(0.1)(0.1) & -11.2(1.0)(0.5)(0.2) & -1.1(0.1)(0.1)(0.0) \\
          (I{=}0,\,1\,\gev^2)                     & 30.7(1.2)(1.0)(0.1) & 118.9(3.4)(3.5)(0.2) & 20.2(0.2)(0.4)(0.0) \\
          (\text{\textvisiblespace},\,1\,\gev^2)  & 182.9(2.6)(2.3)(0.1) & 683.2(7.5)(7.9)(0.2) & 100.4(0.4)(0.8)(0.0) \\
  	  \hline
          (ud,\,2\,\gev^2)                        & 169.2(2.6)(2.3)(0.0) & 630.8(7.4)(7.9)(0.1) & 140.0(0.6)(1.1)(0.0) \\
          (s,\,2\,\gev^2)                         & 13.6(0.0)(0.0)(-) & 53.5(0.0)(0.1)(-) & 15.6(0.0)(0.0)(-) \\
          (c,\,2\,\gev^2)                         & 3.5(0.0)(0.0)(-) & 14.6(0.0)(0.1)(-) & 6.2(0.0)(0.0)(-) \\
          (\text{disc},\,2\,\gev^2)               & -3.3(0.3)(0.1)(0.1) & -11.3(1.0)(0.5)(0.2) & -1.5(0.1)(0.1)(0.0) \\
          (I{=}0,\,2\,\gev^2)                     & 30.7(1.2)(1.0)(0.1) & 119.9(3.4)(3.5)(0.2) & 34.3(0.3)(0.5)(0.0) \\
          (\text{\textvisiblespace},\,2\,\gev^2)  & 182.9(2.6)(2.3)(0.1) & 687.6(7.5)(8.0)(0.2) & 160.4(0.6)(1.1)(0.0) \\
  	  \hline
          (ud,\,3\,\gev^2)                        & 169.2(2.6)(2.3)(0.0) & 631.7(7.5)(7.9)(0.1) & 171.2(0.6)(1.2)(0.0) \\
          (s,\,3\,\gev^2)                         & 13.6(0.0)(0.0)(-) & 53.6(0.0)(0.1)(-) & 19.8(0.0)(0.0)(-) \\
          (c,\,3\,\gev^2)                         & 3.5(0.0)(0.0)(-) & 14.6(0.0)(0.1)(-) & 8.7(0.0)(0.0)(-) \\
          (\text{disc},\,3\,\gev^2)               & -3.3(0.3)(0.1)(0.1) & -11.3(1.0)(0.5)(0.2) & -1.7(0.1)(0.1)(0.0) \\
          (I{=}0,\,3\,\gev^2)                     & 30.7(1.2)(1.0)(0.1) & 120.2(3.4)(3.5)(0.2) & 43.9(0.3)(0.5)(0.0) \\
          (\text{\textvisiblespace},\,3\,\gev^2)  & 182.9(2.6)(2.3)(0.1) & 688.7(7.5)(8.0)(0.2) & 198.0(0.6)(1.2)(0.0) \\
  	  \hline
          (ud,\,4\,\gev^2)                        & 169.2(2.6)(2.3)(0.0) & 632.1(7.5)(7.9)(0.1) & 192.2(0.7)(1.3)(0.0) \\
          (s,\,4\,\gev^2)                         & 13.6(0.0)(0.0)(-) & 53.6(0.0)(0.1)(-) & 22.7(0.0)(0.0)(-) \\
          (c,\,4\,\gev^2)                         & 3.5(0.0)(0.0)(-) & 14.7(0.0)(0.1)(-) & 10.7(0.0)(0.1)(-) \\
          (\text{disc},\,4\,\gev^2)               & -3.3(0.3)(0.1)(0.1) & -11.3(1.0)(0.5)(0.2) & -1.8(0.1)(0.1)(0.0) \\
          (I{=}0,\,4\,\gev^2)                     & 30.7(1.2)(1.0)(0.1) & 120.3(3.4)(3.5)(0.2) & 50.8(0.3)(0.6)(0.0) \\
          (\text{\textvisiblespace},\,4\,\gev^2)  & 182.9(2.6)(2.3)(0.1) & 689.1(7.5)(8.0)(0.2) & 223.8(0.7)(1.3)(0.0) \\
  	  \hline
          (ud,\,5\,\gev^2)                        & 169.2(2.6)(2.3)(0.0) & 632.2(7.5)(8.0)(0.1) & 207.2(0.7)(1.4)(0.0) \\
          (s,\,5\,\gev^2)                         & 13.6(0.0)(0.0)(-) & 53.7(0.0)(0.1)(-) & 24.9(0.0)(0.0)(-) \\
          (c,\,5\,\gev^2)                         & 3.5(0.0)(0.0)(-) & 14.7(0.0)(0.1)(-) & 12.4(0.0)(0.1)(-) \\
          (\text{disc},\,5\,\gev^2)               & -3.3(0.3)(0.1)(0.1) & -11.3(1.0)(0.5)(0.2) & -1.9(0.1)(0.1)(0.0) \\
          (I{=}0,\,5\,\gev^2)                     & 30.7(1.2)(1.0)(0.1) & 120.3(3.4)(3.5)(0.2) & 56.1(0.3)(0.6)(0.0) \\
          (\text{\textvisiblespace},\,5\,\gev^2)  & 182.9(2.6)(2.3)(0.1) & 689.3(7.5)(8.0)(0.2) & 242.5(0.7)(1.4)(0.0) \\
	\end{tabular}
    \end{ruledtabular}
    \caption
    { \label{tab:aelllat} Lattice results for the LO-HVP contribution
      to the anomalous magnetic moments of the $e$, $\mu$ and $\tau$
      leptons obtained from \eqs{eq:aell_int}{eq:pihat} of the main
      text, where the integral over $Q^2$ in \eq{eq:aell_int} is cut
      off at $\Qmax^2$. This quantity is denoted
      $\aelllohvpf{\ell}{f}(Q{\le}\Qmax)$ in \eq{eq:aellsep}, for the
      leptons $\ell=e,\mu,\tau$ and for the flavor contributions
      $f{=}ud,s,c,\text{disc},I{=}0,\text{\textvisiblespace}$. Note
      that
      $\aelllohvpf{\ell}{I{=}1}=\frac9{10}\aelllohvpf{\ell}{ud}$. Results
      for $\Qmax^2=1,\cdots,5\,\gev^2$ are given. In the case of the
      $e$ and $\mu$, this ``low-energy'' contribution is equal to the
      full LO-HVP contribution within errors. Only in the case of the
      $\tau$ is it necessary to add a ``high-energy'' complement at
      the current level of precision. In the results presented, the
      first error bar is statistical, the second is the systematic
      uncertainty associated with the continuum extrapolation and the
      third with the bounding procedure described in
      \sec{sec:tcut}. The latter does not affect the $s$ and $c$
      contributions and is denoted $(-)$ for these contributions. Note
      that finite-volume corrections discussed below in \sec{sec:FV}
      are not included here. }
\end{table*}

To obtain numbers that can be compared with experiment, we must
interpolate our results to the physical mass point and extrapolate
them to the continuum limit. We do so by fitting them to a function
which depends on the Goldstone pion and kaon masses squared, on the
$\eta_c$ mass and on the lattice spacing squared. Since the
simulations are performed close to the physical mass point, a constant
or linear dependence in the mass parameters is always
sufficient. Moreover, for all flavor contributions, good fit qualities
can be achieved with a linear $a^2$ dependence for all three leptons
and all values of $\Qmax$ considered here. Thus, the physical values
of the contributions, $\aelllohvpf{\ell}{f}(Q{\le}\Qmax)$, are
obtained by fits of our simulation results to
\bea
\aelllohvpflat{\ell}{f}(Q &\le& \Qmax)=\aelllohvpf{\ell}{f}(Q{\le} \Qmax)
\biggl[1\nn\\
  && +\gamma_{a,f}a^2+\gamma_{\pi,f}\bigl[(M_\pi^\text{lat})^2-\bar M_\pi^2\bigr]\nn\\
  &&  +\gamma_{K,f}\bigl[(M_{K^\chi}^\text{lat})^2-\bar M_{K^\chi}^2\bigr]\biggr]\\
  &&  +\gamma_{c,f}\bigl[M_{\eta_c}^\text{lat}-M_{\eta_c}\bigr]\biggr]\nn
      \ ,
\eea
with $M_{K^\chi}^2=M_K^2-M_\pi^2/2$, for $\ell=e,\mu,\tau$ and for
$f{=}ud,s,c,\text{disc}$. It turns out that for the statistical
accuracy reached here, the light quark mass in our simulations is
sufficiently well tuned that $\gamma_{\pi,f}$ comes out consistent
with zero for all flavors, except the charm, and omitting it increases
the $p$-values of the fits. Thus we set it to zero for all but the
charm contribution. This can be done for $\gamma_{K,f}$ in the case of
the light, quark-disconnected and charm contributions.  $\gamma_{K,f}$
must be kept as a free parameter for the connected strange
contribution. Moreover, $\gamma_{c,f}$ is needed for the charm
contribution to correct for a slight mistuning of $m_c$.

Here we focus on the results obtained with
$\Qmax{=}2\,\gev$. Projections of these fits to our simulation results
for $\aelllohvpflat{\ell}{f}(Q{\le} 2\,\gev)$, $\ell=e$, $\mu$, $\tau$
and $f{=}ud,s,c,\text{disc}$, onto the
$a^2$-$\aelllohvpf{\ell}{f}(Q{\le} 2\,\gev)$ planes, are shown in
\fig{fig:af_cont_extrap}. The features of these fits are very similar
for all three leptons, as they are for the other values of $\Qmax$ (${=}1$,
$\sqrt2$, $\sqrt3$ and $\sqrt5\,\gev$), so that we treat all of these situations
in the same way and discuss them together.

For the light-quark contribution to $\aelllohvp{\ell}(Q{\le}
2\,\gev)$, the dependence on meson masses is not significant
statistically and the terms associated with this dependence can be
ignored. However, as can be seen in the upper panels of
\fig{fig:af_cont_extrap}, the dependence on $a^2$ is strong, due to
the sensitivity of this contribution to low-energy, two-pion states
which, in turn, are sensitive to taste splittings. The fact that the
anomalous moment of the lighter $e$ is more sensitive to these states
than that of the $\mu$ that is, in turn, more sensitive than that of
the $\tau$, explains the fact that $\aelllohvp{e}(Q{\le} 2\,\gev)$ has
the strongest $a^2$ dependence while $\aelllohvp{\tau}(Q{\le}
2\,\gev)$ has the weakest.

The situation is different for the strange contribution, much less
affected by taste violations. As the second panels of
\fig{fig:af_cont_extrap} show, the continuum limits are very
mild. They are much less so for the charm, as shown in the third
panels, due to the large value of $m_c$ in lattice units. Here it is
the magnetic moments of the more massive leptons which are steeper,
due to their sensitivity to larger $Q$. In addition to the dependence
on $a^2$, a linear dependence on $M_{K^\chi}^2$ is needed for both
contributions and one on $M_{\eta_c}$ is required to correct a slight
mistuning of the charm mass in that quark's contribution.

Our results for $\aelllohvpflat{\ell}{\text{disc}}(Q{\le} 2\,\gev)$ have
large lattice artefacts, as
shown in the bottom panels of \fig{fig:af_cont_extrap}. This is
because the taste violations of the $ud$ contribution enhance the
SU(3)-flavor cancellation against the $s$ contribution in
$\aelllohvpflat{\mu}{\text{disc}}$, as $a^2$ is increased.  In these
results we neglect the charm contribution which we find to be less
than $1\%$ of the total disconnected contribution on our coarsest
lattice, i.e. much smaller than the disconnected, statistical error.
In addition, because statistical errors are quite large, no dependence
on quark mass is required to describe the lattice data.

As explained in the main text, the systematic error associated with
these continuum limits and physical-point interpolations are obtained
by imposing four cuts on the lattice spacing in the quark-connected
case and three for the disconnected contributions. The results of
these cuts are then combined, as detailed in the main text, to give
a central value and statistical and systematic errors. The results
for the various contributions to the magnetic moments of all three leptons
with the four values of the momentum cut $\Qmax$ considered are summarized
in \tab{tab:aelllat}.

\subsection{Determination of $\hat\Pi^f(\Qmax^2)$}
\label{sec:pihatqmax}

\begin{figure}[t]
    \centering
     \includegraphics[width=0.9\columnwidth]{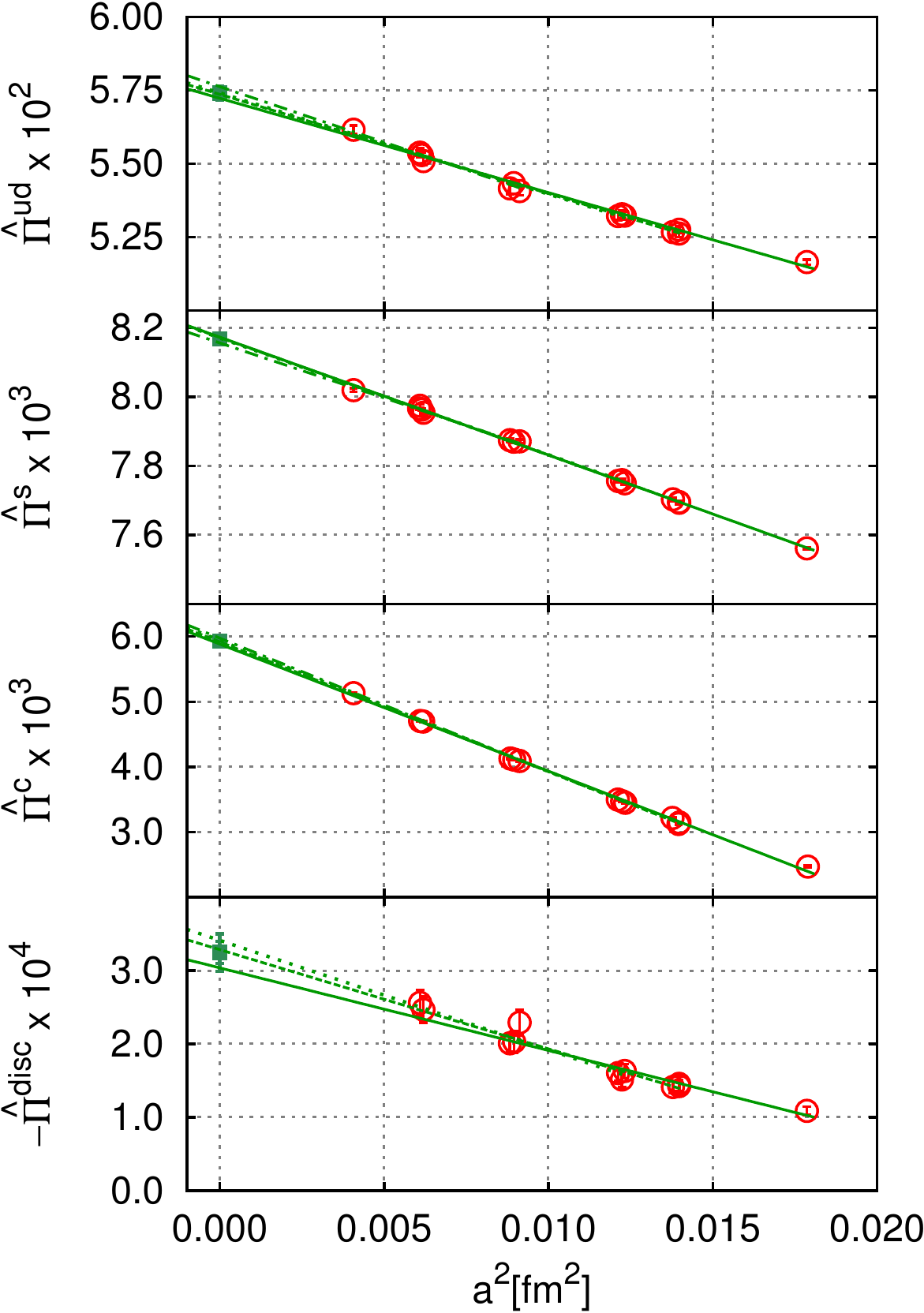}
    \caption
    {\label{fig:Pi5GeV2_cont_extrap} Continuum extrapolation of the
      various flavor contributions to $\hat\Pi(4\,\gev^2)$ obtained
      using $t_c=(3.000\pm 0.134)\,\fm$ for the $ud$ contribution and
      $t_c=(2.600\pm 0.134)\,\fm$ for the disconnected one. From top
      to bottom, the connected light, strange, charm, and disconnected
      contributions.  For the light and disconnected contributions,
      these correspond to the later of the two $t_c$ ranges. The red
      open circles with errors are the results from our 15 simulations
      for the $ud$ and $s$, 13 for the charm and 12 for the
      disconnected contributions, with statistical uncertainties.
      These points have been interpolated to the physical mass point
      using the fits to all lattice spacings (solid lines). The
      different lines represent the fits obtained by imposing cuts in
      $a$ (solid for no cut, dashed for $a\le0.118\,\fm$, dotted for
      $a\le0.111\,\fm$, dot-dashed for $a\le0.095\,\fm$). The fact
      that a few of the lines do not appear to fit the red points is
      due to the dependence on other lattice parameters in those fits,
      which is slightly different from the one corresponding to the
      solid line. The green squares are the continuum extrapolated
      results for the given $t_c$, with statistical and continuum
      extrapolation errors only.}
\end{figure}

In order to determine the various flavor $f$ contributions to the
anomalous magnetic moments of the leptons through \eq{eq:aellsep}, we
also need to compute $\hat\Pi^f(\Qmax^2)$. Even for large $\Qmax^2$,
this is a non-perturbative quantity because it requires a subtraction
at $Q^2=0$. We obtain $\hat\Pi^f(\Qmax^2)$ from our lattice results
for the current-current correlator $C_{ii}^f(t)$ through
\eq{eq:pihat} of the main text. We do so for each one of our simulations, for each
of the flavors, $f{=}ud,s,c,\text{disc}$ and for the five values of
$\Qmax$ considered here. The resulting values
$\hat\Pi_\text{lat}^f(\Qmax^2)$ must then be interpolated to the
physical mass point and extrapolated to the continuum. To this end, we
follow the exact same procedure as for $\aelllohvpf{\ell}{f}$. That
is, we fit our lattice results to a functional form which parametrizes
the dependence on lattice spacing and on quark masses.  As for the
contributions to the leptonic magnetic moments, we find that a constant or linear
dependence on $a^2$, $M_\pi^2$, $M_{K^\chi}^2$ and $M_{\eta_c}$ describes the
lattice results well, for all $f$ and $\Qmax$. Again with the charm we
have to reduce the statistics to get acceptable $\chi^2/\text{dof}$.
Since the features of these fits are very similar to those for
$\aelllohvpflat{\ell}{f}(Q{\le}\Qmax)$, we do not repeat the
discussion here. Instead we plot, in \fig{fig:Pi5GeV2_cont_extrap},
the results of these fits for all four flavors and for the
$\Qmax^2{=}4\,\gev^2$ used in other plots. The results corresponding to
all values of $\Qmax^2$ considered here and to all physically relevant
flavor combinations are summarized in \tab{tab:pivsQmax}. Central
values and statistical and systematic errors are obtained, as for
$\aelllohvpf{\ell}{f}(Q{\le}\Qmax)$, using flat distributions, cuts in
$a^2$ and the bounding procedure described in \sec{sec:tcut}.

\begin{table*}[t]
    \centering
    \begin{ruledtabular}
	\begin{tabular}{CCCCCCC}
	  \Qmax^2\,[\gev^2] & \hat\Pi^{ud}(\Qmax^2) \times 10^2 & \hat\Pi^{s}(\Qmax^2) \times 10^3 & \hat\Pi^{c}(\Qmax^2) \times 10^3 & \hat\Pi^\text{disc}(\Qmax^2) \times 10^4 & \hat\Pi^{I{=}0}(\Qmax^2) \times 10^3 & \hat\Pi(\Qmax^2) \times 10^2\\
          \hline

1.0 & 3.53(1)(2)(0) & 4.20(0)(1)(-) & 1.76(0)(1)(-) & -3.12(14)(19)(3) & 9.18(5)(10)(0) & 4.09(1)(2)(0) \\
2.0 & 4.64(1)(2)(0) & 6.10(0)(1)(-) & 3.31(1)(2)(-) & -3.24(14)(19)(3) & 13.72(5)(10)(0) & 5.54(1)(2)(0) \\
3.0 & 5.28(1)(2)(0) & 7.30(0)(1)(-) & 4.69(1)(3)(-) & -3.26(14)(19)(3) & 16.94(5)(10)(0) & 6.45(1)(2)(0) \\
4.0 & 5.74(1)(2)(0) & 8.17(0)(1)(-) & 5.93(1)(5)(-) & -3.27(14)(19)(3) & 19.51(5)(11)(0) & 7.12(1)(2)(0) \\
5.0 & 6.09(1)(2)(0) & 8.85(0)(1)(-) & 7.05(1)(6)(-) & -3.27(14)(19)(3) & 21.66(5)(12)(0) & 7.65(1)(2)(0) \\
	\end{tabular}
    \end{ruledtabular}
    \caption
        { \label{tab:pivsQmax} Values of the renormalized, scalar
          polarization function, $\hat\Pi^f(\Qmax^2)$, as obtained on
          the lattice through \eq{eq:pihat} of the main text, with a
          continuum extrapolation and an interpolation to the physical
          mass point. In our conventions, $\hat\Pi^f(\Qmax^2)$ is
          smaller by a factor of $4\pi^2$ than it is, for instance, in
          \cite{Chakraborty:2016mwy}. Results for different flavor
          combinations,
          $f{=}ud,s,c,\text{disc},I{=}0,\text{\textvisiblespace}$, are
          given. Note that
          $\hat\Pi^{I{=}1}=\frac9{10}\hat\Pi^{ud}$. In this table, the
          first error is statistical, the second is the systematic
          uncertainty associated with the continuum extrapolation and
          the third with the bounding procedure described in
          \sec{sec:tcut}. The latter does not affect the $s$ and $c$
          contributions and is denoted $(-)$ for those
          quantities. Note that finite-volume corrections discussed
          below in \sec{sec:FV} are not included here.}
\end{table*}

\subsection{Separation of $\aelllohvp{\ell}$ into low- and high-virtuality contributions}
\label{sec:aellsep}

The computation of $\aelllohvp{\ell}$ requires a determination of the
renormalized, scalar polarization function, $\hat\Pi(Q^2)$ for all
values of the Euclidean momentum $Q$, from $0$ to $\infty$. Of course
at finite lattice spacing, momenta up to infinity are not
available. Thus, as explained in the main text, we deal with this
problem by separating the low and high virtuality contributions at a
value of $Q=\Qmax$ in the following way (repeating \eq{eq:aellsep} of the main text):
\begin{eqnarray*}
\aelllohvpf{\ell}{f} &=& \aelllohvpf{\ell}{f}(Q{\le}\Qmax)\nn\\
&&+ \gamma_\ell(\Qmax)\;\hat\Pi^f(\Qmax^2)\\
&&+\dpert\aelllohvpf{\ell}{f}(Q{>}\Qmax)\nn
\ ,\end{eqnarray*}
where the low momentum contribution,
$\aelllohvpf{\ell}{f}(Q{\le}\Qmax)$, is obtained from the lattice as
described in the main text, and where the last term is the
high-momentum, perturbative contribution renormalized at $\Qmax$, i.e.
\bea
\label{eq:aellpert}
&&\dpert\aelllohvpf{\ell}{f}(Q{>}\Qmax)\equiv\\
&&\quad\left(\frac{\alpha}{\pi}\right)^2\int_{\Qmax^2}^\infty \frac{dQ^2}{m_\ell^2}\; \omega\left(\frac{Q^2}{m_\ell^2}\right)\left[\Pi_\text{pert}^f(Q^2)-\Pi_\text{pert}^f(\Qmax^2)\right]\nn
\ ,\eea
where the kinematic function, $\omega(r)$, is defined after
\eq{eq:aell_int} of the main text. The computation of
$\dpert\aelllohvpf{\ell}{f}(Q{>}\Qmax)$ is described in
\sec{sec:ptcontrib}.

The second term in \eq{eq:aellsep} is required to shift the
renormalization point in \eq{eq:aellpert} from $\Qmax$ to $Q{=}0$. It is obtained with
lattice results for $C_{ii}^f(t)$ through \eq{eq:pihat} of the main text, with
$Q{=}\Qmax$. Its determination in the continuum limit is explained in
\sec{sec:pihatqmax}.

The factor $\gamma_\ell(\Qmax)$ in \eq{eq:aellsep} is simply
\be
\label{eq:gamma_ell}
\gamma_\ell(\Qmax^2)=\left(\frac{\alpha}{\pi}\right)^2\int_{\Qmax^2}^\infty
\frac{dQ^2}{m_\ell^2}\; \omega(Q^2/m_\ell^2)\ .
\ee
Although calculating this function
is trivial, for completeness we summarize its values for all three leptons and for the values of $\Qmax^2$ used here in \tab{tab:gamma_ell}.

\begin{table}[h]
    \centering
    \begin{ruledtabular}
	\begin{tabular}{CCCC}
	  \Qmax^2\,[\gev^2] & \gamma_e(\Qmax^2) & \gamma_\mu(\Qmax^2) & \gamma_\tau(\Qmax^2) \\
          \hline
          1 & \text{ 7.26}\times 10^{-18} & \text{ 1.27}\times 10^{-8} & \text{ 3.51}\times 10^{-5} \\ 
          2 & \text{ 1.82}\times 10^{-18} & \text{ 3.25}\times 10^{-9} & \text{ 2.25}\times 10^{-5} \\ 
          3 & \text{ 8.07}\times 10^{-19} & \text{ 1.45}\times 10^{-9} & \text{ 1.63}\times 10^{-5} \\ 
          4 & \text{ 4.54}\times 10^{-19} & \text{ 8.20}\times 10^{-10} & \text{ 1.24}\times 10^{-5} \\
          5 & \text{ 2.90}\times 10^{-19} & \text{ 5.26}\times 10^{-10} & \text{ 9.91}\times 10^{-6} \\
	\end{tabular}
    \end{ruledtabular}
    \caption
        { \label{tab:gamma_ell} Values of the kinematical function
          $\gamma_\ell(\Qmax)$ defined in \eq{eq:gamma_ell}, for
          $\ell{=}e,\mu,\tau$.}
\end{table}

\begin{figure}[t]
    \centering
    \includegraphics[width=0.9\columnwidth]{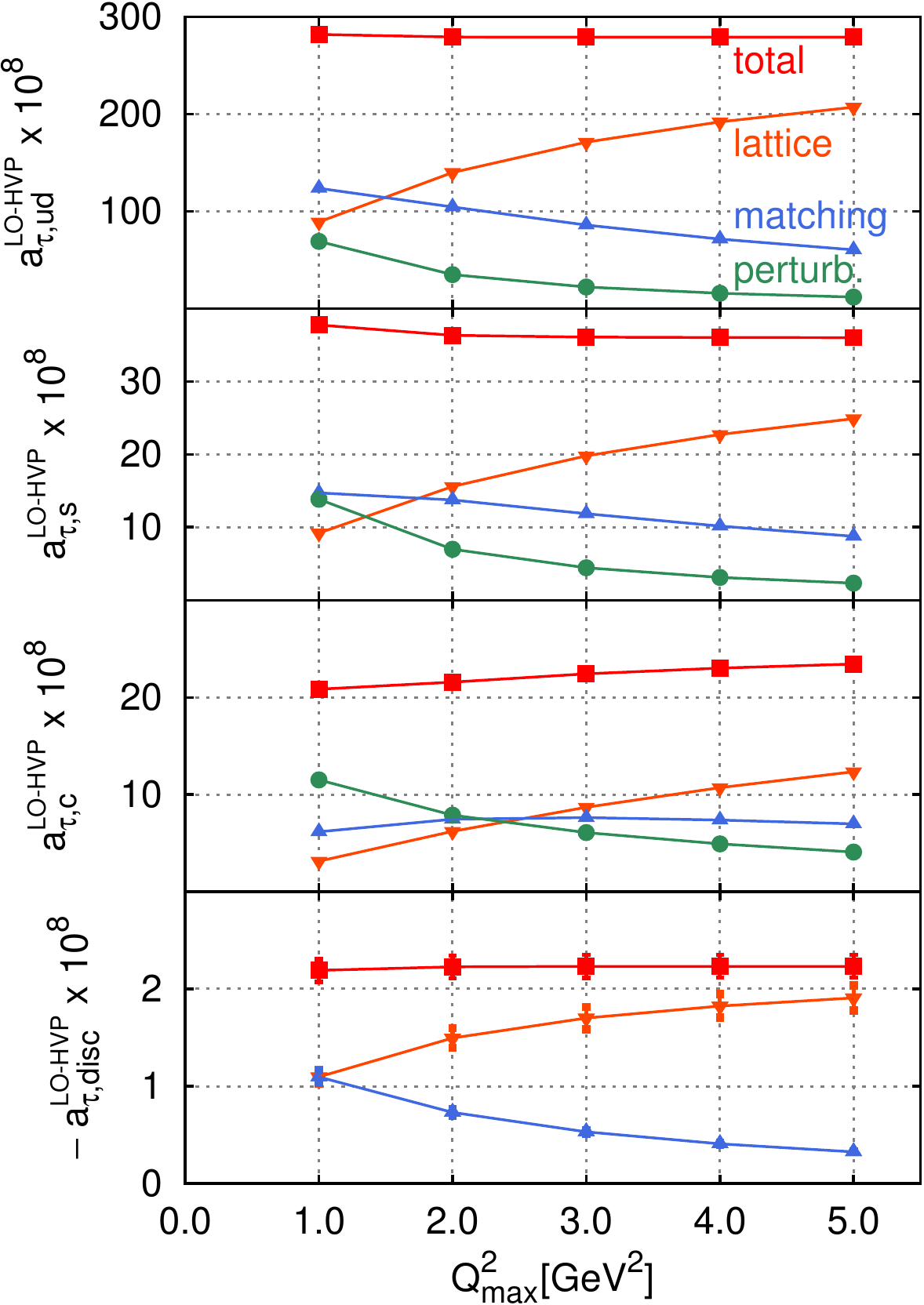}
    \caption
        {\label{fig:atau_f_vs_Qmax} 
          Dependence of our results for
          $\aelllohvpf{\tau}{f}$ on the choice of the value of
          $Q{=}\Qmax$ (from top to bottom, $f{=}ud,s,c,\text{disc}$). All three
          terms of \eq{eq:aellsep} from the main text, as well as their sum, are plotted
          as a function of $\Qmax^2$ for each $f$. The upside-down
          orange triangles correspond to
          $\aelllohvpf{\ell}{f}(Q{\le}\Qmax)$, as obtained on the
          lattice using \eqs{eq:aell_int}{eq:pihat}. The blue
          triangles correspond to the term
          $\gamma_\ell(\Qmax)\;\hat\Pi^f(\Qmax^2)$ of \eq{eq:aellsep},
          with $\hat\Pi^f(\Qmax^2)$ computed on the lattice using
          \eq{eq:pihat}. The green circles represent the
          $O(\alpha_s^4)$ perturbative contribution
          $\dpert\aelllohvpf{\ell}{f}(Q{>}\Qmax)$. For $\aelllohvpf{\tau}{\text{disc}}$ this contribution
          is not shown because it is negligible. $\aelllohvpf{\tau}{f}$,
          which is the sum of the three contributions already
          mentioned, is represented by red squares. On these plots,
          error bars are smaller than the symbols, unless visible.}
\end{figure}

In order for the separation of \eq{eq:aellsep} to be valid, it must be
independent of $\Qmax$ within errors.  We show that this is the case
here. Since the perturbative contributions for $\aelllohvp{\ell}$,
$\ell{=}e,\mu$, are negligible, we focus on $\ell{=}\tau$. The dependence
of $\aelllohvpf{\tau}{f}$ on $\Qmax$ is plotted in
\fig{fig:atau_f_vs_Qmax} for $f{=}ud,s,c,\text{disc}$. As the figure
indicates, the three terms of \eq{eq:aellsep}, for
$f{=}ud,\text{disc}$, add up to a total $\aelllohvpf{\tau}{f}$ that is
independent of $\Qmax^2$ within errors, for $\Qmax\ge \sqrt2\,\gev$. For
$f{=}s,c$, where the lattice uncertainties are below one percent, this
independence is no longer true within errors. However, the observed
variations are accounted for by our matching uncertainty and are small
compared to our total error on $\aelllohvp{\tau}$.

The observed overall independence on $\Qmax$ is an indication that our
continuum-limit, lattice results are consistent with $O(\alpha_s^4)$,
five-loop perturbation theory for $\Qmax\ge \sqrt2\,\gev$.  This result is
highly nontrivial: it provides evidence that we control the
high-virtuality contributions in our computation within our quoted
errors.  For $Q<\sqrt2\,\gev$, agreement with perturbation is less
good. Indeed, the results for $\Qmax{=} 1\,\gev$ suggest that
perturbation theory is beginning to break down at these low scales.

\subsection{Calculation of $\dpert\aelllohvpf{\ell}{f}( Q{>}\Qmax)$}
\label{sec:ptcontrib}

The contributions $\dpert\aelllohvpf{\ell}{f}( Q{>}\Qmax)$, defined in
\eq{eq:aellpert}, can be computed in QCD perturbation theory for
$\Qmax$ sufficiently large, because they are short distance
quantities. For that purpose, we use the $O(\alpha_s^4)$ result for the
$e^+e^-$, $R$-ratio~\cite{Harlander:2002ur} and the optical theorem:
\be
\label{eq:rhad}
\im\Pi^f_\text{pert}(s) = -\frac{R^f_\text{pert}(s)}{12\pi},\quad R^f_\text{pert}(s)\equiv\frac{\sigma(e^+e^-\to f\bar f X)}{4\pi\alpha^2/(3s)}\ ,
\ee
where $s$ is the center-of-mass energy, $f$ in $e^+e^-\to f\bar f X$
stands for a quark of flavor $f$ and $X$ can be gluons with, possibly,
radiated quark pairs. For $f{=}\text{disc}$, which corresponds to the
quark disconnected contribution, there are only gluon final states, which connect
to the electromagnetic current through a quark loop. In the definition of $R^f_\text{pert}$, the
denominator is the tree-level cross-section for $e^+e^-\to \mu^+\mu^-$
in the limit $s\gg m_\mu^2$.  We then obtain the difference
$[\Pi_\text{pert}^f(Q^2)-\Pi_\text{pert}^f(\Qmax^2)]$, needed in
\eq{eq:aellpert}, through the once subtracted dispersion relation:
\bea
&&\Pi_\text{pert}^f(Q^2)-\Pi_\text{pert}^f(\Qmax^2)\\
&&=\frac{Q^2-\Qmax^2}{12\pi^2}\int_0^\infty ds\;\frac1{(s+Q^2)(s+\Qmax^2)}R^f_\text{pert}(s)\ .\nn
\eea
We compute this difference numerically, by feeding into the above
dispersion relation, the four-loop output for $R^f_\text{pert}(s)$
given by the routine \texttt{rhad} of \cite{Harlander:2002ur}. The
result for the difference is then redirected into \eq{eq:aellpert}
and, again, integrated numerically. In \texttt{rhad}, we fix the
strong coupling to its $\msbar$ value at renormalization scale
$\mu{=}\Qmax$ and use the physical values of the pole quark
masses. Because the bottom and top contributions are negligible
compared to our final errors and because we have not accounted for
them in the lattice simulations from which we get the low $Q^2$
contributions, we neglect them here. The disconnected contributions
only appear at $O(\alpha_s^3)$ and vanish for the combined $u$, $d$
and $s$ terms, because the sum of these quarks' charges vanishes and
because the effects of their masses are neglected in
\texttt{rhad}. Moreover, the combined $c$, $b$ and $t$
quark-disconnected contributions are at least four orders of magnitude
smaller than the perturbative corrections which we keep. Thus, we
neglect disconnected contributions altogether. We do not include the
tiny perturbative QED corrections, because they are also significantly
smaller than our final errors and because we account for QED
corrections separately as discussed below in \sec{sec:QEDIB}. Finally,
for $\Qmax\ge 1\,\gev$, the perturbative corrections to
$\aelllohvpf{e}{f}(Q{\le}\Qmax)$ are all smaller than $10^{-19}$ and
can therefore be safely ignored.

The $O(\alpha_s^4)$ results for $\dpert\aelllohvpf{\ell}{f}( Q{>}\Qmax)$
that we obtain are summarized in \tab{tab:aellpert}. Only those for
$f{=}ud,\,c$ and $\ell{=}\mu$, $\tau$ are shown. Because the $u$, $d$
and $s$ masses are neglected in \texttt{rhad}, which is reasonable
given the $\Qmax$ that we consider, we have
$\dpert\aelllohvpf{\ell}{s}\simeq\frac15\dpert\aelllohvpf{\ell}{ud}$,
$\dpert\aelllohvpf{\ell}{I{=}1}{=}\frac9{10}\dpert\aelllohvpf{\ell}{ud}$
and
$\dpert\aelllohvpf{\ell}{I{=}0}\simeq\frac{11}{10}\dpert\aelllohvpf{\ell}{ud}+\dpert\aelllohvpf{\ell}{c}$.
All other perturbative corrections can be neglected.

\begin{table}[t]
    \centering
    \begin{ruledtabular}
	\begin{tabular}{LCC}
	    (f,\, \Qmax^2) & \ell=\mu\; \mbox{(units of $10^{-10}$)} & \ell=\tau\; \mbox{(units of $10^{-8}$)}\\
	    \hline\\[-0.3cm]
            (ud,\, 1,\gev^2) & 1.09  & 69.1\\
            (c,\, 1\,\gev^2) & 0.12   & 11.5\\
	    \hline\\[-0.3cm]
            (ud,\, 2\,\gev^2) & 0.26   & 34.9\\
            (c,\, 2\,\gev^2) & 0.05   & 7.9\\
	    \hline\\[-0.3cm]
            (ud,\, 3\,\gev^2) & 0.11   & 22.2\\
            (c,\, 3\,\gev^2) & 0.03   & 6.1\\
	    \hline\\[-0.3cm]
            (ud,\, 4\,\gev^2) & 0.06   & 15.6\\
            (c,\, 4\,\gev^2) & 0.02   & 4.9\\
	    \hline\\[-0.3cm]
            (ud,\, 5\,\gev^2) & 0.04   & 11.7\\
            (c,\, 5,\gev^2) & 0.01   & 4.1\\
	\end{tabular}
    \end{ruledtabular}
    \caption
        { \label{tab:aellpert} Five-loop, $O(\alpha_s^4)$ perturbative
          results for the contribution,
          $\dpert\aelllohvpf{\ell}{f}(Q{>}\Qmax)$ of \eq{eq:aellpert},
          to the anomalous magnetic moments of the leptons $\ell{=}\mu$
          and $\tau$, coming from momenta $Q{>}\Qmax$ for
          $\Qmax^2{=}1,\cdots,5\,\gev^2$. Results are given for the
          connected $ud$ and $c$ contributions. All other
          contributions can either be obtained from those two or are
          completely negligible, as explained in the text.}
\end{table}

\subsection{Finite-volume corrections}
\label{sec:FV}

As explained in the main text and in \cite{Aubin:2015rzx}, the only
finite-volume (FV) effects relevant in our percent-level calculation
on lattices with spatial extents $\gsim 6\,\fm$ are assumed to be
those associated with the two-pion contribution in the $I{=}1$
channel. We compute them to leading order in SU(2) chiral perturbation
theory ($\chi$PT), as suggested in \cite{Aubin:2015rzx}. We study them
for the scalar, $A_1$ representation of the cubic group that is
appropriate for our determination of the HVP scalar function from the
correlator $\frac13\sum_{i=1}^3C_{ii}^{I{=}1}(t)$. We do not take into
account discretization effects and, in particular, taste splittings,
since we correct our continuum-extrapolated results.

\begin{table}[t]
    \centering
    \begin{ruledtabular}
	\begin{tabular}{LC}
	    \Qmax^2\,[\gev^2] & \ell=\tau\,\mbox{(units of $10^{-8}$)} \\
	    \hline\\[-0.3cm]
            1 & 0.94 \\
            2 & 1.24\\
            3 & 1.39\\
            4 & 1.49\\
            5 & 1.55\\
	\end{tabular}
    \end{ruledtabular}
    \caption{ \label{tab:aelllatFV} One-loop, SU(2) $\chi$PT estimate
      of the FV corrections,
      $\dfv\aelllohvpf{\tau}{I{=}1}(Q{\le}\Qmax)$, that have to be added
      to our lattice results for $\aelllohvpf{\tau}{I{=}1}(Q{\le}\Qmax)$
      to correct for the fact that our simulations are performed in a
      finite volume. The results for
      $\dfv\aelllohvpf{\ell}{I{=}1}(Q{\le}\Qmax)$, $\ell{=}e$, $\mu$, and
      those for $\hat\Pi^{I{=}1}(\Qmax^2)$, which depend little on the
      momentum cut $\Qmax$, in the range of interest, are given in the
      text.}
\end{table}

\begin{table*}[t]
    \centering
    \begin{ruledtabular}
	\begin{tabular}{lCCC}
	    & \ell=e\; \mbox{(units of $10^{-14})$} & \ell=\mu\; \mbox{(units of $10^{-10}$)} & \ell=\tau\; \mbox{(units of $10^{-8}$)}\\
	  \hline\\[-0.3cm]
          $\pi^0\gamma$ & 1.05\pm 0.04 & 4.64\pm 0.04 & 1.77\pm 0.07\\
          $\eta\gamma$ & 0.14\pm 0.00 & 0.65\pm 0.01 & 0.29\pm 0.01\\
          \hline
          $\rho-\omega$ mixing & 0.74\pm 0.37 & 2.71\pm 1.36 & 0.72\pm 0.36 \\
          FSR & 1.17\pm 0.59  & 4.22\pm 2.11 & 1.40\pm 0.70\\
          $M_\pi$ vs $M_{\pi^\pm}$ & -1.45\pm 1.45 & -4.47\pm4.47 & -0.83\pm 0.83\\
          \hline
          total  & 1.7\pm 1.6 & 7.8\pm 5.1 & 3.4\pm 1.1\\
	\end{tabular}
    \end{ruledtabular}
    \caption
        { \label{tab:IB_corr} QED and $(m_d-m_u)$ corrections which
          must be added to our lattice results for the total LO-HVP
          contribution to the anomalous magnetic moments of the $e$,
          $\mu$ and $\tau$ to be able to compare
          them to those obtained from phenomenology. The individual
          corrections are described in the text. The total QED+$(m_d-m_u)$
          correction is given on the last line. }
\end{table*}

We compute the corrections numerically. We first construct the
lattice, position-space, pion propagator from the
fast-Fourier-transformed, momentum-space, scalar propagator.  Two such
propagators are then appropriately combined to give the $\pi\pi$
contribution to the correlator $C_{ii}^{I{=}1}(t)$.  The resulting
correlator is subsequently treated in the same way as the
corresponding lattice QCD, quark-antiquark correlator, to give the two-pion
contribution to $\aelllohvpf{\ell}{I{=}1}(Q{\le} \Qmax)$ and
$\hat\Pi^{I{=}1}(\Qmax^2)$. In particular, this means that the FV
effects associated with the interpolation in $Q$, that is described in
the paragraph preceding the ``Lattice details'' section of the main
text, are accounted for.  The bounding procedure is not implemented
because uncertainties associated with this procedure are estimated
independently.

To obtain these finite-volume corrections, we take $L{=}6\,\fm$ and
$T{=}3L/2$ which, within the uncertainties of these estimates,
describes all of our lattices accurately. The infinite-volume result
is taken to correspond to $T=L=13\,\fm$ and the relevant finite-volume
differences are extrapolated to the continuum limit in the scalar
theory to eliminate small discretization errors. Since we do not
have the lattice simulations to check the accuracy of these
predictions, we ascribe to them a 100\% error and thus treat them as
an order of magnitude estimate.

Because of the very weak dependence of $\aelllohvp{e}$ on $\Qmax$, the
corrections that must be added to our finite-volume result for $\aelllohvpf{e}{I{=}1}(Q{\le} \Qmax)$ are
$\dfv\aelllohvpf{e}{I{=}1}( Q{\le} \Qmax)=4.6\times 10^{-14}$,
independent of $\Qmax^2$ in the range $1\div 5\,\gev^2$. The FV
corrections to $\aelllohvpf{\mu}{I{=}1}(Q{\le} \Qmax)$ have a very
slight $\Qmax$ dependence. However, within the 100\% uncertainty that
we ascribe to these corrections, it is safe to take
$\dfv\aelllohvpf{\mu}{I{=}1}( Q{\le} \Qmax)=13.5\times 10^{-10}$ for
$\Qmax^2$ in the range $1\div 5\,\gev^2$. Only for those on
$\aelllohvpf{\tau}{I{=}1}(Q{\le} \Qmax)$ is the $\Qmax$ dependence
significant. Those corrections are given in \tab{tab:aelllatFV} for
the $\Qmax$ of interest. Note that these same, two-pion, FV
corrections also affect the $f{=}ud,\text{disc}$ contributions to
$\aelllohvp{\ell}(Q{\le} \Qmax)$. We have $\dfv\aelllohvpf{\ell}{ud}(
Q{\le} \Qmax){=}\frac{10}{9}\dfv\aelllohvpf{\ell}{I{=}1}( Q{\le} \Qmax)$
and $\dfv\aelllohvpf{\ell}{\text{disc}}(
Q{\le} \Qmax){=}-\frac{1}{9}\dfv\aelllohvpf{\ell}{I{=}1}( Q{\le} \Qmax)$.

Our results for $\hat\Pi^{I{=}1}(\Qmax^2)$ also suffer from FV
corrections associated with the two-pion contribution. The dependence
on $\Qmax^2$ of the corrections, that have to be added to our FV
results for $\hat\Pi^{I{=}1}(\Qmax^2)$, is mild in the range $1\div
5\,\gev^2$, with $\dfv\hat\Pi^{I{=}1}(\Qmax^2)=(2.3\div 2.5)\times
10^{-4}$. These same effects contribute to $\hat\Pi^{ud}(\Qmax^2)$
with a factor $\frac{10}9$ and to $\hat\Pi^{\text{disc}}(\Qmax^2)$
with a coefficient $-\frac19$. Here again we ascribe a 100\% error to
the estimates of these corrections.

Not surprisingly, the size of these FV corrections is larger, in
relative terms, for the lighter leptons, since their anomalous moments
are more sensitive to longer-distance physics. For the same reason,
they are also slightly larger for smaller values of $\Qmax$, mostly in
the case of the $\tau$ which is more sensitive to $\Qmax$.

\subsection{QED and isospin-breaking corrections}
\label{sec:QEDIB}

Our calculation is performed in the isospin limit, with $m_u{=}m_d$ and
$\alpha{=}0$.  Moreover, as explained in the main text, we tune our
lattice parameters to reproduce observables from an isospin-symmetric
world in which $m_u{=}m_d$ and $\alpha{=}0$.  Thus, our results for the
LO-HVP contribution to the lepton anomalous magnetic moments are
correct up to QED and $(m_d-m_u)$ effects that are proper to the HVP
function. To compare our results with those determined using
experimental cross sections, we must account for those isospin
breaking (IB) effects. Eventually, this should be done {\em ab initio}
with a lattice calculation. However, for the moment only exploratory calculations
are available \cite{Boyle:2017gzv,Giusti:2017jof,Chakraborty:2017tqp} and we resort here to phenomenology.

There are a variety of IB corrections to the lepton anomalous magnetic
moments.  The first, most easily quantifiable ones, are contributions
from final states in the dispersive approach which are obviously
absent in our calculation. These are the $\pi^0\gamma$ and
$\eta\gamma$ contributions, which can be taken directly from the
dispersive approach. Here we take the results from \cite{jeger201706}.

A second set of IB contributions, which we are obviously missing, is
final-state radiation (FSR). This can be determined by a combination
of data and point-particle, QED corrections. Because of the positivity
of the spectral function, this correction must be positive. Here we
use the results from \cite{jeger201706} and attribute to it a 50\% error.

The third set of IB contributions requires hadronic models to
estimate. The first of these is $\rho$-$\omega$ mixing, which we take
from \cite{jeger201706} and conservatively attribute to it a 50\%
error. The second has to do with other QED and $(m_d-m_u)$ effects on
the hadrons which contribute to the HVP. One that is clearly
identifiable is the fact that our charged pions have a mass of
$134.8(3)\,\mev$, which is lower than the physical
$M_{\pi^{\pm}}{=}139.57\,\mev$. In the dispersive language, this means
that in our calculation, the $\pi^+\pi^-$ threshold is lower than it
is in nature. The net effect is that our calculation overestimates the
HVP contribution to the lepton anomalous magnetic moments. This effect
is expected to be particularly pronounced in the case of the $e$ and
$\mu$, which are sensitive to low energies. To estimate this effect,
we take the difference between the $\pi^+\pi^-$ contributions to
$\aelllohvp{\ell}$ obtained to LO in $\chi$PT using $M_{\pi^\pm}$ and
$M_\pi$. We ascribe to it a 100\% uncertainty to cover other,
neglected effects.

In \tab{tab:IB_corr}, we give quantitative estimates of these
corrections to the LO-HVP contributions to the anomalous magnetic moments of all three
leptons. The resulting total IB correction is given on the last line
of \tab{tab:IB_corr} and included in our final results for the
LO-HVP contributions to the anomalous magnetic moments of leptons of \tab{tab:final} from the main text.

\begin{table*}[t]
    \centering
    \begin{ruledtabular}
	\begin{tabular}{LCCC}
	    f & \aelllohvpf{e}{f}\times 10^{14} & \aelllohvpf{\mu}{f}\times 10^{10} & \aelllohvpf{\tau}{f}\times 10^{8}\\
  	  \hline
          ud          & 174.3(2.6)(2.3)(0.0)(0.0)(1.4)(5.1) & 647.6(7.5)(8.0)(0.1)(0.0)(5.1)(15.0) & 281.3(0.8)(1.6)(0.0)(0.2)(1.3)(2.0) \\
          s           & 13.6(0.0)(0.1)(-)(0.0)(0.1)(-) & 53.7(0.0)(0.2)(-)(0.0)(0.4)(-)   & 36.1(0.0)(0.1)(-)(0.2)(0.2)(-) \\
          c           & 3.5(0.0)(0.0)(-)(0.0)(0.0)(-) & 14.7(0.0)(0.1)(-)(0.0)(0.1)(-)   & 22.6(0.0)(0.2)(-)(1.0)(0.1)(-) \\
          \text{disc} & -3.8(0.3)(0.1)(0.1)(0.0)(0.0)(0.1)  & -12.8(1.1)(0.5)(0.2)(0.0)(0.1)(1.5) & -2.4(0.1)(0.1)(0.0)(0.0)(0.0)(0.2) \\
	\end{tabular}
    \end{ruledtabular}
    \caption
        { \label{tab:aellf_final} Final lattice results for individual
          flavor contributions to the LO-HVP componenents of the
          anomalous magnetic moments of the $\ell{=}e$, $\mu$ and $\tau$
          leptons. Results for $\aelllohvpf{\ell}{I{=}0,1}$ and
          $\aelllohvp{\ell}$ are given in \tab{tab:final} of the main
          text. In the results presented, the first error bar is
          statistical, the second is the systematic uncertainty
          associated with the continuum extrapolation, the third with
          the bounding procedure described in \sec{sec:tcut} (where
          applicable), the fourth with the matching to perturbation
          theory discussed in \sec{sec:aellsep}, the fifth with the lattice spacing uncertainty
          discussed in \sec{sec:phypt} and the sixth, where
          applicable, with FV corrections. Dashes in error brackets
          indicate that the corresponding systematic error does not
          affect the result in question.}
\end{table*}

\subsection{Individual flavor contributions to $\aelllohvp{\ell}$ and comparison with other lattice QCD calculations}
\label{sec:latcomp}

We are now in a position to put together all of the ingredients of our
calculation, according to \eq{eq:aellsep} from the main text, to obtain the individual
flavor contributions to $\aelllohvp{\ell}$, for
$\ell{=}e,\mu,\tau$. Thus, we take from \tab{tab:aelllat} of
\sec{sec:aell_contlim} the lower vituality contributions,
$\aelllohvpf{\ell}{f}(Q{<}\Qmax)$, obtained through the continuum limit
of our lattice results. For the matching term,
$\gamma_\ell(\Qmax)\;\hat\Pi^f(\Qmax^2)$, we take the phase-space
factor, $\gamma_\ell(\Qmax)$ of \eq{eq:gamma_ell}, from
\tab{tab:gamma_ell}.  We obtain the nonperturbative quantity
$\Pi^f(\Qmax^2)$ as described in \sec{sec:pihatqmax} and take the
results from \tab{tab:pivsQmax}. Finally, the high virtuality,
perturbative contributions $\dpert\aelllohvpf{\ell}{f}(Q{>}\Qmax)$ are
computed as explained in \sec{sec:ptcontrib} and given in
\tab{tab:aellpert}. To the $f{=}ud,\text{disc}$ contributions, we have
to add the FV corrections discussed in \sec{sec:FV}.

Our final results for the individual flavor contributions to the LO
HVP component of the lepton anomalous magnetic moments are given in
\tab{tab:aellf_final}.  These results include systematic errors
associated with the continuum extrapolation, with our bounding
procedure for the $ud$ and disconnected contributions, with the
matching to perturbation theory and with FV effects. These
contributions are meant to be isospin limit quantities. Thus, we do
not apply any QED or $(m_d-m_u)$ corrections to them. These are
reserved for the total LO-HVP contribution, $\aelllohvp{\ell}$, given
in \tab{tab:final} of the main text.

\begin{figure}[h]
  \centering
  \includegraphics[width=0.9\columnwidth]{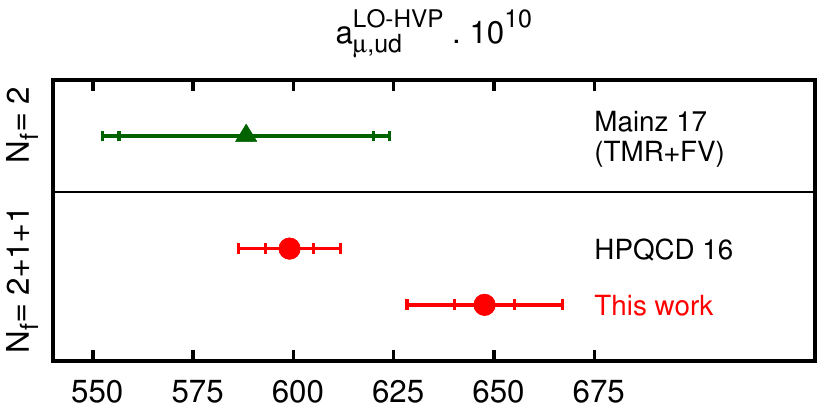}
  \includegraphics[width=0.9\columnwidth]{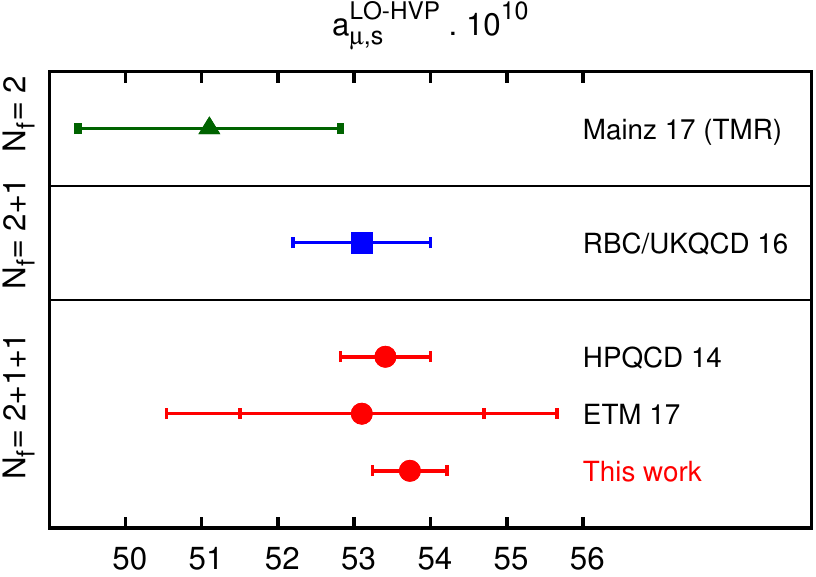}
  \includegraphics[width=0.9\columnwidth]{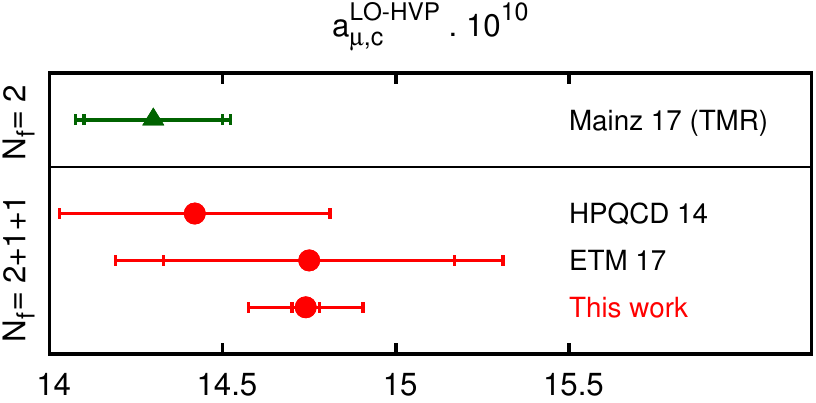}
  \includegraphics[width=0.9\columnwidth]{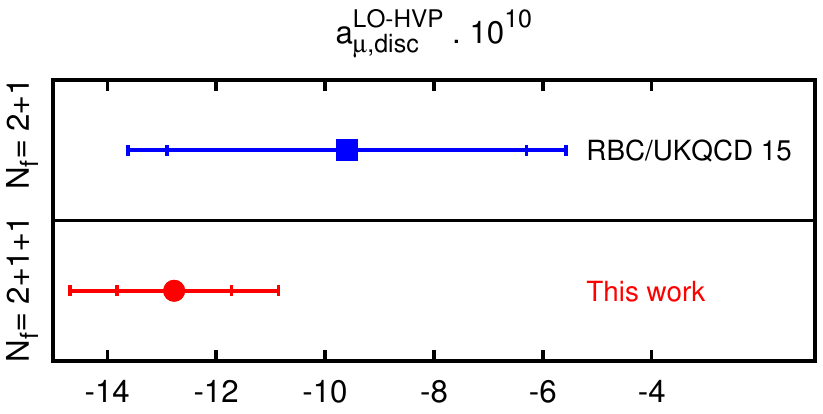}
    \caption
        {\label{fig:aellf_compare} Comparison of our results for the
          individual flavor contributions to $\aelllohvp{\mu}$ with
          those from other lattice collaborations. The reference for
          the results plotted are HPQCD 14 \cite{Chakraborty:2014mwa},
          RBC/UKQCD 15 \cite{Blum:2015you}, RBC/UKQCD 16
          \cite{Blum:2016xpd}, HPQCD 16 \cite{Chakraborty:2016mwy},
          Mainz 17 \cite{DellaMorte:2017dyu} and ETM
          \cite{Giusti:2017jof}. For Mainz 17 latter, TMR and TMR+FV
          stand for the particular results of Table~7 in that paper.}
\end{figure}

In \fig{fig:aellf_compare} we plot our results for the individual
flavor contributions $\aelllohvpf{\mu}{f}$, $f{=}ud,s,c,\text{disc}$,
together with the only other ones available from the lattice
\cite{Chakraborty:2014mwa,Blum:2015you,Chakraborty:2016mwy,Blum:2016xpd,DellaMorte:2017dyu}. Of
those, only the results of
\cite{Chakraborty:2014mwa,Chakraborty:2016mwy,Giusti:2017jof} are
obtained from $N_f{=}2+1+1$ simulations. Those of
\cite{DellaMorte:2017dyu} come from $N_f{=}2$ and those of
\cite{Blum:2015you,Blum:2016xpd} from $N_f{=}2+1$ simulations. For
some reason, \cite{Blum:2015you} do not include FV errors which are
2\% of the $I{=}0$ contribution in our calculation, and should be at
least as large in that reference.

As the figure shows, our $ud$ contribution to $\aelllohvp{\mu}$ is
significantly larger than the results of
\cite{Chakraborty:2016mwy,DellaMorte:2017dyu}. In particular, the
difference with the only other $N_f{=}2+1+1$ result published for this contribution is $2.2$ combined
standard deviations.  Our result for the charm contribution is
fully compatible with the two other lattice
results. Finally, our result for $\aelllohvpf{\mu}{\text{disc}}$ is
compatible with the only other determination \cite{Blum:2015you} and,
even with the inclusion of a FV uncertainty, it has a total error of
15\%.  This error represents 0.26\% of $\aelllohvp{\mu}$ which means
that it barely needs to be improved for determining $\aelllohvp{\mu}$
at the 0.2\% level, as will be required by future experiments.

\bibliography{a_ell.bbl}

\begin{thebibliography}{72}%
\makeatletter
\providecommand \@ifxundefined [1]{%
 \@ifx{#1\undefined}
}%
\providecommand \@ifnum [1]{%
 \ifnum #1\expandafter \@firstoftwo
 \else \expandafter \@secondoftwo
 \fi
}%
\providecommand \@ifx [1]{%
 \ifx #1\expandafter \@firstoftwo
 \else \expandafter \@secondoftwo
 \fi
}%
\providecommand \natexlab [1]{#1}%
\providecommand \enquote  [1]{``#1''}%
\providecommand \bibnamefont  [1]{#1}%
\providecommand \bibfnamefont [1]{#1}%
\providecommand \citenamefont [1]{#1}%
\providecommand \href@noop [0]{\@secondoftwo}%
\providecommand \href [0]{\begingroup \@sanitize@url \@href}%
\providecommand \@href[1]{\@@startlink{#1}\@@href}%
\providecommand \@@href[1]{\endgroup#1\@@endlink}%
\providecommand \@sanitize@url [0]{\catcode `\\12\catcode `\$12\catcode
  `\&12\catcode `\#12\catcode `\^12\catcode `\_12\catcode `\%12\relax}%
\providecommand \@@startlink[1]{}%
\providecommand \@@endlink[0]{}%
\providecommand \url  [0]{\begingroup\@sanitize@url \@url }%
\providecommand \@url [1]{\endgroup\@href {#1}{\urlprefix }}%
\providecommand \urlprefix  [0]{URL }%
\providecommand \Eprint [0]{\href }%
\providecommand \doibase [0]{http://dx.doi.org/}%
\providecommand \selectlanguage [0]{\@gobble}%
\providecommand \bibinfo  [0]{\@secondoftwo}%
\providecommand \bibfield  [0]{\@secondoftwo}%
\providecommand \translation [1]{[#1]}%
\providecommand \BibitemOpen [0]{}%
\providecommand \bibitemStop [0]{}%
\providecommand \bibitemNoStop [0]{.\EOS\space}%
\providecommand \EOS [0]{\spacefactor3000\relax}%
\providecommand \BibitemShut  [1]{\csname bibitem#1\endcsname}%
\let\auto@bib@innerbib\@empty
\bibitem [{\citenamefont {Gerlach}\ and\ \citenamefont
  {Stern}(1922)}]{Gerlach:1922ur}%
  \BibitemOpen
  \bibfield  {author} {\bibinfo {author} {\bibfnamefont {W.}~\bibnamefont
  {Gerlach}}\ and\ \bibinfo {author} {\bibfnamefont {O.}~\bibnamefont
  {Stern}},\ }\href {\doibase 10.1007/BF01329580} {\bibfield  {journal}
  {\bibinfo  {journal} {Z. Phys.}\ }\textbf {\bibinfo {volume} {8}},\ \bibinfo
  {pages} {110} (\bibinfo {year} {1922})}\BibitemShut {NoStop}%
\bibitem [{\citenamefont {Goudschmidt}\ and\ \citenamefont
  {Uhlenbeck}(1926)}]{Goudschmidt:1926ea}%
  \BibitemOpen
  \bibfield  {author} {\bibinfo {author} {\bibfnamefont {S.~A.}\ \bibnamefont
  {Goudschmidt}}\ and\ \bibinfo {author} {\bibfnamefont {G.~H.}\ \bibnamefont
  {Uhlenbeck}},\ }\href {\doibase 10.1038/117264a0} {\bibfield  {journal}
  {\bibinfo  {journal} {Nature}\ }\textbf {\bibinfo {volume} {117}},\ \bibinfo
  {pages} {264} (\bibinfo {year} {1926})}\BibitemShut {NoStop}%
\bibitem [{\citenamefont {Jegerlehner}\ and\ \citenamefont
  {Nyffeler}(2009)}]{Jegerlehner:2009ry}%
  \BibitemOpen
  \bibfield  {author} {\bibinfo {author} {\bibfnamefont {F.}~\bibnamefont
  {Jegerlehner}}\ and\ \bibinfo {author} {\bibfnamefont {A.}~\bibnamefont
  {Nyffeler}},\ }\href {\doibase 10.1016/j.physrep.2009.04.003} {\bibfield
  {journal} {\bibinfo  {journal} {Phys. Rept.}\ }\textbf {\bibinfo {volume}
  {477}},\ \bibinfo {pages} {1} (\bibinfo {year} {2009})},\ \Eprint
  {http://arxiv.org/abs/0902.3360} {arXiv:0902.3360 [hep-ph]} \BibitemShut
  {NoStop}%
\bibitem [{\citenamefont {Hanneke}\ \emph {et~al.}(2008)\citenamefont
  {Hanneke}, \citenamefont {Fogwell},\ and\ \citenamefont
  {Gabrielse}}]{Hanneke:2008tm}%
  \BibitemOpen
  \bibfield  {author} {\bibinfo {author} {\bibfnamefont {D.}~\bibnamefont
  {Hanneke}}, \bibinfo {author} {\bibfnamefont {S.}~\bibnamefont {Fogwell}}, \
  and\ \bibinfo {author} {\bibfnamefont {G.}~\bibnamefont {Gabrielse}},\ }\href
  {\doibase 10.1103/PhysRevLett.100.120801} {\bibfield  {journal} {\bibinfo
  {journal} {Phys. Rev. Lett.}\ }\textbf {\bibinfo {volume} {100}},\ \bibinfo
  {pages} {120801} (\bibinfo {year} {2008})},\ \Eprint
  {http://arxiv.org/abs/0801.1134} {arXiv:0801.1134 [physics.atom-ph]}
  \BibitemShut {NoStop}%
\bibitem [{\citenamefont {Aoyama}\ \emph {et~al.}(2012)\citenamefont {Aoyama},
  \citenamefont {Hayakawa}, \citenamefont {Kinoshita},\ and\ \citenamefont
  {Nio}}]{Aoyama:2012wk}%
  \BibitemOpen
  \bibfield  {author} {\bibinfo {author} {\bibfnamefont {T.}~\bibnamefont
  {Aoyama}}, \bibinfo {author} {\bibfnamefont {M.}~\bibnamefont {Hayakawa}},
  \bibinfo {author} {\bibfnamefont {T.}~\bibnamefont {Kinoshita}}, \ and\
  \bibinfo {author} {\bibfnamefont {M.}~\bibnamefont {Nio}},\ }\href {\doibase
  10.1103/PhysRevLett.109.111808} {\bibfield  {journal} {\bibinfo  {journal}
  {Phys. Rev. Lett.}\ }\textbf {\bibinfo {volume} {109}},\ \bibinfo {pages}
  {111808} (\bibinfo {year} {2012})},\ \Eprint {http://arxiv.org/abs/1205.5370}
  {arXiv:1205.5370 [hep-ph]} \BibitemShut {NoStop}%
\bibitem [{\citenamefont {Aoyama}\ \emph {et~al.}(2015)\citenamefont {Aoyama},
  \citenamefont {Hayakawa}, \citenamefont {Kinoshita},\ and\ \citenamefont
  {Nio}}]{Aoyama:2014sxa}%
  \BibitemOpen
  \bibfield  {author} {\bibinfo {author} {\bibfnamefont {T.}~\bibnamefont
  {Aoyama}}, \bibinfo {author} {\bibfnamefont {M.}~\bibnamefont {Hayakawa}},
  \bibinfo {author} {\bibfnamefont {T.}~\bibnamefont {Kinoshita}}, \ and\
  \bibinfo {author} {\bibfnamefont {M.}~\bibnamefont {Nio}},\ }\href {\doibase
  10.1103/PhysRevD.91.033006} {\bibfield  {journal} {\bibinfo  {journal} {Phys.
  Rev.}\ }\textbf {\bibinfo {volume} {D91}},\ \bibinfo {pages} {033006}
  (\bibinfo {year} {2015})},\ \Eprint {http://arxiv.org/abs/1412.8284}
  {arXiv:1412.8284 [hep-ph]} \BibitemShut {NoStop}%
\bibitem [{\citenamefont {Bennett}\ \emph {et~al.}(2006)\citenamefont {Bennett}
  \emph {et~al.}}]{Bennett:2006fi}%
  \BibitemOpen
  \bibfield  {author} {\bibinfo {author} {\bibfnamefont {G.~W.}\ \bibnamefont
  {Bennett}} \emph {et~al.} (\bibinfo {collaboration} {Muon g-2}),\ }\href
  {\doibase 10.1103/PhysRevD.73.072003} {\bibfield  {journal} {\bibinfo
  {journal} {Phys. Rev.}\ }\textbf {\bibinfo {volume} {D73}},\ \bibinfo {pages}
  {072003} (\bibinfo {year} {2006})},\ \Eprint
  {http://arxiv.org/abs/hep-ex/0602035} {arXiv:hep-ex/0602035 [hep-ex]}
  \BibitemShut {NoStop}%
\bibitem [{\citenamefont {Davier}\ \emph {et~al.}(2017)\citenamefont {Davier},
  \citenamefont {Hoecker}, \citenamefont {Malaescu},\ and\ \citenamefont
  {Zhang}}]{Davier:2017zfy}%
  \BibitemOpen
  \bibfield  {author} {\bibinfo {author} {\bibfnamefont {M.}~\bibnamefont
  {Davier}}, \bibinfo {author} {\bibfnamefont {A.}~\bibnamefont {Hoecker}},
  \bibinfo {author} {\bibfnamefont {B.}~\bibnamefont {Malaescu}}, \ and\
  \bibinfo {author} {\bibfnamefont {Z.}~\bibnamefont {Zhang}},\ }\href
  {\doibase 10.1140/epjc/s10052-017-5161-6} {\bibfield  {journal} {\bibinfo
  {journal} {Eur. Phys. J.}\ }\textbf {\bibinfo {volume} {C77}},\ \bibinfo
  {pages} {827} (\bibinfo {year} {2017})},\ \Eprint
  {http://arxiv.org/abs/1706.09436} {arXiv:1706.09436 [hep-ph]} \BibitemShut
  {NoStop}%
\bibitem [{\citenamefont {Holzbauer}(2016)}]{Holzbauer:2016cnd}%
  \BibitemOpen
  \bibfield  {author} {\bibinfo {author} {\bibfnamefont {J.~L.}\ \bibnamefont
  {Holzbauer}},\ }\bibfield  {booktitle} {\emph {\bibinfo {booktitle}
  {{Proceedings, 12th International Conference on Beauty, Charm, and Hyperons
  in Hadronic Interactions (BEACH 2016): Fairfax, Virginia, USA, June 12-18,
  2016}}},\ }\href {\doibase 10.1088/1742-6596/770/1/012038} {\bibfield
  {journal} {\bibinfo  {journal} {J. Phys. Conf. Ser.}\ }\textbf {\bibinfo
  {volume} {770}},\ \bibinfo {pages} {012038} (\bibinfo {year} {2016})},\
  \Eprint {http://arxiv.org/abs/1610.10069} {arXiv:1610.10069
  [physics.ins-det]} \BibitemShut {NoStop}%
\bibitem [{\citenamefont {Otani}(2015)}]{Otani:2015jra}%
  \BibitemOpen
  \bibfield  {author} {\bibinfo {author} {\bibfnamefont {M.}~\bibnamefont
  {Otani}} (\bibinfo {collaboration} {E34}),\ }\bibfield  {booktitle} {\emph
  {\bibinfo {booktitle} {{Proceedings, 2nd International Symposium on Science
  at J-PARC: Unlocking the Mysteries of Life, Matter and the Universe (J-PARC
  2014): Tsukuba, Japan, July 12-15, 2014}}},\ }\href {\doibase
  10.7566/JPSCP.8.025008} {\bibfield  {journal} {\bibinfo  {journal} {JPS Conf.
  Proc.}\ }\textbf {\bibinfo {volume} {8}},\ \bibinfo {pages} {025008}
  (\bibinfo {year} {2015})}\BibitemShut {NoStop}%
\bibitem [{\citenamefont {Fael}\ \emph {et~al.}(2014)\citenamefont {Fael},
  \citenamefont {Mercolli},\ and\ \citenamefont {Passera}}]{Fael:2013ij}%
  \BibitemOpen
  \bibfield  {author} {\bibinfo {author} {\bibfnamefont {M.}~\bibnamefont
  {Fael}}, \bibinfo {author} {\bibfnamefont {L.}~\bibnamefont {Mercolli}}, \
  and\ \bibinfo {author} {\bibfnamefont {M.}~\bibnamefont {Passera}},\
  }\bibfield  {booktitle} {\emph {\bibinfo {booktitle} {{Proceedings, 12th
  International Workshop on Tau Lepton Physics (TAU 2012): Nagoya, Japan,
  September 17-21, 2012}}},\ }\href {\doibase
  10.1016/j.nuclphysbps.2014.09.025} {\bibfield  {journal} {\bibinfo  {journal}
  {Nucl. Phys. Proc. Suppl.}\ }\textbf {\bibinfo {volume} {253-255}},\ \bibinfo
  {pages} {103} (\bibinfo {year} {2014})},\ \Eprint
  {http://arxiv.org/abs/1301.5302} {arXiv:1301.5302 [hep-ph]} \BibitemShut
  {NoStop}%
\bibitem [{\citenamefont {Oberhof}(2015)}]{Oberhof:2015hea}%
  \BibitemOpen
  \bibfield  {author} {\bibinfo {author} {\bibfnamefont {B.}~\bibnamefont
  {Oberhof}} (\bibinfo {collaboration} {BaBar}),\ }\bibfield  {booktitle}
  {\emph {\bibinfo {booktitle} {{Proceedings, 13th International Workshop on
  Tau Lepton Physics (TAU 2014): Aachen, Germany, September 15-19, 2014}}},\
  }\href {\doibase 10.1016/j.nuclphysbps.2015.02.003} {\bibfield  {journal}
  {\bibinfo  {journal} {Nucl. Part. Phys. Proc.}\ }\textbf {\bibinfo {volume}
  {260}},\ \bibinfo {pages} {12} (\bibinfo {year} {2015})},\ \Eprint
  {http://arxiv.org/abs/1502.01810} {arXiv:1502.01810 [hep-ex]} \BibitemShut
  {NoStop}%
\bibitem [{\citenamefont {Eidelman}\ and\ \citenamefont
  {Passera}(2007)}]{Eidelman:2007sb}%
  \BibitemOpen
  \bibfield  {author} {\bibinfo {author} {\bibfnamefont {S.}~\bibnamefont
  {Eidelman}}\ and\ \bibinfo {author} {\bibfnamefont {M.}~\bibnamefont
  {Passera}},\ }\href {\doibase 10.1142/S0217732307022694} {\bibfield
  {journal} {\bibinfo  {journal} {Mod. Phys. Lett.}\ }\textbf {\bibinfo
  {volume} {A22}},\ \bibinfo {pages} {159} (\bibinfo {year} {2007})},\ \Eprint
  {http://arxiv.org/abs/hep-ph/0701260} {arXiv:hep-ph/0701260 [hep-ph]}
  \BibitemShut {NoStop}%
\bibitem [{\citenamefont {Jegerlehner}(2016)}]{Jegerlehner:2015stw}%
  \BibitemOpen
  \bibfield  {author} {\bibinfo {author} {\bibfnamefont {F.}~\bibnamefont
  {Jegerlehner}},\ }\bibfield  {booktitle} {\emph {\bibinfo {booktitle}
  {{Proceedings, Workshop on Flavour changing and conserving processes 2015
  (FCCP2015)}}},\ }\href {\doibase 10.1051/epjconf/201611801016} {\bibfield
  {journal} {\bibinfo  {journal} {EPJ Web Conf.}\ }\textbf {\bibinfo {volume}
  {118}},\ \bibinfo {pages} {01016} (\bibinfo {year} {2016})},\ \Eprint
  {http://arxiv.org/abs/1511.04473} {arXiv:1511.04473 [hep-ph]} \BibitemShut
  {NoStop}%
\bibitem [{\citenamefont {Eidelman}\ and\ \citenamefont
  {Jegerlehner}(1995)}]{Eidelman:1995ny}%
  \BibitemOpen
  \bibfield  {author} {\bibinfo {author} {\bibfnamefont {S.}~\bibnamefont
  {Eidelman}}\ and\ \bibinfo {author} {\bibfnamefont {F.}~\bibnamefont
  {Jegerlehner}},\ }\href {\doibase 10.1007/BF01553984} {\bibfield  {journal}
  {\bibinfo  {journal} {Z. Phys.}\ }\textbf {\bibinfo {volume} {C67}},\
  \bibinfo {pages} {585} (\bibinfo {year} {1995})},\ \Eprint
  {http://arxiv.org/abs/hep-ph/9502298} {arXiv:hep-ph/9502298 [hep-ph]}
  \BibitemShut {NoStop}%
\bibitem [{\citenamefont {Davier}\ \emph {et~al.}(2011)\citenamefont {Davier},
  \citenamefont {Hoecker}, \citenamefont {Malaescu},\ and\ \citenamefont
  {Zhang}}]{Davier:2010nc}%
  \BibitemOpen
  \bibfield  {author} {\bibinfo {author} {\bibfnamefont {M.}~\bibnamefont
  {Davier}}, \bibinfo {author} {\bibfnamefont {A.}~\bibnamefont {Hoecker}},
  \bibinfo {author} {\bibfnamefont {B.}~\bibnamefont {Malaescu}}, \ and\
  \bibinfo {author} {\bibfnamefont {Z.}~\bibnamefont {Zhang}},\ }\href
  {\doibase 10.1140/epjc/s10052-012-1874-8, 10.1140/epjc/s10052-010-1515-z}
  {\bibfield  {journal} {\bibinfo  {journal} {Eur. Phys. J.}\ }\textbf
  {\bibinfo {volume} {C71}},\ \bibinfo {pages} {1515} (\bibinfo {year}
  {2011})},\ \bibinfo {note} {[Erratum: Eur. Phys. J.C72,1874(2012)]},\ \Eprint
  {http://arxiv.org/abs/1010.4180} {arXiv:1010.4180 [hep-ph]} \BibitemShut
  {NoStop}%
\bibitem [{\citenamefont {Hagiwara}\ \emph {et~al.}(2011)\citenamefont
  {Hagiwara}, \citenamefont {Liao}, \citenamefont {Martin}, \citenamefont
  {Nomura},\ and\ \citenamefont {Teubner}}]{Hagiwara:2011af}%
  \BibitemOpen
  \bibfield  {author} {\bibinfo {author} {\bibfnamefont {K.}~\bibnamefont
  {Hagiwara}}, \bibinfo {author} {\bibfnamefont {R.}~\bibnamefont {Liao}},
  \bibinfo {author} {\bibfnamefont {A.~D.}\ \bibnamefont {Martin}}, \bibinfo
  {author} {\bibfnamefont {D.}~\bibnamefont {Nomura}}, \ and\ \bibinfo {author}
  {\bibfnamefont {T.}~\bibnamefont {Teubner}},\ }\href {\doibase
  10.1088/0954-3899/38/8/085003} {\bibfield  {journal} {\bibinfo  {journal} {J.
  Phys.}\ }\textbf {\bibinfo {volume} {G38}},\ \bibinfo {pages} {085003}
  (\bibinfo {year} {2011})},\ \Eprint {http://arxiv.org/abs/1105.3149}
  {arXiv:1105.3149 [hep-ph]} \BibitemShut {NoStop}%
\bibitem [{\citenamefont {Miller}\ \emph {et~al.}(2012)\citenamefont {Miller},
  \citenamefont {de~Rafael}, \citenamefont {Roberts},\ and\ \citenamefont
  {Stöckinger}}]{Miller:2012opa}%
  \BibitemOpen
  \bibfield  {author} {\bibinfo {author} {\bibfnamefont {J.~P.}\ \bibnamefont
  {Miller}}, \bibinfo {author} {\bibfnamefont {E.}~\bibnamefont {de~Rafael}},
  \bibinfo {author} {\bibfnamefont {B.~L.}\ \bibnamefont {Roberts}}, \ and\
  \bibinfo {author} {\bibfnamefont {D.}~\bibnamefont {Stöckinger}},\ }\href
  {\doibase 10.1146/annurev-nucl-031312-120340} {\bibfield  {journal} {\bibinfo
   {journal} {Ann. Rev. Nucl. Part. Sci.}\ }\textbf {\bibinfo {volume} {62}},\
  \bibinfo {pages} {237} (\bibinfo {year} {2012})}\BibitemShut {NoStop}%
\bibitem [{\citenamefont {Jegerlehner}(2017)}]{Jegerlehner:2017lbd}%
  \BibitemOpen
  \bibfield  {author} {\bibinfo {author} {\bibfnamefont {F.}~\bibnamefont
  {Jegerlehner}},\ }in\ \href@noop {} {\emph {\bibinfo {booktitle} {{KLOE-2
  workshop on $e^+e^-$ collider physics at 1 GeV, INFN-Laboratori Nazionali di
  Frascati, Italy, 26-28 October 2016}}}}\ (\bibinfo {year} {2017})\ \Eprint
  {http://arxiv.org/abs/1705.00263} {arXiv:1705.00263 [hep-ph]} \BibitemShut
  {NoStop}%
\bibitem [{\citenamefont {Keshavarzi}\ \emph {et~al.}(2018)\citenamefont
  {Keshavarzi}, \citenamefont {Nomura},\ and\ \citenamefont
  {Teubner}}]{Keshavarzi:2018mgv}%
  \BibitemOpen
  \bibfield  {author} {\bibinfo {author} {\bibfnamefont {A.}~\bibnamefont
  {Keshavarzi}}, \bibinfo {author} {\bibfnamefont {D.}~\bibnamefont {Nomura}},
  \ and\ \bibinfo {author} {\bibfnamefont {T.}~\bibnamefont {Teubner}},\ }\href
  {\doibase 10.1103/PhysRevD.97.114025} {\bibfield  {journal} {\bibinfo
  {journal} {Phys. Rev.}\ }\textbf {\bibinfo {volume} {D97}},\ \bibinfo {pages}
  {114025} (\bibinfo {year} {2018})},\ \Eprint
  {http://arxiv.org/abs/1802.02995} {arXiv:1802.02995 [hep-ph]} \BibitemShut
  {NoStop}%
\bibitem [{\citenamefont {Blum}(2003)}]{Blum:2002ii}%
  \BibitemOpen
  \bibfield  {author} {\bibinfo {author} {\bibfnamefont {T.}~\bibnamefont
  {Blum}},\ }\href {\doibase 10.1103/PhysRevLett.91.052001} {\bibfield
  {journal} {\bibinfo  {journal} {Phys. Rev. Lett.}\ }\textbf {\bibinfo
  {volume} {91}},\ \bibinfo {pages} {052001} (\bibinfo {year} {2003})},\
  \Eprint {http://arxiv.org/abs/hep-lat/0212018} {arXiv:hep-lat/0212018
  [hep-lat]} \BibitemShut {NoStop}%
\bibitem [{\citenamefont {Aubin}\ and\ \citenamefont
  {Blum}(2006)}]{Aubin:2006xv}%
  \BibitemOpen
  \bibfield  {author} {\bibinfo {author} {\bibfnamefont {C.}~\bibnamefont
  {Aubin}}\ and\ \bibinfo {author} {\bibfnamefont {T.}~\bibnamefont {Blum}},\
  }\href@noop {} {\bibfield  {journal} {\bibinfo  {journal} {Phys.Rev.D}\
  }\textbf {\bibinfo {volume} {75:114502,2007}} (\bibinfo {year} {2006})},\
  \Eprint {http://arxiv.org/abs/hep-lat/0608011} {hep-lat/0608011} \BibitemShut
  {NoStop}%
\bibitem [{\citenamefont {Feng}\ \emph {et~al.}(2011)\citenamefont {Feng},
  \citenamefont {Jansen}, \citenamefont {Petschlies},\ and\ \citenamefont
  {Renner}}]{Feng:2011zk}%
  \BibitemOpen
  \bibfield  {author} {\bibinfo {author} {\bibfnamefont {X.}~\bibnamefont
  {Feng}}, \bibinfo {author} {\bibfnamefont {K.}~\bibnamefont {Jansen}},
  \bibinfo {author} {\bibfnamefont {M.}~\bibnamefont {Petschlies}}, \ and\
  \bibinfo {author} {\bibfnamefont {D.~B.}\ \bibnamefont {Renner}},\ }\href
  {\doibase 10.1103/PhysRevLett.107.081802} {\bibfield  {journal} {\bibinfo
  {journal} {Phys.Rev.Lett.}\ }\textbf {\bibinfo {volume} {107}},\ \bibinfo
  {pages} {081802} (\bibinfo {year} {2011})},\ \Eprint
  {http://arxiv.org/abs/1103.4818} {arXiv:1103.4818 [hep-lat]} \BibitemShut
  {NoStop}%
\bibitem [{\citenamefont {Della~Morte}\ \emph {et~al.}(2012)\citenamefont
  {Della~Morte}, \citenamefont {Jager}, \citenamefont {Juttner},\ and\
  \citenamefont {Wittig}}]{DellaMorte:2011aa}%
  \BibitemOpen
  \bibfield  {author} {\bibinfo {author} {\bibfnamefont {M.}~\bibnamefont
  {Della~Morte}}, \bibinfo {author} {\bibfnamefont {B.}~\bibnamefont {Jager}},
  \bibinfo {author} {\bibfnamefont {A.}~\bibnamefont {Juttner}}, \ and\
  \bibinfo {author} {\bibfnamefont {H.}~\bibnamefont {Wittig}},\ }\href
  {\doibase 10.1007/JHEP03(2012)055} {\bibfield  {journal} {\bibinfo  {journal}
  {JHEP}\ }\textbf {\bibinfo {volume} {03}},\ \bibinfo {pages} {055} (\bibinfo
  {year} {2012})},\ \Eprint {http://arxiv.org/abs/1112.2894} {arXiv:1112.2894
  [hep-lat]} \BibitemShut {NoStop}%
\bibitem [{\citenamefont {Burger}\ \emph {et~al.}(2014)\citenamefont {Burger}
  \emph {et~al.}}]{Burger:2013jya}%
  \BibitemOpen
  \bibfield  {author} {\bibinfo {author} {\bibfnamefont {F.}~\bibnamefont
  {Burger}} \emph {et~al.} (\bibinfo {collaboration} {ETM}),\ }\href {\doibase
  10.1007/JHEP02(2014)099} {\bibfield  {journal} {\bibinfo  {journal} {JHEP}\
  }\textbf {\bibinfo {volume} {1402}},\ \bibinfo {pages} {099} (\bibinfo {year}
  {2014})},\ \Eprint {http://arxiv.org/abs/1308.4327} {arXiv:1308.4327
  [hep-lat]} \BibitemShut {NoStop}%
\bibitem [{\citenamefont {Blum}\ \emph {et~al.}(2012)\citenamefont {Blum},
  \citenamefont {Hayakawa},\ and\ \citenamefont {Izubuchi}}]{Blum:2013qu}%
  \BibitemOpen
  \bibfield  {author} {\bibinfo {author} {\bibfnamefont {T.}~\bibnamefont
  {Blum}}, \bibinfo {author} {\bibfnamefont {M.}~\bibnamefont {Hayakawa}}, \
  and\ \bibinfo {author} {\bibfnamefont {T.}~\bibnamefont {Izubuchi}},\
  }\bibfield  {booktitle} {\emph {\bibinfo {booktitle} {{Proceedings, 30th
  International Symposium on Lattice Field Theory (Lattice 2012): Cairns,
  Australia, June 24-29, 2012}}},\ }\href@noop {} {\bibfield  {journal}
  {\bibinfo  {journal} {PoS}\ }\textbf {\bibinfo {volume} {LATTICE2012}},\
  \bibinfo {pages} {022} (\bibinfo {year} {2012})},\ \Eprint
  {http://arxiv.org/abs/1301.2607} {arXiv:1301.2607 [hep-lat]} \BibitemShut
  {NoStop}%
\bibitem [{\citenamefont {Gregory}\ \emph {et~al.}(2014)\citenamefont
  {Gregory}, \citenamefont {Fodor}, \citenamefont {Hoelbling}, \citenamefont
  {Krieg}, \citenamefont {Lellouch}, \citenamefont {Malak}, \citenamefont
  {McNeile},\ and\ \citenamefont {Szabo}}]{Gregory:2013taa}%
  \BibitemOpen
  \bibfield  {author} {\bibinfo {author} {\bibfnamefont {E.~B.}\ \bibnamefont
  {Gregory}}, \bibinfo {author} {\bibfnamefont {Z.}~\bibnamefont {Fodor}},
  \bibinfo {author} {\bibfnamefont {C.}~\bibnamefont {Hoelbling}}, \bibinfo
  {author} {\bibfnamefont {S.}~\bibnamefont {Krieg}}, \bibinfo {author}
  {\bibfnamefont {L.}~\bibnamefont {Lellouch}}, \bibinfo {author}
  {\bibfnamefont {R.}~\bibnamefont {Malak}}, \bibinfo {author} {\bibfnamefont
  {C.}~\bibnamefont {McNeile}}, \ and\ \bibinfo {author} {\bibfnamefont
  {K.}~\bibnamefont {Szabo}} (\bibinfo {collaboration}
  {Budapest-Marseille-Wuppertal}),\ }\bibfield  {booktitle} {\emph {\bibinfo
  {booktitle} {{Proceedings, 31st International Symposium on Lattice Field
  Theory (Lattice 2013)}}},\ }\href@noop {} {\bibfield  {journal} {\bibinfo
  {journal} {PoS}\ }\textbf {\bibinfo {volume} {LATTICE2013}},\ \bibinfo
  {pages} {302} (\bibinfo {year} {2014})},\ \Eprint
  {http://arxiv.org/abs/1311.4446} {arXiv:1311.4446 [hep-lat]} \BibitemShut
  {NoStop}%
\bibitem [{\citenamefont {Malak}\ \emph {et~al.}(2015)\citenamefont {Malak},
  \citenamefont {Fodor}, \citenamefont {Hoelbling}, \citenamefont {Lellouch},
  \citenamefont {Sastre},\ and\ \citenamefont {Szabo}}]{Malak:2015sla}%
  \BibitemOpen
  \bibfield  {author} {\bibinfo {author} {\bibfnamefont {R.}~\bibnamefont
  {Malak}}, \bibinfo {author} {\bibfnamefont {Z.}~\bibnamefont {Fodor}},
  \bibinfo {author} {\bibfnamefont {C.}~\bibnamefont {Hoelbling}}, \bibinfo
  {author} {\bibfnamefont {L.}~\bibnamefont {Lellouch}}, \bibinfo {author}
  {\bibfnamefont {A.}~\bibnamefont {Sastre}}, \ and\ \bibinfo {author}
  {\bibfnamefont {K.}~\bibnamefont {Szabo}} (\bibinfo {collaboration}
  {Budapest-Marseille-Wuppertal}),\ }\bibfield  {booktitle} {\emph {\bibinfo
  {booktitle} {{Proceedings, 32nd International Symposium on Lattice Field
  Theory (Lattice 2014)}}},\ }\href@noop {} {\bibfield  {journal} {\bibinfo
  {journal} {PoS}\ }\textbf {\bibinfo {volume} {LATTICE2014}},\ \bibinfo
  {pages} {161} (\bibinfo {year} {2015})},\ \Eprint
  {http://arxiv.org/abs/1502.02172} {arXiv:1502.02172 [hep-lat]} \BibitemShut
  {NoStop}%
\bibitem [{\citenamefont {Chakraborty}\ \emph {et~al.}(2014)\citenamefont
  {Chakraborty}, \citenamefont {Davies}, \citenamefont {Donald}, \citenamefont
  {Dowdall}, \citenamefont {Koponen}, \citenamefont {Lepage},\ and\
  \citenamefont {and}}]{Chakraborty:2014mwa}%
  \BibitemOpen
  \bibfield  {author} {\bibinfo {author} {\bibfnamefont {B.}~\bibnamefont
  {Chakraborty}}, \bibinfo {author} {\bibfnamefont {C.}~\bibnamefont {Davies}},
  \bibinfo {author} {\bibfnamefont {G.}~\bibnamefont {Donald}}, \bibinfo
  {author} {\bibfnamefont {R.}~\bibnamefont {Dowdall}}, \bibinfo {author}
  {\bibfnamefont {J.}~\bibnamefont {Koponen}}, \bibinfo {author} {\bibfnamefont
  {G.}~\bibnamefont {Lepage}}, \ and\ \bibinfo {author} {\bibfnamefont {T.~T.}\
  \bibnamefont {and}} (\bibinfo {collaboration} {HPQCD}),\ }\href {\doibase
  10.1103/PhysRevD.89.114501} {\bibfield  {journal} {\bibinfo  {journal}
  {Phys.Rev.}\ }\textbf {\bibinfo {volume} {D89}},\ \bibinfo {pages} {114501}
  (\bibinfo {year} {2014})},\ \Eprint {http://arxiv.org/abs/1403.1778}
  {arXiv:1403.1778 [hep-lat]} \BibitemShut {NoStop}%
\bibitem [{\citenamefont {Blum}\ \emph
  {et~al.}(2016{\natexlab{a}})\citenamefont {Blum}, \citenamefont {Boyle},
  \citenamefont {Izubuchi}, \citenamefont {Jin}, \citenamefont {Jüttner},
  \citenamefont {Lehner}, \citenamefont {Maltman}, \citenamefont {Marinkovic},
  \citenamefont {Portelli},\ and\ \citenamefont {Spraggs}}]{Blum:2015you}%
  \BibitemOpen
  \bibfield  {author} {\bibinfo {author} {\bibfnamefont {T.}~\bibnamefont
  {Blum}}, \bibinfo {author} {\bibfnamefont {P.~A.}\ \bibnamefont {Boyle}},
  \bibinfo {author} {\bibfnamefont {T.}~\bibnamefont {Izubuchi}}, \bibinfo
  {author} {\bibfnamefont {L.}~\bibnamefont {Jin}}, \bibinfo {author}
  {\bibfnamefont {A.}~\bibnamefont {Jüttner}}, \bibinfo {author}
  {\bibfnamefont {C.}~\bibnamefont {Lehner}}, \bibinfo {author} {\bibfnamefont
  {K.}~\bibnamefont {Maltman}}, \bibinfo {author} {\bibfnamefont
  {M.}~\bibnamefont {Marinkovic}}, \bibinfo {author} {\bibfnamefont
  {A.}~\bibnamefont {Portelli}}, \ and\ \bibinfo {author} {\bibfnamefont
  {M.}~\bibnamefont {Spraggs}},\ }\href {\doibase
  10.1103/PhysRevLett.116.232002} {\bibfield  {journal} {\bibinfo  {journal}
  {Phys. Rev. Lett.}\ }\textbf {\bibinfo {volume} {116}},\ \bibinfo {pages}
  {232002} (\bibinfo {year} {2016}{\natexlab{a}})},\ \Eprint
  {http://arxiv.org/abs/1512.09054} {arXiv:1512.09054 [hep-lat]} \BibitemShut
  {NoStop}%
\bibitem [{\citenamefont {Chakraborty}\ \emph {et~al.}(2016)\citenamefont
  {Chakraborty}, \citenamefont {Davies}, \citenamefont {Koponen}, \citenamefont
  {Lepage}, \citenamefont {Peardon},\ and\ \citenamefont
  {Ryan}}]{Chakraborty:2015ugp}%
  \BibitemOpen
  \bibfield  {author} {\bibinfo {author} {\bibfnamefont {B.}~\bibnamefont
  {Chakraborty}}, \bibinfo {author} {\bibfnamefont {C.~T.~H.}\ \bibnamefont
  {Davies}}, \bibinfo {author} {\bibfnamefont {J.}~\bibnamefont {Koponen}},
  \bibinfo {author} {\bibfnamefont {G.~P.}\ \bibnamefont {Lepage}}, \bibinfo
  {author} {\bibfnamefont {M.~J.}\ \bibnamefont {Peardon}}, \ and\ \bibinfo
  {author} {\bibfnamefont {S.~M.}\ \bibnamefont {Ryan}},\ }\href {\doibase
  10.1103/PhysRevD.93.074509} {\bibfield  {journal} {\bibinfo  {journal} {Phys.
  Rev.}\ }\textbf {\bibinfo {volume} {D93}},\ \bibinfo {pages} {074509}
  (\bibinfo {year} {2016})},\ \Eprint {http://arxiv.org/abs/1512.03270}
  {arXiv:1512.03270 [hep-lat]} \BibitemShut {NoStop}%
\bibitem [{\citenamefont {Bali}\ and\ \citenamefont
  {Endrődi}(2015)}]{Bali:2015msa}%
  \BibitemOpen
  \bibfield  {author} {\bibinfo {author} {\bibfnamefont {G.}~\bibnamefont
  {Bali}}\ and\ \bibinfo {author} {\bibfnamefont {G.}~\bibnamefont
  {Endrődi}},\ }\href {\doibase 10.1103/PhysRevD.92.054506} {\bibfield
  {journal} {\bibinfo  {journal} {Phys. Rev.}\ }\textbf {\bibinfo {volume}
  {D92}},\ \bibinfo {pages} {054506} (\bibinfo {year} {2015})},\ \Eprint
  {http://arxiv.org/abs/1506.08638} {arXiv:1506.08638 [hep-lat]} \BibitemShut
  {NoStop}%
\bibitem [{\citenamefont {Blum}\ \emph
  {et~al.}(2016{\natexlab{b}})\citenamefont {Blum} \emph
  {et~al.}}]{Blum:2016xpd}%
  \BibitemOpen
  \bibfield  {author} {\bibinfo {author} {\bibfnamefont {T.}~\bibnamefont
  {Blum}} \emph {et~al.} (\bibinfo {collaboration} {RBC/UKQCD}),\ }\href
  {\doibase 10.1007/JHEP05(2017)034, 10.1007/JHEP04(2016)063} {\bibfield
  {journal} {\bibinfo  {journal} {JHEP}\ }\textbf {\bibinfo {volume} {04}},\
  \bibinfo {pages} {063} (\bibinfo {year} {2016}{\natexlab{b}})},\ \bibinfo
  {note} {[Erratum: JHEP05,034(2017)]},\ \Eprint
  {http://arxiv.org/abs/1602.01767} {arXiv:1602.01767 [hep-lat]} \BibitemShut
  {NoStop}%
\bibitem [{\citenamefont {Chakraborty}\ \emph {et~al.}(2017)\citenamefont
  {Chakraborty}, \citenamefont {Davies}, \citenamefont {de~Oliviera},
  \citenamefont {Koponen}, \citenamefont {Lepage},\ and\ \citenamefont {Van~de
  Water}}]{Chakraborty:2016mwy}%
  \BibitemOpen
  \bibfield  {author} {\bibinfo {author} {\bibfnamefont {B.}~\bibnamefont
  {Chakraborty}}, \bibinfo {author} {\bibfnamefont {C.~T.~H.}\ \bibnamefont
  {Davies}}, \bibinfo {author} {\bibfnamefont {P.~G.}\ \bibnamefont
  {de~Oliviera}}, \bibinfo {author} {\bibfnamefont {J.}~\bibnamefont
  {Koponen}}, \bibinfo {author} {\bibfnamefont {G.~P.}\ \bibnamefont {Lepage}},
  \ and\ \bibinfo {author} {\bibfnamefont {R.~S.}\ \bibnamefont {Van~de
  Water}},\ }\href {\doibase 10.1103/PhysRevD.96.034516} {\bibfield  {journal}
  {\bibinfo  {journal} {Phys. Rev.}\ }\textbf {\bibinfo {volume} {D96}},\
  \bibinfo {pages} {034516} (\bibinfo {year} {2017})},\ \Eprint
  {http://arxiv.org/abs/1601.03071} {arXiv:1601.03071 [hep-lat]} \BibitemShut
  {NoStop}%
\bibitem [{\citenamefont {Borsanyi}\ \emph {et~al.}(2017)\citenamefont
  {Borsanyi}, \citenamefont {Fodor}, \citenamefont {Kawanai}, \citenamefont
  {Krieg}, \citenamefont {Lellouch}, \citenamefont {Malak}, \citenamefont
  {Miura}, \citenamefont {Szabo}, \citenamefont {Torrero},\ and\ \citenamefont
  {Toth}}]{Borsanyi:2016lpl}%
  \BibitemOpen
  \bibfield  {author} {\bibinfo {author} {\bibfnamefont {S.}~\bibnamefont
  {Borsanyi}}, \bibinfo {author} {\bibfnamefont {Z.}~\bibnamefont {Fodor}},
  \bibinfo {author} {\bibfnamefont {T.}~\bibnamefont {Kawanai}}, \bibinfo
  {author} {\bibfnamefont {S.}~\bibnamefont {Krieg}}, \bibinfo {author}
  {\bibfnamefont {L.}~\bibnamefont {Lellouch}}, \bibinfo {author}
  {\bibfnamefont {R.}~\bibnamefont {Malak}}, \bibinfo {author} {\bibfnamefont
  {K.}~\bibnamefont {Miura}}, \bibinfo {author} {\bibfnamefont {K.~K.}\
  \bibnamefont {Szabo}}, \bibinfo {author} {\bibfnamefont {C.}~\bibnamefont
  {Torrero}}, \ and\ \bibinfo {author} {\bibfnamefont {B.}~\bibnamefont {Toth}}
  (\bibinfo {collaboration} {Budapest-Marseille-Wuppertal}),\ }\href {\doibase
  10.1103/PhysRevD.96.074507} {\bibfield  {journal} {\bibinfo  {journal} {Phys.
  Rev.}\ }\textbf {\bibinfo {volume} {D96}},\ \bibinfo {pages} {074507}
  (\bibinfo {year} {2017})},\ \Eprint {http://arxiv.org/abs/1612.02364}
  {arXiv:1612.02364 [hep-lat]} \BibitemShut {NoStop}%
\bibitem [{\citenamefont {Della~Morte}\ \emph {et~al.}(2017)\citenamefont
  {Della~Morte}, \citenamefont {Francis}, \citenamefont {Gülpers},
  \citenamefont {Herdoíza}, \citenamefont {von Hippel}, \citenamefont {Horch},
  \citenamefont {Jäger}, \citenamefont {Meyer}, \citenamefont {Nyffeler},\
  and\ \citenamefont {Wittig}}]{DellaMorte:2017dyu}%
  \BibitemOpen
  \bibfield  {author} {\bibinfo {author} {\bibfnamefont {M.}~\bibnamefont
  {Della~Morte}}, \bibinfo {author} {\bibfnamefont {A.}~\bibnamefont
  {Francis}}, \bibinfo {author} {\bibfnamefont {V.}~\bibnamefont {Gülpers}},
  \bibinfo {author} {\bibfnamefont {G.}~\bibnamefont {Herdoíza}}, \bibinfo
  {author} {\bibfnamefont {G.}~\bibnamefont {von Hippel}}, \bibinfo {author}
  {\bibfnamefont {H.}~\bibnamefont {Horch}}, \bibinfo {author} {\bibfnamefont
  {B.}~\bibnamefont {Jäger}}, \bibinfo {author} {\bibfnamefont {H.~B.}\
  \bibnamefont {Meyer}}, \bibinfo {author} {\bibfnamefont {A.}~\bibnamefont
  {Nyffeler}}, \ and\ \bibinfo {author} {\bibfnamefont {H.}~\bibnamefont
  {Wittig}},\ }\href {\doibase 10.1007/JHEP10(2017)020} {\bibfield  {journal}
  {\bibinfo  {journal} {Journal of High Energy Physics}\ } (\bibinfo {year}
  {2017}),\ 10.1007/JHEP10(2017)020},\ \Eprint
  {http://arxiv.org/abs/1705.01775v2} {arXiv:1705.01775v2 [hep-lat]}
  \BibitemShut {NoStop}%
\bibitem [{\citenamefont {Giusti}\ \emph {et~al.}()\citenamefont {Giusti},
  \citenamefont {Lubicz}, \citenamefont {Martinelli}, \citenamefont
  {Sanfilippo},\ and\ \citenamefont {Simula}}]{Giusti:2017jof}%
  \BibitemOpen
  \bibfield  {author} {\bibinfo {author} {\bibfnamefont {D.}~\bibnamefont
  {Giusti}}, \bibinfo {author} {\bibfnamefont {V.}~\bibnamefont {Lubicz}},
  \bibinfo {author} {\bibfnamefont {G.}~\bibnamefont {Martinelli}}, \bibinfo
  {author} {\bibfnamefont {F.}~\bibnamefont {Sanfilippo}}, \ and\ \bibinfo
  {author} {\bibfnamefont {S.}~\bibnamefont {Simula}},\ }\href@noop {} {\
  }\Eprint {http://arxiv.org/abs/1707.03019v2} {1707.03019v2} \BibitemShut
  {NoStop}%
\bibitem [{\citenamefont {Boyle}\ \emph {et~al.}(2017)\citenamefont {Boyle},
  \citenamefont {Gülpers}, \citenamefont {Harrison}, \citenamefont {Jüttner},
  \citenamefont {Lehner}, \citenamefont {Portelli},\ and\ \citenamefont
  {Sachrajda}}]{Boyle:2017gzv}%
  \BibitemOpen
  \bibfield  {author} {\bibinfo {author} {\bibfnamefont {P.}~\bibnamefont
  {Boyle}}, \bibinfo {author} {\bibfnamefont {V.}~\bibnamefont {Gülpers}},
  \bibinfo {author} {\bibfnamefont {J.}~\bibnamefont {Harrison}}, \bibinfo
  {author} {\bibfnamefont {A.}~\bibnamefont {Jüttner}}, \bibinfo {author}
  {\bibfnamefont {C.}~\bibnamefont {Lehner}}, \bibinfo {author} {\bibfnamefont
  {A.}~\bibnamefont {Portelli}}, \ and\ \bibinfo {author} {\bibfnamefont
  {C.~T.}\ \bibnamefont {Sachrajda}},\ }\href {\doibase
  10.1007/JHEP09(2017)153} {\bibfield  {journal} {\bibinfo  {journal} {Journal
  of High Energy Physics}\ }\textbf {\bibinfo {volume} {2017}} (\bibinfo {year}
  {2017}),\ 10.1007/JHEP09(2017)153},\ \Eprint
  {http://arxiv.org/abs/1706.05293v1} {1706.05293v1} \BibitemShut {NoStop}%
\bibitem [{\citenamefont {Chakraborty}\ \emph {et~al.}(2018)\citenamefont
  {Chakraborty} \emph {et~al.}}]{Chakraborty:2017tqp}%
  \BibitemOpen
  \bibfield  {author} {\bibinfo {author} {\bibfnamefont {B.}~\bibnamefont
  {Chakraborty}} \emph {et~al.} (\bibinfo {collaboration} {Fermilab Lattice,
  LATTICE-HPQCD, MILC}),\ }\href {\doibase 10.1103/PhysRevLett.120.152001}
  {\bibfield  {journal} {\bibinfo  {journal} {Phys. Rev. Lett.}\ }\textbf
  {\bibinfo {volume} {120}},\ \bibinfo {pages} {152001} (\bibinfo {year}
  {2018})},\ \Eprint {http://arxiv.org/abs/1710.11212} {arXiv:1710.11212
  [hep-lat]} \BibitemShut {NoStop}%
\bibitem [{\citenamefont {Colquhoun}\ \emph {et~al.}(2015)\citenamefont
  {Colquhoun}, \citenamefont {Dowdall}, \citenamefont {Davies}, \citenamefont
  {Hornbostel},\ and\ \citenamefont {Lepage}}]{Colquhoun:2014ica}%
  \BibitemOpen
  \bibfield  {author} {\bibinfo {author} {\bibfnamefont {B.}~\bibnamefont
  {Colquhoun}}, \bibinfo {author} {\bibfnamefont {R.~J.}\ \bibnamefont
  {Dowdall}}, \bibinfo {author} {\bibfnamefont {C.~T.~H.}\ \bibnamefont
  {Davies}}, \bibinfo {author} {\bibfnamefont {K.}~\bibnamefont {Hornbostel}},
  \ and\ \bibinfo {author} {\bibfnamefont {G.~P.}\ \bibnamefont {Lepage}},\
  }\href {\doibase 10.1103/PhysRevD.91.074514} {\bibfield  {journal} {\bibinfo
  {journal} {Phys. Rev.}\ }\textbf {\bibinfo {volume} {D91}},\ \bibinfo {pages}
  {074514} (\bibinfo {year} {2015})},\ \Eprint {http://arxiv.org/abs/1408.5768}
  {arXiv:1408.5768 [hep-lat]} \BibitemShut {NoStop}%
\bibitem [{\citenamefont {Burger}\ \emph {et~al.}(2016)\citenamefont {Burger},
  \citenamefont {Jansen}, \citenamefont {Petschlies},\ and\ \citenamefont
  {Pientka}}]{Burger:2015oya}%
  \BibitemOpen
  \bibfield  {author} {\bibinfo {author} {\bibfnamefont {F.}~\bibnamefont
  {Burger}}, \bibinfo {author} {\bibfnamefont {K.}~\bibnamefont {Jansen}},
  \bibinfo {author} {\bibfnamefont {M.}~\bibnamefont {Petschlies}}, \ and\
  \bibinfo {author} {\bibfnamefont {G.}~\bibnamefont {Pientka}},\ }\href
  {\doibase 10.1140/epjc/s10052-016-4307-2} {\bibfield  {journal} {\bibinfo
  {journal} {Eur. Phys. J.}\ }\textbf {\bibinfo {volume} {C76}},\ \bibinfo
  {pages} {464} (\bibinfo {year} {2016})},\ \Eprint
  {http://arxiv.org/abs/1501.05110} {arXiv:1501.05110 [hep-lat]} \BibitemShut
  {NoStop}%
\bibitem [{\citenamefont {Bernecker}\ and\ \citenamefont
  {Meyer}(2011)}]{Bernecker:2011gh}%
  \BibitemOpen
  \bibfield  {author} {\bibinfo {author} {\bibfnamefont {D.}~\bibnamefont
  {Bernecker}}\ and\ \bibinfo {author} {\bibfnamefont {H.~B.}\ \bibnamefont
  {Meyer}},\ }\href {\doibase 10.1140/epja/i2011-11148-6} {\bibfield  {journal}
  {\bibinfo  {journal} {Eur.Phys.J.}\ }\textbf {\bibinfo {volume} {A47}},\
  \bibinfo {pages} {148} (\bibinfo {year} {2011})},\ \Eprint
  {http://arxiv.org/abs/1107.4388} {1107.4388} \BibitemShut {NoStop}%
\bibitem [{\citenamefont {Spraggs}\ \emph {et~al.}(2016)\citenamefont
  {Spraggs}, \citenamefont {Boyle}, \citenamefont {Del~Debbio}, \citenamefont
  {Jüttner}, \citenamefont {Lehner}, \citenamefont {Maltman}, \citenamefont
  {Marinkovic},\ and\ \citenamefont {Portelli}}]{Spraggs:2016jcx}%
  \BibitemOpen
  \bibfield  {author} {\bibinfo {author} {\bibfnamefont {M.}~\bibnamefont
  {Spraggs}}, \bibinfo {author} {\bibfnamefont {P.}~\bibnamefont {Boyle}},
  \bibinfo {author} {\bibfnamefont {L.}~\bibnamefont {Del~Debbio}}, \bibinfo
  {author} {\bibfnamefont {A.}~\bibnamefont {Jüttner}}, \bibinfo {author}
  {\bibfnamefont {C.}~\bibnamefont {Lehner}}, \bibinfo {author} {\bibfnamefont
  {K.}~\bibnamefont {Maltman}}, \bibinfo {author} {\bibfnamefont
  {M.}~\bibnamefont {Marinkovic}}, \ and\ \bibinfo {author} {\bibfnamefont
  {A.}~\bibnamefont {Portelli}},\ }\bibfield  {booktitle} {\emph {\bibinfo
  {booktitle} {{Proceedings, 33rd International Symposium on Lattice Field
  Theory (Lattice 2015): Kobe, Japan, July 14-18, 2015}}},\ }\href@noop {}
  {\bibfield  {journal} {\bibinfo  {journal} {PoS}\ }\textbf {\bibinfo {volume}
  {LATTICE2015}},\ \bibinfo {pages} {106} (\bibinfo {year} {2016})},\ \Eprint
  {http://arxiv.org/abs/1601.00537} {arXiv:1601.00537 [hep-lat]} \BibitemShut
  {NoStop}%
\bibitem [{\citenamefont {Aubin}\ \emph {et~al.}(2012)\citenamefont {Aubin},
  \citenamefont {Blum}, \citenamefont {Golterman},\ and\ \citenamefont
  {Peris}}]{Aubin:2012me}%
  \BibitemOpen
  \bibfield  {author} {\bibinfo {author} {\bibfnamefont {C.}~\bibnamefont
  {Aubin}}, \bibinfo {author} {\bibfnamefont {T.}~\bibnamefont {Blum}},
  \bibinfo {author} {\bibfnamefont {M.}~\bibnamefont {Golterman}}, \ and\
  \bibinfo {author} {\bibfnamefont {S.}~\bibnamefont {Peris}},\ }\href
  {\doibase 10.1103/PhysRevD.86.054509} {\bibfield  {journal} {\bibinfo
  {journal} {Phys. Rev.}\ }\textbf {\bibinfo {volume} {D86}},\ \bibinfo {pages}
  {054509} (\bibinfo {year} {2012})},\ \Eprint {http://arxiv.org/abs/1205.3695}
  {arXiv:1205.3695 [hep-lat]} \BibitemShut {NoStop}%
\bibitem [{\citenamefont {Lautrup}\ \emph {et~al.}(1972)\citenamefont
  {Lautrup}, \citenamefont {Peterman},\ and\ \citenamefont
  {de~Rafael}}]{Lautrup:1971jf}%
  \BibitemOpen
  \bibfield  {author} {\bibinfo {author} {\bibfnamefont {B.~e.}\ \bibnamefont
  {Lautrup}}, \bibinfo {author} {\bibfnamefont {A.}~\bibnamefont {Peterman}}, \
  and\ \bibinfo {author} {\bibfnamefont {E.}~\bibnamefont {de~Rafael}},\ }\href
  {\doibase 10.1016/0370-1573(72)90011-7} {\bibfield  {journal} {\bibinfo
  {journal} {Phys. Rept.}\ }\textbf {\bibinfo {volume} {3}},\ \bibinfo {pages}
  {193} (\bibinfo {year} {1972})}\BibitemShut {NoStop}%
\bibitem [{\citenamefont {de~Rafael}(1994)}]{deRafael:1993za}%
  \BibitemOpen
  \bibfield  {author} {\bibinfo {author} {\bibfnamefont {E.}~\bibnamefont
  {de~Rafael}},\ }\href {\doibase 10.1016/0370-2693(94)91114-2} {\bibfield
  {journal} {\bibinfo  {journal} {Phys. Lett.}\ }\textbf {\bibinfo {volume}
  {B322}},\ \bibinfo {pages} {239} (\bibinfo {year} {1994})},\ \Eprint
  {http://arxiv.org/abs/hep-ph/9311316} {arXiv:hep-ph/9311316 [hep-ph]}
  \BibitemShut {NoStop}%
\bibitem [{SMP()}]{SMPRL17}%
  \BibitemOpen
  \href@noop {} {}\bibinfo {note} {Please see Supplemental
  Material}\BibitemShut {NoStop}%
\bibitem [{\citenamefont {de~Rafael}(2017)}]{deRafael:2017gay}%
  \BibitemOpen
  \bibfield  {author} {\bibinfo {author} {\bibfnamefont {E.}~\bibnamefont
  {de~Rafael}},\ }\href {\doibase 10.1103/PhysRevD.96.014510} {\bibfield
  {journal} {\bibinfo  {journal} {Phys. Rev.}\ }\textbf {\bibinfo {volume}
  {D96}},\ \bibinfo {pages} {014510} (\bibinfo {year} {2017})},\ \Eprint
  {http://arxiv.org/abs/1702.06783} {arXiv:1702.06783 [hep-ph]} \BibitemShut
  {NoStop}%
\bibitem [{\citenamefont {Dominguez}\ \emph {et~al.}(2017)\citenamefont
  {Dominguez}, \citenamefont {Schilcher},\ and\ \citenamefont
  {Spiesberger}}]{Dominguez:2017omw}%
  \BibitemOpen
  \bibfield  {author} {\bibinfo {author} {\bibfnamefont {C.~A.}\ \bibnamefont
  {Dominguez}}, \bibinfo {author} {\bibfnamefont {K.}~\bibnamefont
  {Schilcher}}, \ and\ \bibinfo {author} {\bibfnamefont {H.}~\bibnamefont
  {Spiesberger}},\ }\bibfield  {booktitle} {\emph {\bibinfo {booktitle}
  {{Proceedings, 31st Rencontres de Physique de La Vallée d'Aoste (La Thuile):
  La Thuile, Aosta , Italy, March 5-11, 2017}}},\ }\href {\doibase
  10.1393/ncc/i2017-17179-1} {\bibfield  {journal} {\bibinfo  {journal} {Nuovo
  Cim.}\ }\textbf {\bibinfo {volume} {C40}},\ \bibinfo {pages} {179} (\bibinfo
  {year} {2017})},\ \Eprint {http://arxiv.org/abs/1704.02843} {arXiv:1704.02843
  [hep-ph]} \BibitemShut {NoStop}%
\bibitem [{\citenamefont {Luscher}\ and\ \citenamefont
  {Weisz}(1985)}]{Luscher:1984xn}%
  \BibitemOpen
  \bibfield  {author} {\bibinfo {author} {\bibfnamefont {M.}~\bibnamefont
  {Luscher}}\ and\ \bibinfo {author} {\bibfnamefont {P.}~\bibnamefont
  {Weisz}},\ }\href {\doibase 10.1007/BF01206178} {\bibfield  {journal}
  {\bibinfo  {journal} {Commun. Math. Phys.}\ }\textbf {\bibinfo {volume}
  {97}},\ \bibinfo {pages} {59} (\bibinfo {year} {1985})},\ \bibinfo {note}
  {[Erratum: Commun. Math. Phys.98,433(1985)]}\BibitemShut {NoStop}%
\bibitem [{\citenamefont {Morningstar}\ and\ \citenamefont
  {Peardon}(2004)}]{Morningstar:2003gk}%
  \BibitemOpen
  \bibfield  {author} {\bibinfo {author} {\bibfnamefont {C.}~\bibnamefont
  {Morningstar}}\ and\ \bibinfo {author} {\bibfnamefont {M.~J.}\ \bibnamefont
  {Peardon}},\ }\href {\doibase 10.1103/PhysRevD.69.054501} {\bibfield
  {journal} {\bibinfo  {journal} {Phys. Rev.}\ }\textbf {\bibinfo {volume}
  {D69}},\ \bibinfo {pages} {054501} (\bibinfo {year} {2004})},\ \Eprint
  {http://arxiv.org/abs/hep-lat/0311018} {arXiv:hep-lat/0311018 [hep-lat]}
  \BibitemShut {NoStop}%
\bibitem [{\citenamefont {Davies}\ \emph {et~al.}(2010)\citenamefont {Davies},
  \citenamefont {McNeile}, \citenamefont {Wong}, \citenamefont {Follana},
  \citenamefont {Horgan}, \citenamefont {Hornbostel}, \citenamefont {Lepage},
  \citenamefont {Shigemitsu},\ and\ \citenamefont {Trottier}}]{Davies:2009ih}%
  \BibitemOpen
  \bibfield  {author} {\bibinfo {author} {\bibfnamefont {C.~T.~H.}\
  \bibnamefont {Davies}}, \bibinfo {author} {\bibfnamefont {C.}~\bibnamefont
  {McNeile}}, \bibinfo {author} {\bibfnamefont {K.~Y.}\ \bibnamefont {Wong}},
  \bibinfo {author} {\bibfnamefont {E.}~\bibnamefont {Follana}}, \bibinfo
  {author} {\bibfnamefont {R.}~\bibnamefont {Horgan}}, \bibinfo {author}
  {\bibfnamefont {K.}~\bibnamefont {Hornbostel}}, \bibinfo {author}
  {\bibfnamefont {G.~P.}\ \bibnamefont {Lepage}}, \bibinfo {author}
  {\bibfnamefont {J.}~\bibnamefont {Shigemitsu}}, \ and\ \bibinfo {author}
  {\bibfnamefont {H.}~\bibnamefont {Trottier}},\ }\href {\doibase
  10.1103/PhysRevLett.104.132003} {\bibfield  {journal} {\bibinfo  {journal}
  {Phys. Rev. Lett.}\ }\textbf {\bibinfo {volume} {104}},\ \bibinfo {pages}
  {132003} (\bibinfo {year} {2010})},\ \Eprint {http://arxiv.org/abs/0910.3102}
  {arXiv:0910.3102 [hep-ph]} \BibitemShut {NoStop}%
\bibitem [{\citenamefont {Clark}\ and\ \citenamefont
  {Kennedy}(2007)}]{Clark:2006fx}%
  \BibitemOpen
  \bibfield  {author} {\bibinfo {author} {\bibfnamefont {M.~A.}\ \bibnamefont
  {Clark}}\ and\ \bibinfo {author} {\bibfnamefont {A.~D.}\ \bibnamefont
  {Kennedy}},\ }\href {\doibase 10.1103/PhysRevLett.98.051601} {\bibfield
  {journal} {\bibinfo  {journal} {Phys. Rev. Lett.}\ }\textbf {\bibinfo
  {volume} {98}},\ \bibinfo {pages} {051601} (\bibinfo {year} {2007})},\
  \Eprint {http://arxiv.org/abs/hep-lat/0608015} {arXiv:hep-lat/0608015
  [hep-lat]} \BibitemShut {NoStop}%
\bibitem [{\citenamefont {Bellwied}\ \emph {et~al.}(2015)\citenamefont
  {Bellwied}, \citenamefont {Borsanyi}, \citenamefont {Fodor}, \citenamefont
  {Katz}, \citenamefont {Pasztor}, \citenamefont {Ratti},\ and\ \citenamefont
  {Szabo}}]{Bellwied:2015lba}%
  \BibitemOpen
  \bibfield  {author} {\bibinfo {author} {\bibfnamefont {R.}~\bibnamefont
  {Bellwied}}, \bibinfo {author} {\bibfnamefont {S.}~\bibnamefont {Borsanyi}},
  \bibinfo {author} {\bibfnamefont {Z.}~\bibnamefont {Fodor}}, \bibinfo
  {author} {\bibfnamefont {S.~D.}\ \bibnamefont {Katz}}, \bibinfo {author}
  {\bibfnamefont {A.}~\bibnamefont {Pasztor}}, \bibinfo {author} {\bibfnamefont
  {C.}~\bibnamefont {Ratti}}, \ and\ \bibinfo {author} {\bibfnamefont {K.~K.}\
  \bibnamefont {Szabo}},\ }\href {\doibase 10.1103/PhysRevD.92.114505}
  {\bibfield  {journal} {\bibinfo  {journal} {Phys. Rev.}\ }\textbf {\bibinfo
  {volume} {D92}},\ \bibinfo {pages} {114505} (\bibinfo {year} {2015})},\
  \Eprint {http://arxiv.org/abs/1507.04627} {arXiv:1507.04627 [hep-lat]}
  \BibitemShut {NoStop}%
\bibitem [{\citenamefont {Blum}\ \emph {et~al.}(2013)\citenamefont {Blum},
  \citenamefont {Izubuchi},\ and\ \citenamefont {Shintani}}]{Blum:2012uh}%
  \BibitemOpen
  \bibfield  {author} {\bibinfo {author} {\bibfnamefont {T.}~\bibnamefont
  {Blum}}, \bibinfo {author} {\bibfnamefont {T.}~\bibnamefont {Izubuchi}}, \
  and\ \bibinfo {author} {\bibfnamefont {E.}~\bibnamefont {Shintani}},\ }\href
  {\doibase 10.1103/PhysRevD.88.094503} {\bibfield  {journal} {\bibinfo
  {journal} {Phys.Rev.}\ }\textbf {\bibinfo {volume} {D88}},\ \bibinfo {pages}
  {094503} (\bibinfo {year} {2013})},\ \Eprint {http://arxiv.org/abs/1208.4349}
  {arXiv:1208.4349 [hep-lat]} \BibitemShut {NoStop}%
\bibitem [{\citenamefont {Francis}\ \emph {et~al.}(2014)\citenamefont
  {Francis}, \citenamefont {Gülpers}, \citenamefont {Jäger}, \citenamefont
  {Meyer}, \citenamefont {von Hippel} \emph {et~al.}}]{Francis:2014hoa}%
  \BibitemOpen
  \bibfield  {author} {\bibinfo {author} {\bibfnamefont {A.}~\bibnamefont
  {Francis}}, \bibinfo {author} {\bibfnamefont {V.}~\bibnamefont {Gülpers}},
  \bibinfo {author} {\bibfnamefont {B.}~\bibnamefont {Jäger}}, \bibinfo
  {author} {\bibfnamefont {H.}~\bibnamefont {Meyer}}, \bibinfo {author}
  {\bibfnamefont {G.}~\bibnamefont {von Hippel}},  \emph {et~al.},\ }\href@noop
  {} {\bibfield  {journal} {\bibinfo  {journal} {PoS}\ }\textbf {\bibinfo
  {volume} {LATTICE2014}},\ \bibinfo {pages} {128} (\bibinfo {year} {2014})},\
  \Eprint {http://arxiv.org/abs/1411.7592} {arXiv:1411.7592 [hep-lat]}
  \BibitemShut {NoStop}%
\bibitem [{\citenamefont {Harlander}\ and\ \citenamefont
  {Steinhauser}(2003)}]{Harlander:2002ur}%
  \BibitemOpen
  \bibfield  {author} {\bibinfo {author} {\bibfnamefont {R.~V.}\ \bibnamefont
  {Harlander}}\ and\ \bibinfo {author} {\bibfnamefont {M.}~\bibnamefont
  {Steinhauser}},\ }\href {\doibase 10.1016/S0010-4655(03)00204-2} {\bibfield
  {journal} {\bibinfo  {journal} {Comput. Phys. Commun.}\ }\textbf {\bibinfo
  {volume} {153}},\ \bibinfo {pages} {244} (\bibinfo {year} {2003})},\ \Eprint
  {http://arxiv.org/abs/hep-ph/0212294} {arXiv:hep-ph/0212294 [hep-ph]}
  \BibitemShut {NoStop}%
\bibitem [{\citenamefont {Aubin}\ \emph {et~al.}(2016)\citenamefont {Aubin},
  \citenamefont {Blum}, \citenamefont {Chau}, \citenamefont {Golterman},
  \citenamefont {Peris},\ and\ \citenamefont {Tu}}]{Aubin:2015rzx}%
  \BibitemOpen
  \bibfield  {author} {\bibinfo {author} {\bibfnamefont {C.}~\bibnamefont
  {Aubin}}, \bibinfo {author} {\bibfnamefont {T.}~\bibnamefont {Blum}},
  \bibinfo {author} {\bibfnamefont {P.}~\bibnamefont {Chau}}, \bibinfo {author}
  {\bibfnamefont {M.}~\bibnamefont {Golterman}}, \bibinfo {author}
  {\bibfnamefont {S.}~\bibnamefont {Peris}}, \ and\ \bibinfo {author}
  {\bibfnamefont {C.}~\bibnamefont {Tu}},\ }\href {\doibase
  10.1103/PhysRevD.93.054508} {\bibfield  {journal} {\bibinfo  {journal} {Phys.
  Rev.}\ }\textbf {\bibinfo {volume} {D93}},\ \bibinfo {pages} {054508}
  (\bibinfo {year} {2016})},\ \Eprint {http://arxiv.org/abs/1512.07555}
  {arXiv:1512.07555 [hep-lat]} \BibitemShut {NoStop}%
\bibitem [{\citenamefont {Francis}\ \emph {et~al.}(2013)\citenamefont
  {Francis}, \citenamefont {Jaeger}, \citenamefont {Meyer},\ and\ \citenamefont
  {Wittig}}]{Francis:2013qna}%
  \BibitemOpen
  \bibfield  {author} {\bibinfo {author} {\bibfnamefont {A.}~\bibnamefont
  {Francis}}, \bibinfo {author} {\bibfnamefont {B.}~\bibnamefont {Jaeger}},
  \bibinfo {author} {\bibfnamefont {H.~B.}\ \bibnamefont {Meyer}}, \ and\
  \bibinfo {author} {\bibfnamefont {H.}~\bibnamefont {Wittig}},\ }\href
  {\doibase 10.1103/PhysRevD.88.054502} {\bibfield  {journal} {\bibinfo
  {journal} {Phys.Rev.}\ }\textbf {\bibinfo {volume} {D88}},\ \bibinfo {pages}
  {054502} (\bibinfo {year} {2013})},\ \Eprint {http://arxiv.org/abs/1306.2532}
  {arXiv:1306.2532 [hep-lat]} \BibitemShut {NoStop}%
\bibitem [{\citenamefont {Blum}\ \emph {et~al.}(2018)\citenamefont {Blum},
  \citenamefont {Boyle}, \citenamefont {Gülpers}, \citenamefont {Izubuchi},
  \citenamefont {Jin}, \citenamefont {Jung}, \citenamefont {Jüttner},
  \citenamefont {Lehner}, \citenamefont {Portelli},\ and\ \citenamefont
  {Tsang}}]{Blum:2018mom}%
  \BibitemOpen
  \bibfield  {author} {\bibinfo {author} {\bibfnamefont {T.}~\bibnamefont
  {Blum}}, \bibinfo {author} {\bibfnamefont {P.~A.}\ \bibnamefont {Boyle}},
  \bibinfo {author} {\bibfnamefont {V.}~\bibnamefont {Gülpers}}, \bibinfo
  {author} {\bibfnamefont {T.}~\bibnamefont {Izubuchi}}, \bibinfo {author}
  {\bibfnamefont {L.}~\bibnamefont {Jin}}, \bibinfo {author} {\bibfnamefont
  {C.}~\bibnamefont {Jung}}, \bibinfo {author} {\bibfnamefont {A.}~\bibnamefont
  {Jüttner}}, \bibinfo {author} {\bibfnamefont {C.}~\bibnamefont {Lehner}},
  \bibinfo {author} {\bibfnamefont {A.}~\bibnamefont {Portelli}}, \ and\
  \bibinfo {author} {\bibfnamefont {J.~T.}\ \bibnamefont {Tsang}},\ }\href@noop
  {} {\  (\bibinfo {year} {2018})},\ \Eprint {http://arxiv.org/abs/1801.07224}
  {arXiv:1801.07224 [hep-lat]} \BibitemShut {NoStop}%
\bibitem [{\citenamefont {L\protect{\"u}scher}(1991)}]{Luscher:1991cf}%
  \BibitemOpen
  \bibfield  {author} {\bibinfo {author} {\bibfnamefont {M.}~\bibnamefont
  {L\protect{\"u}scher}},\ }\href {\doibase 10.1016/0550-3213(91)90584-K}
  {\bibfield  {journal} {\bibinfo  {journal} {Nucl.Phys.}\ }\textbf {\bibinfo
  {volume} {B364}},\ \bibinfo {pages} {237} (\bibinfo {year}
  {1991})}\BibitemShut {NoStop}%
\bibitem [{\citenamefont {Aoki}\ \emph {et~al.}(2017)\citenamefont {Aoki} \emph
  {et~al.}}]{Aoki:2016frl}%
  \BibitemOpen
  \bibfield  {author} {\bibinfo {author} {\bibfnamefont {S.}~\bibnamefont
  {Aoki}} \emph {et~al.},\ }\href {\doibase 10.1140/epjc/s10052-016-4509-7}
  {\bibfield  {journal} {\bibinfo  {journal} {Eur. Phys. J.}\ }\textbf
  {\bibinfo {volume} {C77}},\ \bibinfo {pages} {112} (\bibinfo {year}
  {2017})},\ \Eprint {http://arxiv.org/abs/1607.00299} {arXiv:1607.00299
  [hep-lat]} \BibitemShut {NoStop}%
\bibitem [{\citenamefont {Chakraborty}\ \emph {et~al.}(2015)\citenamefont
  {Chakraborty}, \citenamefont {Davies}, \citenamefont {Galloway},
  \citenamefont {Knecht}, \citenamefont {Koponen} \emph
  {et~al.}}]{Chakraborty:2014aca}%
  \BibitemOpen
  \bibfield  {author} {\bibinfo {author} {\bibfnamefont {B.}~\bibnamefont
  {Chakraborty}}, \bibinfo {author} {\bibfnamefont {C.}~\bibnamefont {Davies}},
  \bibinfo {author} {\bibfnamefont {B.}~\bibnamefont {Galloway}}, \bibinfo
  {author} {\bibfnamefont {P.}~\bibnamefont {Knecht}}, \bibinfo {author}
  {\bibfnamefont {J.}~\bibnamefont {Koponen}},  \emph {et~al.},\ }\href
  {\doibase 10.1103/PhysRevD.91.054508} {\bibfield  {journal} {\bibinfo
  {journal} {Phys.Rev.}\ }\textbf {\bibinfo {volume} {D91}},\ \bibinfo {pages}
  {054508} (\bibinfo {year} {2015})},\ \Eprint {http://arxiv.org/abs/1408.4169}
  {arXiv:1408.4169 [hep-lat]} \BibitemShut {NoStop}%
\bibitem [{\citenamefont {Gasser}\ and\ \citenamefont
  {Zarnauskas}(2010)}]{Gasser:2010wz}%
  \BibitemOpen
  \bibfield  {author} {\bibinfo {author} {\bibfnamefont {J.}~\bibnamefont
  {Gasser}}\ and\ \bibinfo {author} {\bibfnamefont {G.~R.~S.}\ \bibnamefont
  {Zarnauskas}},\ }\href {\doibase 10.1016/j.physletb.2010.08.021} {\bibfield
  {journal} {\bibinfo  {journal} {Phys. Lett.}\ }\textbf {\bibinfo {volume}
  {B693}},\ \bibinfo {pages} {122} (\bibinfo {year} {2010})},\ \Eprint
  {http://arxiv.org/abs/1008.3479} {arXiv:1008.3479 [hep-ph]} \BibitemShut
  {NoStop}%
\bibitem [{\citenamefont {Lüscher}(2010)}]{Luscher:2010iy}%
  \BibitemOpen
  \bibfield  {author} {\bibinfo {author} {\bibfnamefont {M.}~\bibnamefont
  {Lüscher}},\ }\href {\doibase 10.1007/JHEP08(2010)071,
  10.1007/JHEP03(2014)092} {\bibfield  {journal} {\bibinfo  {journal} {JHEP}\
  }\textbf {\bibinfo {volume} {1008}},\ \bibinfo {pages} {071} (\bibinfo {year}
  {2010})},\ \Eprint {http://arxiv.org/abs/1006.4518} {arXiv:1006.4518
  [hep-lat]} \BibitemShut {NoStop}%
\bibitem [{\citenamefont {Borsanyi}\ \emph {et~al.}(2012)\citenamefont
  {Borsanyi} \emph {et~al.}}]{Borsanyi:2012zs}%
  \BibitemOpen
  \bibfield  {author} {\bibinfo {author} {\bibfnamefont {S.}~\bibnamefont
  {Borsanyi}} \emph {et~al.} (\bibinfo {collaboration}
  {Budapest-Marseille-Wuppertal}),\ }\href {\doibase 10.1007/JHEP09(2012)010}
  {\bibfield  {journal} {\bibinfo  {journal} {JHEP}\ }\textbf {\bibinfo
  {volume} {09}},\ \bibinfo {pages} {010} (\bibinfo {year} {2012})},\ \Eprint
  {http://arxiv.org/abs/1203.4469} {arXiv:1203.4469 [hep-lat]} \BibitemShut
  {NoStop}%
\bibitem [{\citenamefont {Bell}\ and\ \citenamefont
  {de~Rafael}(1969)}]{Bell:1996md}%
  \BibitemOpen
  \bibfield  {author} {\bibinfo {author} {\bibfnamefont {J.~S.}\ \bibnamefont
  {Bell}}\ and\ \bibinfo {author} {\bibfnamefont {E.}~\bibnamefont
  {de~Rafael}},\ }\href {\doibase 10.1016/0550-3213(69)90250-8} {\bibfield
  {journal} {\bibinfo  {journal} {Nucl. Phys.}\ }\textbf {\bibinfo {volume}
  {B11}},\ \bibinfo {pages} {611} (\bibinfo {year} {1969})}\BibitemShut
  {NoStop}%
\bibitem [{\citenamefont {Rosner}\ \emph {et~al.}(2015)\citenamefont {Rosner},
  \citenamefont {Stone},\ and\ \citenamefont {Van~de Water}}]{Rosner:2015wva}%
  \BibitemOpen
  \bibfield  {author} {\bibinfo {author} {\bibfnamefont {J.~L.}\ \bibnamefont
  {Rosner}}, \bibinfo {author} {\bibfnamefont {S.}~\bibnamefont {Stone}}, \
  and\ \bibinfo {author} {\bibfnamefont {R.~S.}\ \bibnamefont {Van~de Water}},\
  }\href@noop {} {\bibfield  {journal} {\bibinfo  {journal} {Submitted to:
  Particle Data Book}\ } (\bibinfo {year} {2015})},\ \Eprint
  {http://arxiv.org/abs/1509.02220} {arXiv:1509.02220 [hep-ph]} \BibitemShut
  {NoStop}%
\bibitem [{\citenamefont {Patrignani}\ \emph {et~al.}(2016)\citenamefont
  {Patrignani} \emph {et~al.}}]{Patrignani:2016xqp}%
  \BibitemOpen
  \bibfield  {author} {\bibinfo {author} {\bibfnamefont {C.}~\bibnamefont
  {Patrignani}} \emph {et~al.} (\bibinfo {collaboration} {Particle Data
  Group}),\ }\href {\doibase 10.1088/1674-1137/40/10/100001} {\bibfield
  {journal} {\bibinfo  {journal} {Chin. Phys.}\ }\textbf {\bibinfo {volume}
  {C40}},\ \bibinfo {pages} {100001} (\bibinfo {year} {2016})}\BibitemShut
  {NoStop}%
\bibitem [{Note1()}]{Note1}%
  \BibitemOpen
  \bibinfo {note} {This analysis yields a value of $w_0$ which is compatible
  with the value used by HPQCD in \cite {Chakraborty:2016mwy},
  $w_0=0.1715(9)\protect \tmspace +\thinmuskip {.1667em}\protect \mathrm {fm}$
  \cite {Dowdall:2013rya}, within less than one combined standard deviation.
  Thus, the origin of the tension with HPQCD on $a_{\mu ,ud}^{\protect \text
  {LO-HVP}}$, discussed in Sec.~X of the SM, cannot attributed to a different
  value of $w_0$. In fact, this is confirmed by our excellent agreement with
  HPQCD on the connected strange and charm contributions to the HVP, which have
  the same sensitivity to lattice spacing as $a_{\mu ,ud}^{\protect \text
  {LO-HVP}}$. The reporting of a complete analysis of $w_0$ is left for a
  future publication.}\BibitemShut {Stop}%
\bibitem [{jeg()}]{jeger201706}%
  \BibitemOpen
  \href@noop {} {}\bibinfo {note} {F. Jegerlehner, {\em private communication,
  June 2017.}}\BibitemShut {Stop}%
\bibitem [{\citenamefont {Dowdall}\ \emph {et~al.}(2013)\citenamefont
  {Dowdall}, \citenamefont {Davies}, \citenamefont {Lepage},\ and\
  \citenamefont {McNeile}}]{Dowdall:2013rya}%
  \BibitemOpen
  \bibfield  {author} {\bibinfo {author} {\bibfnamefont {R.~J.}\ \bibnamefont
  {Dowdall}}, \bibinfo {author} {\bibfnamefont {C.~T.~H.}\ \bibnamefont
  {Davies}}, \bibinfo {author} {\bibfnamefont {G.~P.}\ \bibnamefont {Lepage}},
  \ and\ \bibinfo {author} {\bibfnamefont {C.}~\bibnamefont {McNeile}},\ }\href
  {\doibase 10.1103/PhysRevD.88.074504} {\bibfield  {journal} {\bibinfo
  {journal} {Phys. Rev.}\ }\textbf {\bibinfo {volume} {D88}},\ \bibinfo {pages}
  {074504} (\bibinfo {year} {2013})},\ \Eprint {http://arxiv.org/abs/1303.1670}
  {arXiv:1303.1670 [hep-lat]} \BibitemShut {NoStop}%
\end{thebibliography}%
\bibliographystyle{apsrev4-1}


\end{document}